\documentclass[longauth]{aa}

\usepackage{graphicx}
\usepackage{txfonts}
\usepackage[colorlinks=true,
            linkcolor=red,
            urlcolor=blue,
            citecolor=blue]{hyperref}

\begin{document}

    \title{Large Interferometer For Exoplanets (\emph{LIFE}):}%
    \subtitle{I. Improved exoplanet detection yield estimates for a large mid-infrared space-interferometer mission}
    \titlerunning{\emph{LIFE}: I. Improved exoplanet detection yield estimates for a MIR space interferometer}

    \author{S.P. Quanz\inst{\ref{eth},\ref{nccr}}\fnmsep\thanks{Correspondence: \href{mailto:sascha.quanz@phys.ethz.ch}{sascha.quanz@phys.ethz.ch}.} \and
            M.~Ottiger\inst{\ref{eth}} \and
            E.~Fontanet\inst{\ref{eth}} \and
            J.~Kammerer\inst{\ref{eso},\ref{anu},\ref{stsci}} \and
            F.~Menti\inst{\ref{eth}} \and
            F.~Dannert\inst{\ref{eth}} \and
            A.~Gheorghe\inst{\ref{eth}} \and
            O.~Absil\inst{\ref{liege}}\and
            V.S.~Airapetian\inst{\ref{goddard}}\and
            E.~Alei\inst{\ref{eth},\ref{nccr}}\and
            R.~Allart\inst{\ref{montreal}}\and
            D.~Angerhausen\inst{\ref{eth},\ref{nccr}} \and
            S.~Blumenthal\inst{\ref{oxford}}\and
            L.A.~Buchhave\inst{\ref{denmark}}\and
            J.~Cabrera\inst{\ref{dlr}}\and
            \'O.~Carri\'on-Gonz\'alez\inst{\ref{tuberlin}}\and
            G.~Chauvin\inst{\ref{grenoble}}\and
            W.C.~Danchi\inst{\ref{goddard}}\and
            C.~Dandumont\inst{\ref{csl}}\and
            D.~Defr\`ere\inst{\ref{leuven}}\and
            C.~Dorn\inst{\ref{uzh}}\and
            D.~Ehrenreich\inst{\ref{geneva}}\and
            S.~Ertel\inst{\ref{lbti},\ref{tucson}}
            M.~Fridlund\inst{\ref{leiden},\ref{onsala}}\and
            A.~Garc\'ia Mu\~noz\inst{\ref{tuberlin}}\and
            C.~Gasc\'on\inst{\ref{bellaterra}}\and
            J.~H.~Girard\inst{\ref{stsci}}\and
            A.~Glauser\inst{\ref{eth}} \and
            J.L.~Grenfell\inst{\ref{dlr}}\and
            G.~Guidi\inst{\ref{eth},\ref{nccr}} \and
            J.~Hagelberg\inst{\ref{geneva}}\and
            R.~Helled\inst{\ref{uzh}}\and
            M.J.~Ireland\inst{\ref{anu}}\and
            M.~Janson\inst{\ref{stockholm}}\and
            R.K.~Kopparapu\inst{\ref{goddard}}\and
            J.~Korth\inst{\ref{cologne}}\and
            T.~Kozakis\inst{\ref{denmark}}\and
            S.~Kraus\inst{\ref{exeter}}\and
            A.~L\'eger\inst{\ref{orsay}}\and
            L.~Leedj{\"a}rv\inst{\ref{estonia}}\and
            T.~Lichtenberg\inst{\ref{oxford}}\and
            J.~Lillo-Box\inst{\ref{esac}}\and
            H.~Linz\inst{\ref{mpia}}\and
            R. ~Liseau\inst{\ref{onsala}}\and
            J.~Loicq\inst{\ref{csl}}\and
            V.~Mahendra\inst{\ref{chennai}}\and
            F.~Malbet\inst{\ref{grenoble}}\and
            J.~Mathew\inst{\ref{anu}}\and
            B.~Mennesson\inst{\ref{jpl}}\and
            M.R.~Meyer\inst{\ref{michigan}}\and
            L.~Mishra\inst{\ref{bern},\ref{geneva},\ref{nccr}}\and
            K.~Molaverdikhani\inst{\ref{mpia},\ref{lsw}}\and
            L.~Noack\inst{\ref{fuberlin}} \and
            A.V.~Oza\inst{\ref{jpl},\ref{bern}}\and
            E.~Pall\'e\inst{\ref{tenerife1},\ref{tenerife2}}\and
            H.~Parviainen\inst{\ref{tenerife1},\ref{tenerife2}}\and
            A.~Quirrenbach\inst{\ref{lsw}}\and
            H.~Rauer\inst{\ref{dlr}}\and
            I.~Ribas\inst{\ref{bellaterra},\ref{barcelona}}\and
            M.~Rice\inst{\ref{yale}}\and
            A.~Romagnolo\inst{\ref{warsaw}}\and
            S.~Rugheimer\inst{\ref{oxford}}\and
            E.W.~Schwieterman\inst{\ref{riverside}}\and
            E.~Serabyn\inst{\ref{jpl}}\and
            S.~Sharma\inst{\ref{delhi}}\and
            K.G.~Stassun\inst{\ref{vanderbilt}}\and
            J.~Szul\'agyi\inst{\ref{eth}}\and
            H.S.~Wang\inst{\ref{eth},\ref{nccr}} \and
            F.~Wunderlich\inst{\ref{dlr}}\and
            M.C.~Wyatt\inst{\ref{cambridge}}\and
            the \emph{LIFE} Collaboration\inst{\ref{life}}
            }
    \authorrunning{Quanz et al.}
    \newpage
     \institute{
    ETH Zurich, Institute for Particle Physics \& Astrophysics, Wolfgang-Pauli-Str. 27, 8093 Zurich, Switzerland\label{eth}
    \and
    National Center of Competence in Research PlanetS
    (www.nccr-planets.ch)\label{nccr}
    \and
    European Southern Observatory, Karl-Schwarzschild-Str. 2, 85748 Garching, Germany\label{eso}
    \and
    Research School of Astronomy \& Astrophysics, Australian National University, ACT 2611, Australia\label{anu}
    \and
   STAR Institute, University of Li\`ege, 19C all\'ee du Six Ao\^ut, 4000 Li\`ege, Belgium\label{liege}
   \and
   NASA Goddard Space Flight Center, 8800 Greenbelt Rd, Greenbelt, MD, 20771, USA\label{goddard}
   \and
   Department of Physics, and Institute for Research on Exoplanets, Universit\'e de Montr\'eal, Montr\'eal, H3T 1J4, Canada\label{montreal}
   \and       
   University of Oxford, Department of Atmospheric, Oceanic and Planetary Physics, Clarendon Laboratory, Sherrington Road, Oxford OX1 3PU, United Kingdom\label{oxford}
   \and
   DTU Space, National Space Institute, Technical University of Denmark, Elektrovej 328, DK-2800 Kgs. Lyngby, Denmark\label{denmark}
   \and
   Department of Extrasolar Planets and Atmospheres, Institute for Planetary Research, German Aerospace Centre, Rutherfordstr. 2, 12489 Berlin\label{dlr}
   \and
   Zentrum f\"ur Astronomie und Astrophysik, Technische Universit\"at Berlin, Hardenbergstrasse 36, D-10623 Berlin, Germany\label{tuberlin}
   \and
   Univ. Grenoble Alpes, CNRS, IPAG, F-38000 Grenoble, France\label{grenoble}
   \and
   Centre Spatial de Li\`ege, Universit\'e de Li\`ege, Avenue Pr\'e-Aily, 4031 Angleur, Belgium\label{csl}
   \and
   Institute of Astronomy, KU Leuven, Celestijnenlaan 200D, 3001, Leuven, Belgium\label{leuven}
   \and
   University of Zurich, Institute of Computational Sciences, Winterthurerstrasse 190, 8057 Zurich, Switzerland\label{uzh}
   \and
   Observatoire astronomique de l'Universit\'e de Gen\`eve, chemin Pegasi 51b, 1290 Versoix, Switzerland\label{geneva}
   \and
   Large Binocular Telescope Observatory, 933 North Cherry Avenue, Tucson, AZ 85721, USA\label{lbti}
   \and
   Steward Observatory, Department of Astronomy, University of Arizona, 993 N. Cherry Ave, Tucson, AZ, 85721, USA\label{tucson}
   \and
   Leiden Observatory, Leiden University, 2333CA Leiden, The Netherlands\label{leiden}
   \and
   Department of Space, Earth \& Environment, Chalmers University of Technology, Onsala Space Observatory, 439 92 Onsala, Sweden\label{onsala}
   \and
   Institut de Ci\`encies de l'Espai (ICE, CSIC), Campus UAB, C/Can Magrans s/n, 08193 Bellaterra, Spain\label{bellaterra}
   \and
   Space Telescope Science Institute, 3700 San Martin Drive, Baltimore, MD 21218, USA\label{stsci}
   \and
   Department of Astronomy, Stockholm University, Alba Nova University Center, 10691 Stockholm, Sweden\label{stockholm}
   \and
   Department of Space, Earth and Environment, Astronomy and Plasma Physics, Chalmers University of Technology, SE-412 96 Gothenburg, Sweden\label{cologne}
   \and 
   University of Exeter, School of Physics and Astronomy, Stocker Road, Exeter, EX4 4QL, UK\label{exeter}
   \and
   IAS, CNRS (UMR 8617), bat 121, Univ. Paris-Sud, F-91405 Orsay, France\label{orsay}
   \and
   University of Tartu, Tartu Observatory, 1 Observatooriumi Str., 61602 T{\~o}ravere, Tartumaa, Estonia\label{estonia}
   \and
   Centro de Astrobiolog\'ia (CAB, CSIC-INTA), Depto. de Astrof\'isica, ESAC campus 28692 Villanueva de la Ca\~nada (Madrid), Spain\label{esac} 
   \and 
   Max-Planck-Institut f\"ur Astronomie, K\"onigstuhl 17, 69117 Heidelberg, Germany\label{mpia}
   \and
   SRM Institute of Science and Technology, Chennai, India\label{chennai}
   \and
   Jet Propulsion Laboratory, California Institute of Technology, 4800 Oak Grove Dr., Pasadena, CA 91109, USA\label{jpl}
   \and
   Department of Astronomy, University of Michigan, Ann Arbor, MI 48109, USA\label{michigan}
   \and
   Physikalisches Institut, Universit\"at Bern, Gesellschaftsstrasse 6, 3012 Bern, Switzerland\label{bern}  
   \and
    Landessternwarte, Zentrum f\"ur Astronomie der Universit\"at Heidelberg, K\"onigstuhl 12, 69117 Heidelberg, Germany\label{lsw}
    \and
    Freie Universit\"at Berlin, Department of Earth Sciences, Malteserstr. 74-100, 12249 Berlin, Germany\label{fuberlin}\newpage
    \and
    Instituto de Astrof\'isica de Canarias (IAC), E-38200 La Laguna, Tenerife, Spain\label{tenerife1}
    \and 
    Dept. Astrof\'isica, Universidad de La Laguna (ULL), E-38206 La Laguna, Tenerife, Spain\label{tenerife2}
    \and
    Institut d'Estudis Espacials de Catalunya (IEEC), C/Gran Capit\`a 2-4, 08034 Barcelona, Spain\label{barcelona}
    \and
    Department of Astronomy, Yale University, New Haven, CT 06511, USA\label{yale}
    \and
    Nicolaus Copernicus Astronomical Center, Polish Academy of Sciences, ul. Bartycka 18, 00-716 Warsaw, Poland\label{warsaw}
    \and
    Department of Earth and Planetary Sciences, University of California, Riverside, 900 University Ave. Riverside, CA, USA 92521\label{riverside}
    \and
    Deshbandhu College, University of Delhi 110019 Delhi, India\label{delhi}
    \and
    Vanderbilt University, Department of Physics \& Astronomy, 6301 Stevenson Center Ln., Nashville, TN 37235, USA\label{vanderbilt}
    \and
    Institute of Astronomy, University of Cambridge, Madingley Road, Cambridge CB3 0HA, UK\label{cambridge}
    \and
   \url{www.life-space-mission.com}\label{life}
    }
    
   \date{Received: <date> / Accepted: <date>}
    
    \abstract
    {One of the long-term goals of exoplanet science is the atmospheric characterization of dozens of small exoplanets in order to understand their diversity and search for habitable worlds and potential biosignatures. Achieving this goal requires a space mission of sufficient scale that can spatially separate the signals from exoplanets and their host stars and thus directly scrutinize the exoplanets and their atmospheres.}
    {We seek to quantify the exoplanet detection performance of a space-based mid-infrared (MIR) nulling interferometer that measures  the thermal emission of exoplanets. We study the impact of various parameters and compare the performance with that of large single-aperture mission concepts that detect exoplanets in reflected light.}
    {We have developed an instrument simulator that considers all major astrophysical noise sources and coupled it with Monte Carlo simulations of a synthetic exoplanet population around main-sequence stars within 20 pc of the Sun. This allows us to quantify the number (and types) of exoplanets that our mission concept could detect. Considering single visits only, we discuss two different scenarios for distributing 2.5 years of an initial search phase among the stellar targets. Different apertures sizes and wavelength ranges are investigated.}
    {An interferometer consisting of four 2~m apertures working in the 4--18.5~$\mu$m wavelength range with a total instrument throughput of 5\% could detect up to $\approx$550 exoplanets with radii between 0.5 and 6 R$_\oplus$ with an integrated S/N$\ge$7. At least  $\approx$160 of the detected exoplanets have radii $\le$1.5 R$_\oplus$. Depending on the observing scenario, $\approx$25--45 rocky exoplanets (objects with radii between 0.5 and 1.5 $_{\oplus}$) orbiting within the empirical habitable zone (eHZ) of their host stars are among the detections. With four 3.5~m apertures, the total number of detections can increase to up to $\approx$770, including $\approx$60--80 rocky eHZ planets. With four times 1~m apertures, the maximum detection yield is $\approx$315 exoplanets, including $\le$20 rocky eHZ planets. The vast majority of small, temperate exoplanets are detected around M dwarfs. The impact of changing the wavelength range to 3--20 $\mu$m or 6--17 $\mu$m on the detection yield is negligible.}
    {A large space-based MIR nulling interferometer will be able to directly detect hundreds of small, nearby exoplanets, tens of which would be habitable world candidates. This shows that such a mission can compete with large single-aperture reflected light missions. Further increasing the number of habitable world candidates, in particular around solar-type stars, appears possible via the implementation of a multi-visit strategy during the search phase. The high median S/N of most of the detected planets will allow for first estimates of their radii and effective temperatures and will help prioritize the targets for a second mission phase to obtain high-S/N thermal emission spectra, leveraging the superior diagnostic power of the MIR regime compared to shorter wavelengths.} 
    

   \maketitle

\section{Introduction}
\label{sec:introduction}
One of the major objectives of exoplanet science is the atmospheric characterization of a statistically relevant sample of small exoplanets. Specific emphasis will be on temperate terrestrial planets to investigate whether there are other worlds similar to Earth that may harbor life. While occurrence rates of Earth-like planets around solar-type stars are still somewhat uncertain \citep[e.g.,][]{bryson2020b}, thanks to transiting exoplanet discovery missions such as \emph{Kepler} \citep{borucki2010} and the \emph{Transiting Exoplanet Survey Satellite} \citep[TESS;][]{ricker2015} and ongoing long-term radial velocity (RV) surveys, we know that, statistically, planets with radii and masses comparable to or slightly larger than those of Earth and with shorter orbital periods are very abundant \citep[e.g.,][]{mayor2011,tuomi2019,kunimoto2020,bryson2020b}. Some major detections were even made within 20 pc of the Sun with both transit searches \citep[e.g.,][]{bertathompson2015,vanderspek2019,gillon2017,gillon2017b} and RV surveys \citep[e.g.,][]{angladaescude2016,ribas2016,jeffers2020,Astudillo2017,diaz2019,zechmeister2019}, with RV planets typically being closer to the Sun because of the geometric bias of transiting planets. 

Going forward, and focusing on the atmospheric characterization of small planets, the \emph{James Webb Space Telescope} (\emph{JWST}) might reveal whether some of these objects transiting nearby M dwarfs have managed to retain their atmospheres \citep[e.g.,][]{koll2019} despite the high level of activity of their host stars, in particular at younger ages \citep[e.g.,][]{ribas2016,macGregor2018,johnstone2019}; an in-depth investigation of atmospheric constituents with \emph{JWST} seems, however, rather challenging \citep[e.g.,][]{kreidberg2016,morley2017,krissansentotton2018}. The \emph{Atmospheric Remote-sensing Infrared Exoplanet Large-survey} mission \citep[Ariel; ][]{tinetti2018} of the European Space Agency (ESA) will provide transmission and emission spectra of hundreds of exoplanets, but the focus will be on objects with hot or warm hydrogen-dominated atmospheres; only a few small, relatively hot exoplanets will be studied. Upgraded or new fully optimized instruments at existing 8-meter-class ground-based telescopes may have a chance of directly detecting the nearest small exoplanet, Proxima~Cen~b \citep[e.g.,][]{lovis2017}. Due to their unprecedented spatial resolution and sensitivity, the upcoming 30--40~m ground-based extremely large telescopes (ELTs) will be powerful enough to directly detect small planets around the nearest stars. Instruments working at mid-infrared (MIR) wavelengths, such as the \emph{Mid-infrared ELT Imager and Spectrograph} \citep[METIS;][]{brandl2018,brandl2021}, will detect the thermal emission of the planets \citep{quanz2015,bowens2021}. Instruments working at optical or near-infrared (NIR) wavelengths and featuring high-resolution spectrographs coupled with adaptive optics systems, such as the \emph{Planetary Camera and Spectrograph} \citep[PCS;][]{kasper2021} and the \emph{High Resolution Spectrograph} \citep[HIRES;][]{hires2020}, aim at detection in reflected light. 

Unfortunately, none of the currently planned ground-based instrument projects and approved space missions is capable of investigating in detail the atmospheres of several dozen small exoplanets, including a sizable subsample residing in or close to the so-called habitable zone (HZ) of their host stars \citep{kasting1993,kopparapu2013}. This is one of the reasons why, in the context of the Astrophysics Decadal Survey in the United States, new flagship missions, the \emph{Habitable Exoplanet Observatory} \citep[HabEx;][]{habex2019} and the \emph{Large UV/Optical/IR Surveyor} \citep[LUVOIR;][]{luvoir2019}, are currently under assessment; one of their main science drivers is the direct detection and characterization of temperate terrestrial exoplanets in reflected light\footnote{We note that during the refereeing process of this paper, the Consensus Study Report ``Pathways to Discovery in Astronomy and Astrophysics for the 2020s'' was published by the National Academies of Sciences, Engineering, and Medicine recommending a large ($\sim$6 m aperture) infrared/optical/ultraviolet (IR/O/UV) space telescope as a future flagship mission \citep{decadal_report_2021}.}.

Here, we focus on a different observational approach and a new initiative that aims at developing a space-based MIR nulling interferometer capable of detecting and characterizing the thermal emission of (temperate) rocky exoplanets. The characterization of temperate exoplanets in the MIR was recently announced to be a potential science theme for a future science mission within ESA's Voyage 2050 program\footnote{\url{https://www.cosmos.esa.int/web/voyage-2050}}. The idea to employ interferometric nulling for exoplanet science was originally proposed by \cite{bracewell1978} and later followed up in \cite{leger1995} and \cite{angel1997}; in the late 1990s to mid 2000s, concept studies were carried out by both ESA and NASA: the \emph{Darwin} mission and the \emph{Terrestrial Planet Finder - Interferometer} (\emph{TPF-I}) mission, respectively. In the end, these concepts did not go forward for implementation because of technical challenges, but also because our understanding of the exoplanet population was significantly more limited. This has, however, changed. Given the enormous scientific progress in exoplanet research since the mid 2000s and significant advances related to key technologies, it is time to reassess such a mission concept and quantify its potential scientific return. 
In 2018, a first such study was published \citep{kammererquanz2018}, which investigated the exoplanet detection yield of a space-based MIR nulling interferometer based on exoplanet occurrence rates from NASA's \emph{Kepler} mission; it claimed that a few hundred small exoplanets  (radii between 0.5 and $6~\mathrm{R}_\oplus$) could be within reach for such an instrument. These promising results, in combination with  ongoing lab activities related to nulling interferometry, resulted in the creation of the \emph{Large Interferometer For Exoplanets (LIFE)} initiative\footnote{\url{www.life-space-mission.com}}.

The present paper is the first in a series of papers currently in preparation. It is assumed that a mission such as \emph{LIFE} would consist of two main phases: (1) a search phase to directly detect a large sample of exoplanets orbiting nearby stars and (2) a characterization phase to reobserve a subsample of these exoplanets and investigate their properties and atmospheres in detail. In the following, we focus on the search phase and quantify how many exoplanets \emph{LIFE} would be able to detect depending on different mission parameters. Future work will focus more on the characterization phase of the mission, including questions related to atmospheric diversity and evolution, habitability, and the search for indications of biological activity. In this context, the MIR wavelength regime offers complementary information and even several advantages compared to studies at optical or NIR wavelengths. These include more direct constraints on the temperature and size of the objects and a large set of molecular absorption lines of main atmospheric species, including biosignatures \citep[e.g.,][]{schwietermann2018,catling2018}, some of which, for example CH$_4$, might be easier to detect in the MIR than at shorter wavelengths. 

In comparison to previous studies that quantify the detection yield of an MIR nulling interferometer \citep[e.g.,][]{kammererquanz2018,quanz2018,quanz2021}, we have significantly updated and improved our simulation approach as further described below. Similar detection yield analyses were carried out for the reflected light missions mentioned above, enabling a direct comparison of the different mission concepts. 

We note that in the ideal case, exoplanet surveys with existing and future high-precision RV instruments (e.g., CARMENES, NIRPS, ESPRESSO, MAROON-X, HARPS3, and EXPRES) will continue to uncover a significant fraction of the exoplanet population within 20 pc, including rocky, potentially habitable, planets\footnote{As an alternative to ground-based RV searches, space-based high-precision astrometry missions could be envisioned \citep[e.g.,][]{malbet2018,janson2018}.}. In this case, the search phase of future direct detection exoplanet missions would be shortened and more of the limited observing time could be allotted to the characterization of the objects. Whether in the end the low-amplitude RV signals from Earth-like
planets can be separated from astrophysical noise sources, such as stellar jitter \citep[e.g.,][]{oshagh2017}, and to what extent, consequently, those RV surveys will be complete, remains to be seen. 

In Sect.~\ref{sec:simulations} we describe in detail the setup of our yield simulations. The results are presented in Sect.~\ref{sec:results}, and we discuss them in Sect.~\ref{sec:discussion}. We conclude and summarize our main findings in Sect.~\ref{sec:summary}.
\section{Setup of yield simulations}
\label{sec:simulations}
The general approach of our Monte Carlo-based yield simulations is described in  \citet{kammererquanz2018}, but we have implemented several updates as detailed in the following.

\subsection{Stellar target catalog}
\label{sec:star_cat}
We used a new \emph{LIFE} target star catalog that includes single main-sequence stars and wide separation binaries of spectral types FGKM out to 20 pc of the Sun. While the catalog includes a total of 1732 objects, only a subset was considered for the simulations of the search phase depending on the optimization strategy (see Sect.~\ref{sec:scenarios}  for details). The catalog and its creation are further explained in Appendix~\ref{sec:appendix_stars}.

\subsection{Exoplanet population}
\label{sec:planet_pop}
In our simulations we generated an artificial population of exoplanets around the stars in our target catalog and did not consider any known exoplanets. The underlying exoplanet occurrence rates as a function of the radius and orbital period follow the results from NASA's ExoPaG SAG13 \citep{kopparapu2018} for single FGK stars and \cite{dressing2015} for single M stars. This allows for a comparison with the results obtained in the context of the \emph{HabEx} and \emph{LUVOIR} studies mentioned above, which used a very similar underlying exoplanet population. Binary stars with measured (apparent) separations greater than 50 AU were treated as single stars. For binary systems with smaller separations, the occurrence rates were scaled down by a factor of 0.3 over the entire period and radius range \citep[cf.][]{kraus2016}. We focused our analysis on planets with radii, $R_{\textrm p}$, in the range $0.5\,R_{\oplus} \leq R_{\textrm p}\leq 6\,R_{\oplus}$ for FGK stars and $0.5\,R_{\oplus} \leq  R_{\textrm p}\leq 4\,R_{\oplus}$ for M stars. Orbital periods, $P_{\textrm P}$, in the range $0.5\,d \leq P_{\textrm P} \leq 500\,d$ and $0.5\,d \leq P_{\textrm P} \leq 200\,d$ were considered for FGK and M stars, respectively.  Cold, Neptune-, or Jupiter-like objects with separations $\ge$3 au, for which the SAG13 statistics cannot be applied \cite[cf.][]{dulz2020} and which were included in the \emph{HabEx} and \emph{LUVOIR} studies, were not considered as they are typically too cold for a detection within a reasonable amount of integration time. All planets were assumed to have circular orbits and were assigned a random Bond albedo, $A_{\textrm B}\in [0, 0.8)$\footnote{For reference, the Bond albedos of Venus and Mercury are $\approx$0.8 and $\approx$0.1, respectively, bracketing the values for the Solar System planets.}, and a geometric albedo, $A_{\textrm g}\in [0,0.1)$\footnote{These low values are motivated by the MIR wavelength range we are considering.}, that is constant over the wavelength range we consider \citep[cf.][]{kammererquanz2018}. Both albedos were uniformly distributed in the considered intervals. The planets were treated as black bodies with their whole surface area radiating with an equilibrium temperature ($T_{\textrm{eq}}$) determined by the luminosity of their host star, their Bond albedo, and their orbital separation. 

In Table~\ref{table:rocky_planets} we define two types of exoplanets that are important throughout the paper: rocky planets orbiting within the empirical habitable zone (eHZ) and exo-Earth candidates (EECs). We also provide the respective occurrence rates as provided by our exoplanet population. 

\begin{table*}
\caption{Types of exoplanets that are of particular importance throughout the paper. The first row shows our definition of a rocky planet orbiting within the empirical habitable zone (eHZ). The second row defines exo-Earth candidates (EECs) as used in the yield estimates for \emph{HabEx} and \emph{LUVOIR}. The last columns summarize the occurrence rates for these objects as provided by the assumed exoplanet population.}             
\label{table:rocky_planets}      
\centering                          
\begin{tabular}{c c c c c}        
\hline\hline                 
Planet type & $R_\textrm{P}$ [R$_\oplus$] & Stellar flux range [S$_{\oplus}$]  & \multicolumn{2}{c}{Occurrence rates\tablefootmark{a} } \\ 
                &                                               &                                                       &   M stars & FGK stars\\ \hline
Rocky eHZ & 0.5 -- 1.5 & 1.776 -- 0.32\tablefootmark{b}&  0.558 &  0.370  \\
    Exo-Earth Candidates (EECs) & 0.82 / 0.62\tablefootmark{c} -- 1.4 & 1.107 -- 0.356\tablefootmark{d} & 0.312 &  0.164 \\\hline
\hline                                   
\end{tabular}
\tablefoot{
\tablefoottext{a}{Occurrence rates are values for single stars averaged over our input catalog for the given range of spectral types. Because the stellar flux range is spectral type dependent, the number of objects falling within this range, and hence the occurrence rates, varies with spectral type as well.}
\tablefoottext{b}{The flux range is given by the ``recent Venus'' and ``early Mars'' limits and is spectral type dependent \citep{kopparapu2014}. The values given here correspond to 1 M$_\oplus$ planet orbiting a solar twin. We note that both limits take into account the luminosity evolution of the Sun, which was fainter during the epochs when Venus and Mars provided habitable conditions. For present-day solar luminosity, these insolation limits correspond to separations of 0.75 and 1.77 au, respectively, excluding Venus from, but including Mars in, the eHZ.}
\tablefoottext{c}{For EECs, the lower limit of the radius range depends on the separation from the star; closer to the star, planets are required to have a larger radius. This can be described by $R_\textrm{P}^\textrm{min}=0.8\cdot S^{0.25}$, where $S$ is the insolation. The \emph{HabEx} and \emph{LUVOIR} studies focused primarily on solar-type stars and used a corresponding expression based on the semimajor axis, $a$, i.e., $R_\textrm{P}^\textrm{min}=0.8\cdot a^{-0.5}$.  As we are also interested in M stars, we had to convert this into an expression for $S$.}
\tablefoottext{d}{The flux range is given by the ``runaway greenhouse'' and ``maximum greenhouse'' limits and is spectral type dependent \citep{kopparapu2014}. The values given here correspond to a 1 M$_\oplus$ planet orbiting a solar twin.}}
\end{table*}

\subsection{Simulating spacecraft, instrument, and noise sources with \textsc{LIFEsim}}
\label{sec:noise_sims}
In order to estimate the signal-to-noise ratio (S/N) of our simulated exoplanets we have developed the new simulation tool \textsc{LIFEsim} \citep{dannert2022}. This tool enables us to simulate the temporally modulated signal that a planetary system would leave in an observing sequence with a space-based nulling interferometer \citep[cf.][]{lay2005} and further includes the most relevant -- and wavelength-dependent -- astrophysical noise sources. This is an important difference from our earlier exoplanet detection yield estimates, where, instead of explicitly simulating the interferometer transmission and signal modulation for every simulated planet, constraints on the inner working angle of the instrument and sensitivity were used to assess the discovery potential \citep{kammererquanz2018,quanz2018}.

\begin{figure}[h]
    \centering
    \includegraphics[width=\linewidth]{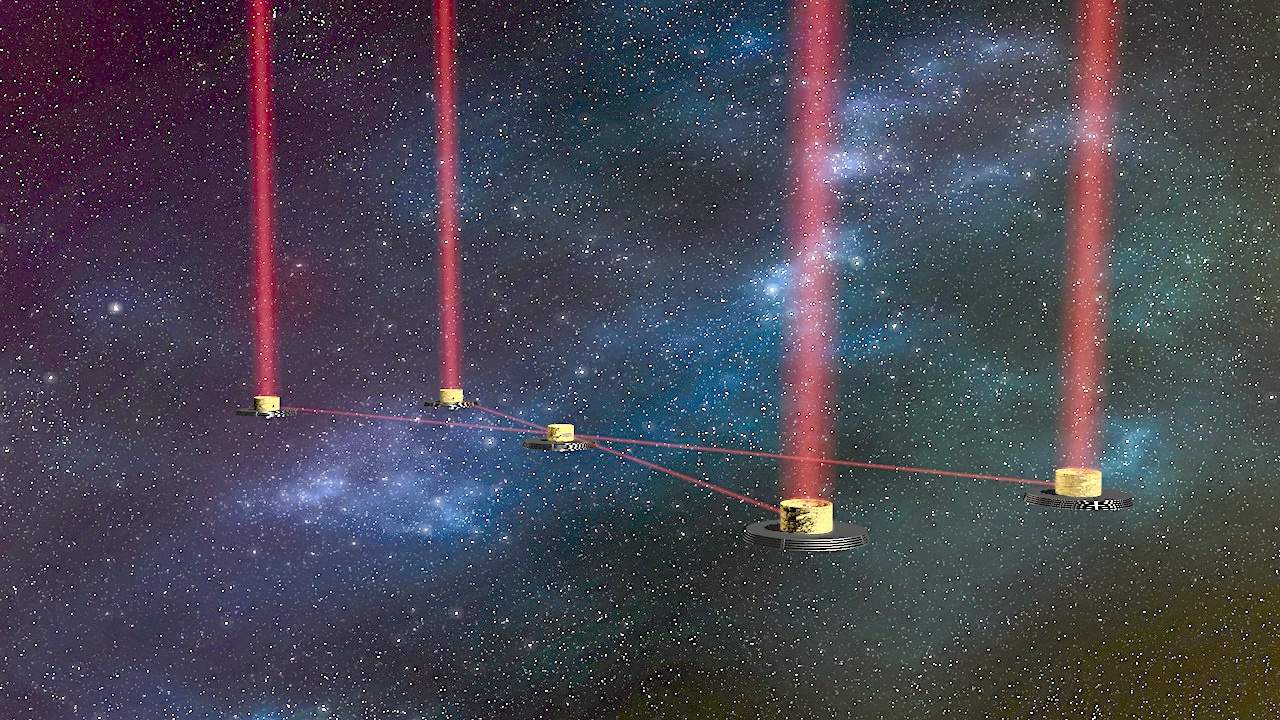}
    \caption{Artist's impression of the \emph{LIFE} nulling-interferometry mission, consisting of four collector spacecraft in a rectangular array configuration sending light to a beam combiner spacecraft in the center. The present analysis assumes an X-array configuration with a baseline ratio of 6:1.}
    \label{fig:artist_impression}
\end{figure}

In the following, we considered an interferometer consisting of four collector spacecraft in a so-called X-array configuration that feed their beams into a fifth beam combiner spacecraft located in their center (see Fig.~\ref{fig:artist_impression}). 

The ratio between the long and the short baseline of the X-array was assumed to be 6:1 for the time being and a $\pi/2$ phase shift was applied between two conjugated transmission maps \citep[cf.][]{defrere2010}. The short baselines of the array are referred to as ``nulling baselines'' and are responsible for creating the central dark fringe that nulls the host star. The long baselines are referred to as ``imaging baselines'' and are responsible for a higher-frequency modulation of the transmission map perpendicular to the fringe pattern created by the nulling baselines. The resulting modulation map, the difference between the two conjugate transmission maps, effectively suppresses the signal coming from any centrally symmetric source (such as emission from optically thin and smooth exozodiacal dusk disks with random inclination) so that only the shot noise of the source contributes to the S/N of the observations. The 6:1 baseline ratio has been shown to be more robust against instability noise compared to a 2:1 baseline ratio \citep{lay2006}, but additional trade-off studies are needed to further validate this choice. A detailed description of \textsc{LIFEsim} is provided in \citet{dannert2022}, but we refer the reader to \citet{defrere2010} for a description of the general nulling and beam combination scheme and the resulting final modulation map for an X-array interferometer. 

In our simulations we assumed that for each new target star the array is reconfigured so that the center of the projected eHZ (cf. Table~\ref{table:rocky_planets}; Sect.~\ref{sec:planet_pop}) falls within the first transmission maximum of the nulling baselines at a reference wavelength of 15~$\mu$m. However, we imposed a minimum separation between two adjacent collector spacecraft of at least 10~m and allow for a maximum separation of 600~m. Initial tests had shown that having the reference wavelength between 15 and 20~$\mu$m resulted in comparable detection yields, but that shorter (e.g., 10~$\mu$m) or longer (e.g., 25~$\mu$m) reference wavelengths provided lower yield numbers. Keeping the baseline lengths of the configuration in mind (from a technical perspective), we decided to use 15~$\mu$m as reference for all analyses presented in the following. 
The aperture diameter $D$ of the collector spacecraft is a free parameter in our instrument model, and we discuss the impact of aperture size on the results in Sect.~\ref{sec:res_apertures}. For the computation of the total photon flux received by the collector spacecraft we ignored any possible obscuration from a secondary mirror. To be conservative, we assumed 5\%\footnote{This 5\% throughput is applied to the modulation maps, which already contain only 50\% of the incoming light.} for the optical throughput of the instrument, but will update this number when the concept for the optical layout is maturing. In earlier studies in the context of the \emph{Darwin/TPF-I} missions a throughput of 10\% was assumed \citep[e.g.,][]{lay2007,defrere2010} and also at the \emph{Large Binocular Telescope Interferometer (LBTI)} the most recent estimate for the optical throughout around $\sim$11$~\mu$m is $\approx$0.11 (S. Ertel, private communication). For the detector quantum efficiency, we assumed 70\% over the full wavelength range, which is identical to the \emph{Darwin} studies mentioned above. Recent experiments with 15-micron-cutoff HgCdTe detector arrays have yielded quantum efficiencies of $\gtrsim$0.8 between 6 and 12 $\mu$m wavelengths \citep{cabrera2020} and the Si:As IBC detectors of the JWST/MIRI instrument achieve $\gtrsim$0.7 between 12 and 20 $\mu$m \citep{rieke2015}.

At the moment, our S/N calculations are photon-based and include all major astrophysical noise terms. We implicitly assumed that our measurements would not be limited by instrumental effects (see Sect.~\ref{sec:detection_criteria} below for the definition of our detection criterion). The impact of phase and/or amplitude variations as major systematic noise sources is currently being assessed. Also, thermal background from the aperture mirrors and the instrument optics, and detector-related noise sources will be included in subsequent work. For the mirrors of the collector spacecraft and the instrument optics not to contribute significantly to the measurement implies a required temperature of $\lesssim$40 K \citep{defrere2010}.
Noise terms that were explicitly included are:

\begin{description}
    \item{\textbf{Photon noise from the simulated planets:}}
    Given the distance, radius, and equilibrium temperature of our simulated planets, their photon flux (assuming black-body emission) and related noise are fully described. 
    \item{\textbf{Photon noise from stellar leakage:}}
    Depending on the distance to the star, its radius, and the length of the nulling baseline, a small fraction of stellar photons may ``leak'' through the central dark fringe and hence contribute to the photon noise. 
     \item{\textbf{Photon noise from exozodi disks:}}
    For each simulated planetary system we randomly assigned a level of emission from a dusty exozodi disk following the observed (nominal) distribution from the Hunt for Observable Signatures of Terrestrial Systems (HOSTS) survey \citep{ertel2018,ertel2020}. 
    To compute the spectral energy distribution (SED) of the exozodi disk we used the publicly available code from \citet{kennedy2015}.  
    All exozodi disks were assumed to be optically thin, smooth (i.e., without any substructure) and seen face-on. We refer the reader to Sect.~\ref{zodidiscussion} for a discussion about these assumptions.
    \item{\textbf{Photon noise from local zodiacal light:}} The optically thin zodiacal dust in our Solar System is a source of significant MIR emission. In \textsc{LIFEsim} the surface brightness is described by a pointing-dependent 2D model originally developed for the \emph{Darwin} simulator
    and based on data from \emph{Cosmic Background Explorer} \citep[COBE;][]{kelsall1998}. Compared to the original \emph{COBE} data, the model slightly over-predicts the flux in the 6-20~$\mu$m range and for a pointing direction with a relative latitude of more than 90$^\circ$ from the Sun by 10-20\%. For wavelengths shorter than 6~$\mu$m the difference increases to a factor of 2-3 at 3~$\mu$m. However, at these shorter wavelengths the total photon noise is strongly dominated (up to several orders of magnitude) by the contribution from stellar leakage. In our simulations we assumed that we always point in anti-sunward direction (\emph{LIFE} will be launched to the Earth-Sun L2 point) but considered the true latitude of the target star. 
\end{description}

In our S/N calculations we implicitly assumed that the combined beams are fed through single-mode fibers before the signal is spectrally dispersed. The effective field-of-view (FoV) of the fibers, and hence of each collector spacecraft, is wavelength dependent and given by $\lambda/D$. 

\begin{figure}[h]
    \centering
    \includegraphics[width=\linewidth]{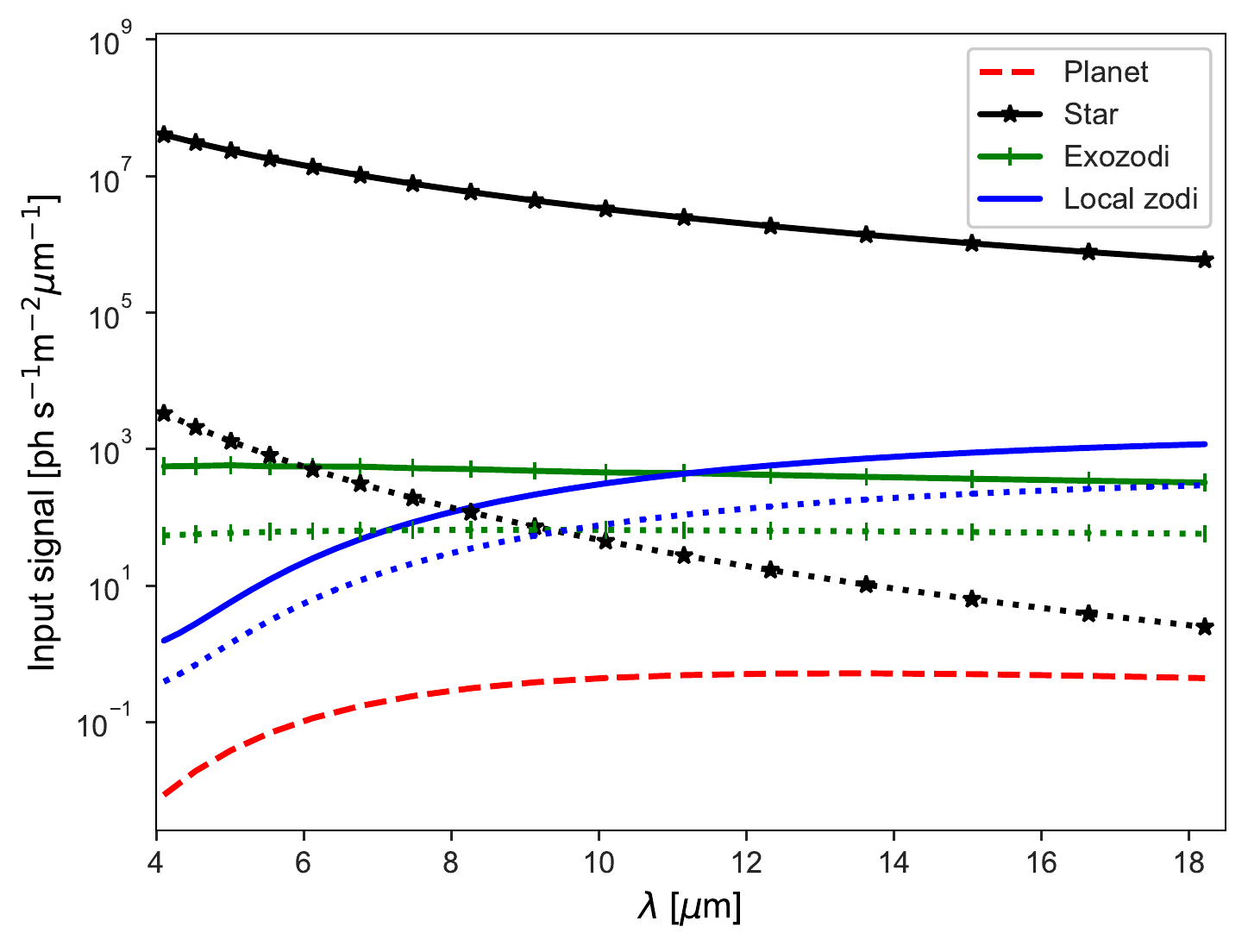}
    \caption{Example illustrating the photon flux and noise contributions from the various astrophysical sources in our nulling-interferometry simulations: exoplanet flux (1 R$_\oplus$ and 276 K effective temperature located at 10 pc; dashed red line), flux from a Sun-like star (black line with stars), local zodiacal light (green line with ticks), and exozodi (1 zodi; solid blue line). The corresponding photon noise contributions (1-$\sigma$) are shown with the same color code, but as dotted lines.}    
    \label{fig:noise_model}
\end{figure}

In Fig.~\ref{fig:noise_model} we show an example of how the various noise terms compare to the incoming photon flux from a 1 R$_\oplus$ exoplanet with an effective temperature of 276 K orbiting at 1 au from a Sun-like star at 10 pc distance. The system is assumed to be located within the ecliptic and contains an exozodi disk with the same brightness as the zodiacal light in the Solar System.

\subsection{Mission parameters}
\label{sec:mission}
Assuming a total mission lifetime of 5-6 years, we assigned an available on-source observing time of 2 years to the initial search phase. This translates into 2.5 years of mission time considering 25\% of general mission overhead; the remaining time of the mission is dedicated to detailed follow-up observations of a subset of the detected exoplanets and possibly an ancillary science program. The slew time from one target to the next was fixed to 10 hours, which is part of the 2 year observing time. For the moment we only considered single visits of target stars during the survey.

\subsection{Setup of Monte Carlo simulations }
\label{sec:mc_sims}
To create the exoplanet population we used the freely available \textsc{P-Pop} Monte Carlo tool\footnote{\url{https://github.com/kammerje/P-pop}}, which for each star of the target catalog randomly draws exoplanets from the distributions described above \citep[cf.][]{kammererquanz2018} and puts them at random positions along their orbits. The orbital inclination was also randomly chosen for each system, but planets in multi-planet systems were assumed to be co-planar. To ensure that multi-planet systems were dynamically stable we applied a stability criterion following the approach by \citet{he2019} that is based on the mutual Hill radius of neighboring planets. Specifically, for circular orbits as assumed here, a system was considered stable if for all planet pairs within the system $$\Delta=\frac{a_{out}-a_{in}}{R_H}>8\quad,$$
where $a_{out}$ and $a_{in}$ are the semimajor axes of the outer and inner planet, respectively, and $R_H$ is the mutual Hill radius given by
$$R_H=\frac{a_{in}+a_{out}}{2}\Bigg[\frac{m_{in}+m_{out}}{3M_*}\Bigg]^{1/3}\quad,$$
with $m_{in}$ and $m_{out}$ being the mass of the inner and outer planet, respectively, and $M_*$ being the mass of the host star. If a system or pair of planets was unstable, we re-drew the system, which happened in less than 2\% of the cases. In total, we generated 500 planetary systems per target star. All planets were  then run through \textsc{LIFEsim} in order to compute their photon fluxes as well as the photon noise from the various sources listed above.  

\begin{figure*}[t!]
    \centering
    \includegraphics[width=0.48\linewidth]{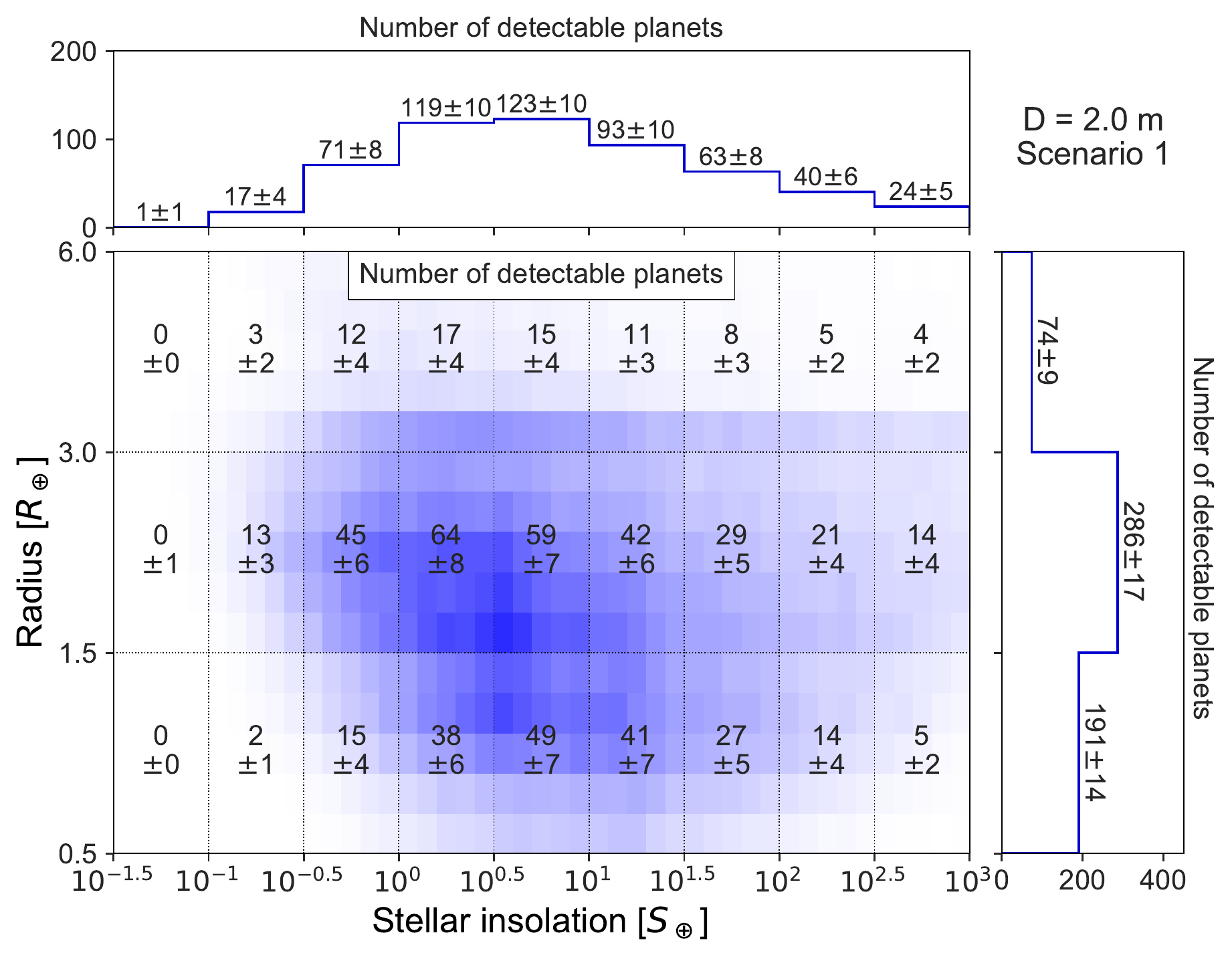}
    \includegraphics[width=0.48\linewidth]{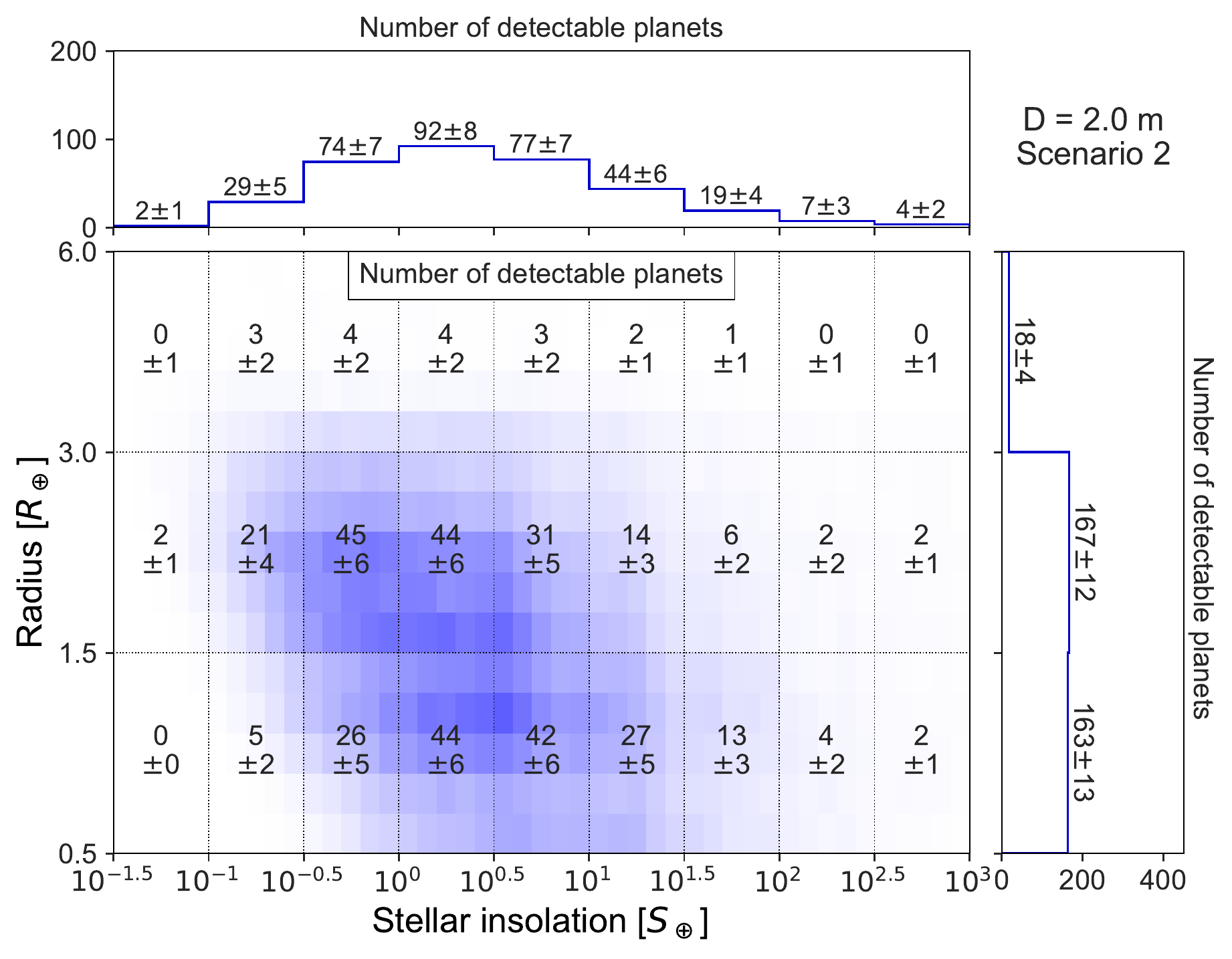}
    \caption{Total exoplanet detection yield from our reference case scenario simulations ($D=2$ m; $\lambda=4-18.5\,\mu m$) in the radius vs. stellar insolation plane. The plots show the number of expected planet detections per grid cell, including the statistical 1-$\sigma$ uncertainty from the Monte Carlo approach but excluding uncertainties in the exoplanet occurrence rates. Left panel: Scenario 1 (search phase optimized for maximizing the total number of exoplanets).\ Right panel: Scenario 2 (search phase optimized for maximizing the number of rocky eHZ exoplanets.)}
    \label{fig:baseline_yields_grid}
\end{figure*}

\subsection{Distribution of observing time: Two scenarios}
\label{sec:scenarios}
We considered two scenarios that determine the distribution of the available on-source observing time of 2 years: maximizing the total number of detected exoplanets (scenario 1) or the number of rocky exoplanets orbiting within the eHZ of their host star (scenario 2; see Table~\ref{table:rocky_planets} for definition). We note that the eHZ  includes a much larger range of insolations than the ``classical'' HZ \citep{kasting1993,kopparapu2013}, but a much smaller range than the ``extended hydrogen'' HZ \citep{pierrehumbert2011}, which is estimated to reach $\approx$10 au ($\approx 0.01\;S_{\oplus}$) for a G-type star. Depending on the main science goals of \emph{LIFE}, one may prefer either of the two scenarios, but, as we will see below, maximizing the number of temperate, rocky exoplanets (scenario 2) leads to a decrease in the total number of detectable planets (scenario 1). 

The algorithm to distribute the observing time was similar to the one discussed in \citet{lay2007} and considered that for each star in a given Monte Carlo run one can compute the detection efficiency (defined as number of detected planets per time interval $\delta t$). The number of detected planets depends on the threshold one puts on the S/N of the planets, which in turn depends on the assumed aperture size of the collector spacecraft and the assumed length of $\delta t$ (in our analysis we assumed $\delta t$=1h). Also, one can decide which subset of planets to focus on (i.e., scenario 1 or scenario 2). By computing the number of detectable planets for all stars and over a sufficiently large range of time intervals, one can identify the star that offers the maximum possible detection yield for the smallest time interval. This star and the corresponding planet(s) as well as the length of the required time interval were saved, and the star offering the second best detection efficiency was searched. We repeated this process until the available  observing time (including the 10 h slew time from one star to the next) was used up. This yielded the total number of detectable planets per star as well as a rank-ordered list of target stars based on their expected contribution to the detection yield. 
We then calculated the gain (i.e., planet yield per time) as an average over all 500 Monte Carlo realizations for each star. This allowed us to construct an observing sequence, which yielded the final average numbers we are quoting below. For completeness we note that in this analysis we implicitly assumed that the X-array of the collector spacecraft did an integer number of full rotations around its center irrespective of the assumed integration time. This allowed for an easier computation of the exoplanets' signals passing through the interferometer's modulation map \citep{dannert2022}.

\subsection{Detection criterion}
\label{sec:detection_criteria}
In the following, we required a S/N$\ge$7 spectrally integrated over the full wavelength range for a planet to be considered a detection. This choice compensates for the lack of an instrumental noise model in the current simulations. Under the assumptions that the instrumental noise contribution is equal to or lower than the astrophysical noise and that the total noise can be written as the square root of the sum of the instrumental and astrophysical noise (i.e., $\sigma_{tot}=\sqrt{(\sigma_{inst})^2+(\sigma_{astro})^2}$), a total S/N$\ge$7 corresponds to an astrophysical signal of S/N$_{astro}\ge$5.

As we were only considering photon noise, we verified that (slightly modified versions of) published signal extraction algorithms for nulling-interferometry data \citep[e.g.,][]{thiebaut2006,mugnier2006} actually achieve a performance close to the ideal photon-noise limited case and can also be applied to multi-planet systems \citep{dannert2022}.
\section{Results}
\label{sec:results}
For the two scenarios outlined in Sect.~\ref{sec:scenarios}, we chose an aperture size of  $D=2$ m as our reference case, but we also investigated cases with $D=1$~m and $D=3.5$~m (the latter corresponding to the aperture of ESA's \emph{Herschel} spacecraft, the largest monolithic infrared space telescope ever launched). Besides using 4--18.5~$\mu$m as wavelength range in the reference case, we also computed detection yields for 3--20~$\mu$m and 6--17~$\mu$m. We note that for determining the final wavelength range not only the expected detection yield during a 2.5-year search phase should be considered, but also the scientific importance of molecular bands for atmospheric characterization at the short and long wavelength end \citep{konrad2022} and technical aspects. We remind the reader that in all cases the assumed instrument throughput is 5\% (see Sect.~\ref{sec:noise_sims}).

\begin{figure*}[h]
    \centering
    \includegraphics[width=0.48\linewidth]{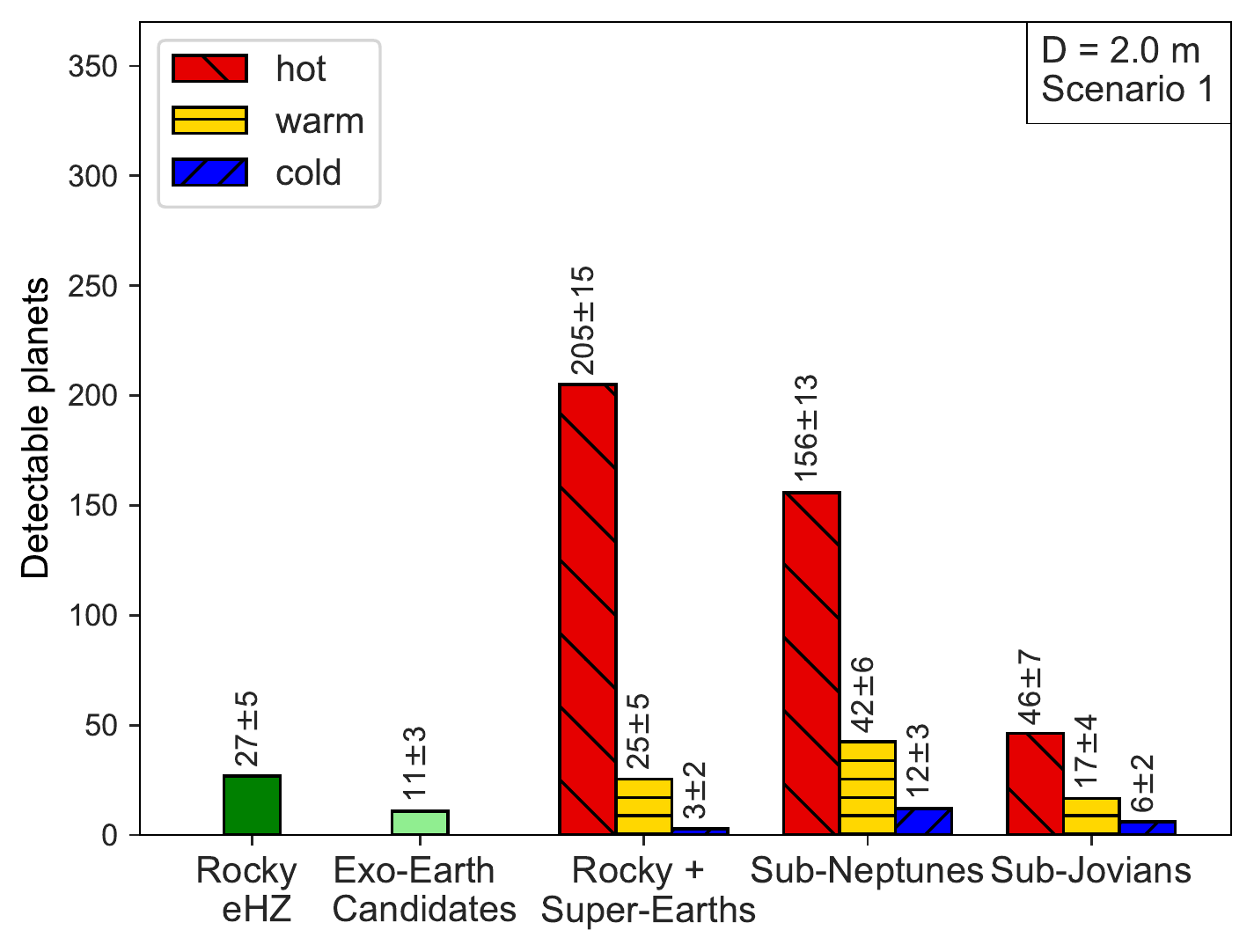}
    \includegraphics[width=0.48\linewidth]{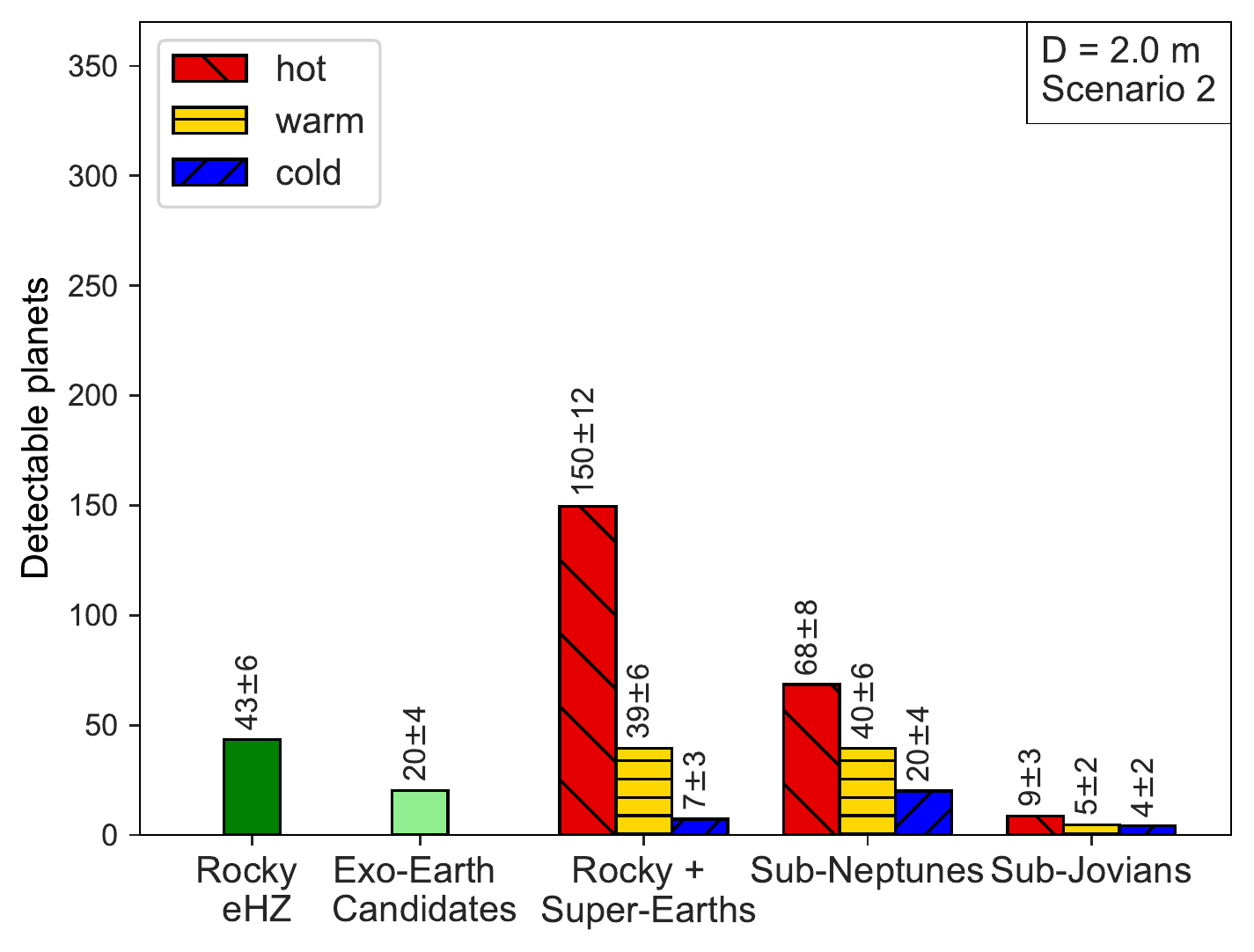}
    \caption{Total exoplanet detection yield from our reference case scenario simulations ($D=2$ m; $\lambda=4-18.5\,\mu m$) using the planet classification scheme introduced by \cite{kopparapu2018} (see Table~\ref{table:1}). For comparison, we also plot the number of terrestrial exoplanets within the eHZ as defined in Sect.~\ref{sec:scenarios} as the leftmost bar labeled ``Rocky eHZ.'' The bars show the number of expected planet detections, including the statistical 1-$\sigma$ uncertainty from the Monte Carlo approach but excluding uncertainties in the exoplanet occurrence rates. Left panel: Scenario 1, i.e., the search phase is optimized for maximizing the total number of exoplanets. Right panel: Scenario 2, i.e., the search phase is optimized for maximizing the number of rocky eHZ exoplanets.}
    \label{fig:baseline_yields_bars}
\end{figure*}
\begin{figure*}[h]
    \centering
    \includegraphics[width=0.48\linewidth]{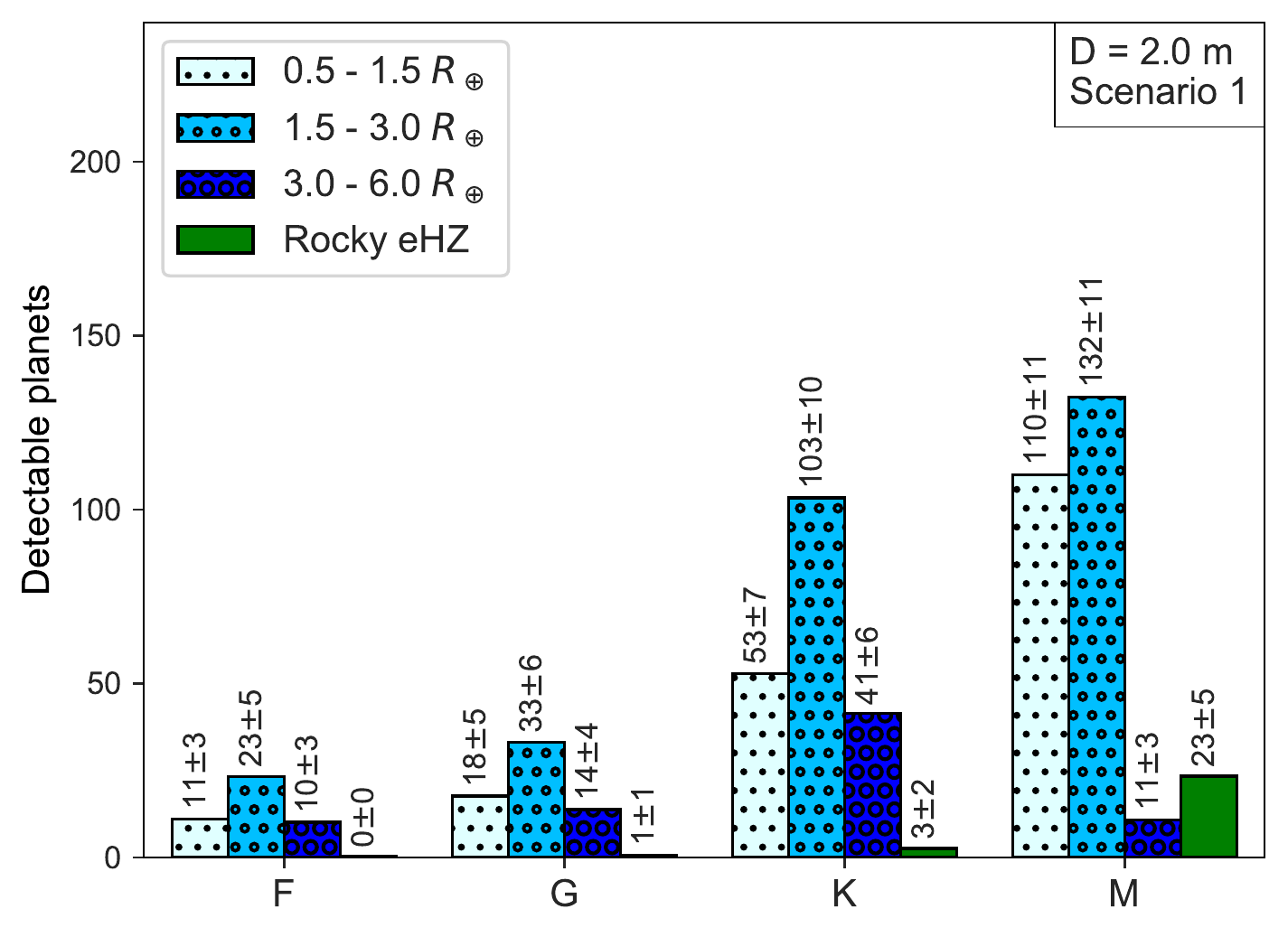}
    \includegraphics[width=0.48\linewidth]{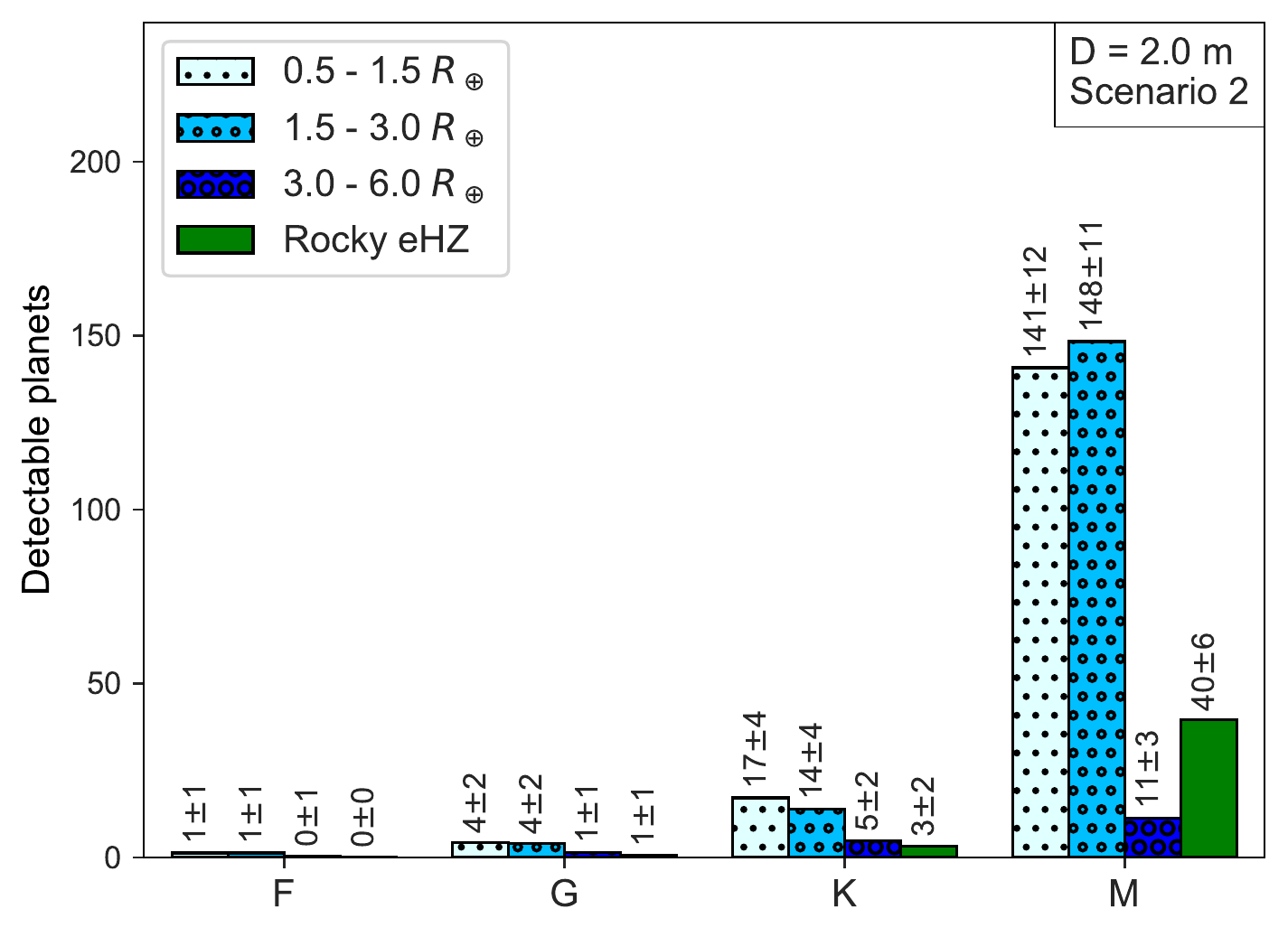}
    \caption{Total exoplanet detection yield from our reference case scenario simulations ($D=2$ m; $\lambda=4-18.5\,\mu m$) shown as a function of spectral type of the host star (left panel: Scenario 1; right panel: Scenario 2). The bars show the number of expected planet detections, including the statistical 1-$\sigma$ uncertainty from the Monte Carlo approach but excluding uncertainties in the exoplanet occurrence rates. In both panels, the green bars (labeled ``Rocky HZ'') show the number of rocky eHZ exoplanets, which are a subset of the planets in the 0.5--1.5 R$_\oplus$ bin in Fig.~\ref{fig:baseline_yields_grid}.}
    \label{fig:stellar_types}
\end{figure*}

\subsection{Exoplanet yield of reference case scenarios}
\label{sec:res_yields}

In Fig.~\ref{fig:baseline_yields_grid} we show the expected number of detectable exoplanets and the standard deviation resulting from our 500 Monte Carlo runs in the radius versus stellar insolation plane for the reference case setup and the two scenarios described in Sect.~\ref{sec:scenarios}. Figure~\ref{fig:baseline_yields_bars} is based on the same information, but this time we follow the exoplanet classification scheme introduced by \cite{kopparapu2018} (see Table~\ref{table:1}). This scheme was also used in the final study reports by the \emph{LUVOIR} and \emph{HabEx} teams \citep{luvoir2019,habex2019}, which allows for an easier comparison between the different mission concepts. A current short-coming is that the scheme assumes a Sun-like host star. Variations in the host star SED could potentially alter the stellar flux condensation boundaries of the considered chemical species by a few percent. A more robust analysis with different host stellar spectral types, including M dwarfs, is needed to correct this. 
We note that in Fig.~\ref{fig:baseline_yields_bars}, we also show the number of terrestrial exoplanets located within the eHZ as defined in Sect.~\ref{sec:scenarios}. for comparison with the other classes of exoplanets. It is important to keep in mind that in all cases the number of detectable exoplanets as a function of their radius and received insolation is influenced by both the assumed underlying exoplanet population and our technical assumptions. 

\begin{table*}
\caption{Exoplanet classification scheme introduced by \cite{kopparapu2018}. Together with the planet types defined in Table~\ref{table:rocky_planets}, we apply this scheme in Figs.~\ref{fig:baseline_yields_bars},~\ref{fig:1m_yields_bars},~\ref{fig:3.5m_yields_bars}, and~\ref{fig:yields_no_m_stars}, as well as in Figs.~\ref{fig:LIFE_yield_aperture} and~\ref{fig:LIFE_yield_lambdarange} in Appendix~\ref{sec:appendix_plots}. For reference, Venus would be classified as a ``hot, rocky'' planet and Earth and Mars as ``warm, rocky'' planets.}             
\label{table:1}      
\centering                          
\begin{tabular}{c c c c c}        
\hline\hline                 
Planet type & $R_\textrm{P}$ [R$_\oplus$] & \multicolumn{3}{c}{Stellar flux range [S$_{\oplus}$]} \\ 
            &                           & Hot & Warm & Cold  \\\hline
\hline                        
  Rocky  &  0.5 -- 1 & 182 -- 1.0 & 1.0 -- 0.28 & 0.28 -- 0.0035  \\      
  Super-Earths  &  1 -- 1.75 & 187 -- 1.12 & 1.12 -- 0.30 & 0.30 -- 0.0030  \\      
  Sub-Neptunes  &  1.75 -- 3.5 & 188 -- 1.15 & 1.15 -- 0.32 & 0.32 -- 0.0030  \\      
  Sub-Jovians  &  3.5 -- 6 & 220 -- 1.65 & 1.65 -- 0.45 & 0.45 -- 0.0030 \\      

\hline                                   
\end{tabular}

\end{table*}

These plots show that within the present simulation framework, \emph{LIFE} would discover hundreds of nearby exoplanets, the vast majority of which have radii between 0.5 and 3 R$_\oplus$. It also shows that the choice of observing scenario has a significant impact on the planet yield: on the one hand the number of rocky exoplanets orbiting within the eHZ can indeed be significantly increased by a factor of $\approx$1.6 if one optimizes the observing strategy accordingly (scenario 2). For EECs (see Table~\ref{table:rocky_planets}) it is even a factor of $\approx$2. However, this comes at a price as the resulting total number of detectable exoplanets is significantly smaller compared to scenario 1 ($\approx$350 vs. $\approx$550). Because of their moderate temperatures, rocky exoplanets in the eHZ are fainter than objects orbiting closer to the star and require larger amounts of integration time to be detected. While for scenario 1 the typical observing time per target is between 15 and 35 hours, it is between 50 and 130 hours for scenario 2.  

\begin{figure}[h]
    \centering
    \includegraphics[width=\linewidth]{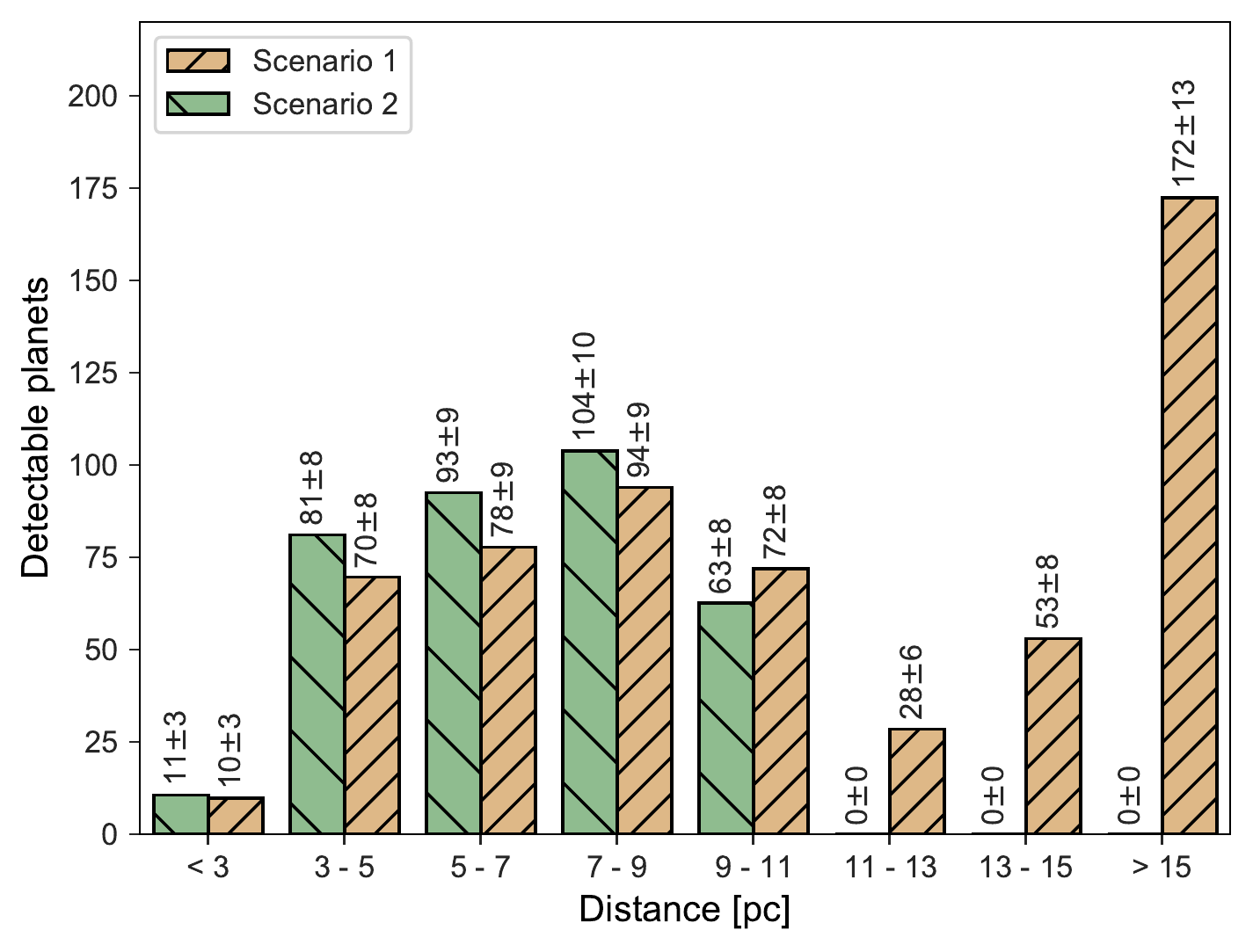}
    \caption{Distance distribution of the detected planet populations shown in Fig.~\ref{fig:baseline_yields_grid}. The bars show the number of expected planet detections, including the statistical 1-$\sigma$ uncertainty from the Monte Carlo approach but excluding uncertainties in the exoplanet occurrence rates.}
    \label{fig:distance_distribution}
\end{figure}

In Fig.~\ref{fig:stellar_types} we show the distribution of detectable exoplanets as a function of the spectral type of their host star. For the rocky eHZ exoplanets there is a strong preference for M dwarfs. This is because M dwarfs are, on average, much more numerous and hence there is a larger number of M dwarfs close to the Sun. In addition, they have a higher occurrence rate of terrestrial exoplanets \citep[cf.][]{dressing2015}. In Fig.~\ref{fig:distance_distribution}  we show the distance distribution of the detected exoplanets for both scenarios. By maximizing the number of rocky eHZ exoplanets one exclusively observes stars within $\sim$10 pc of the Sun.

\begin{figure}[h]
    \centering
    \includegraphics[width=\linewidth]{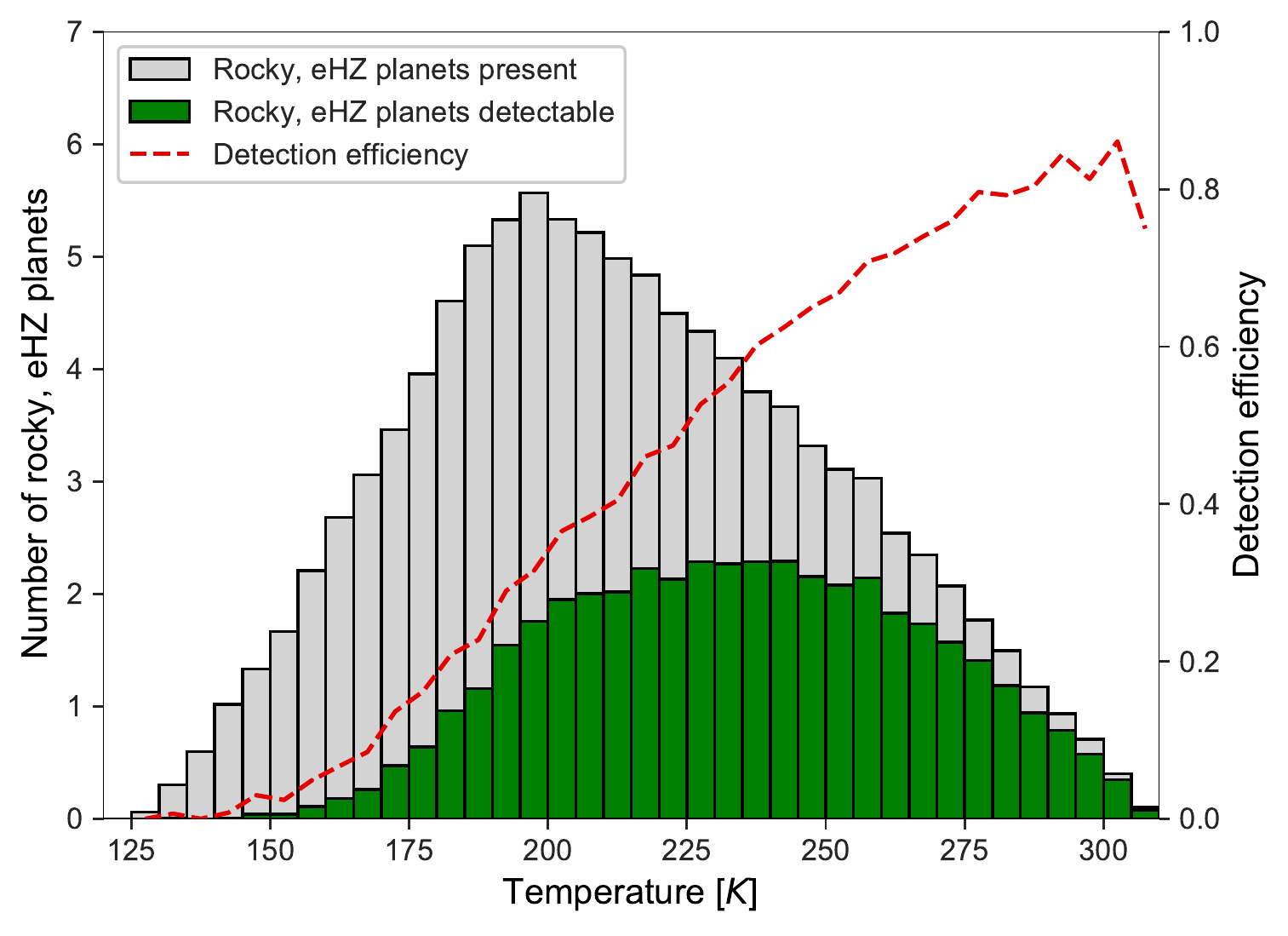}
    \includegraphics[width=\linewidth]{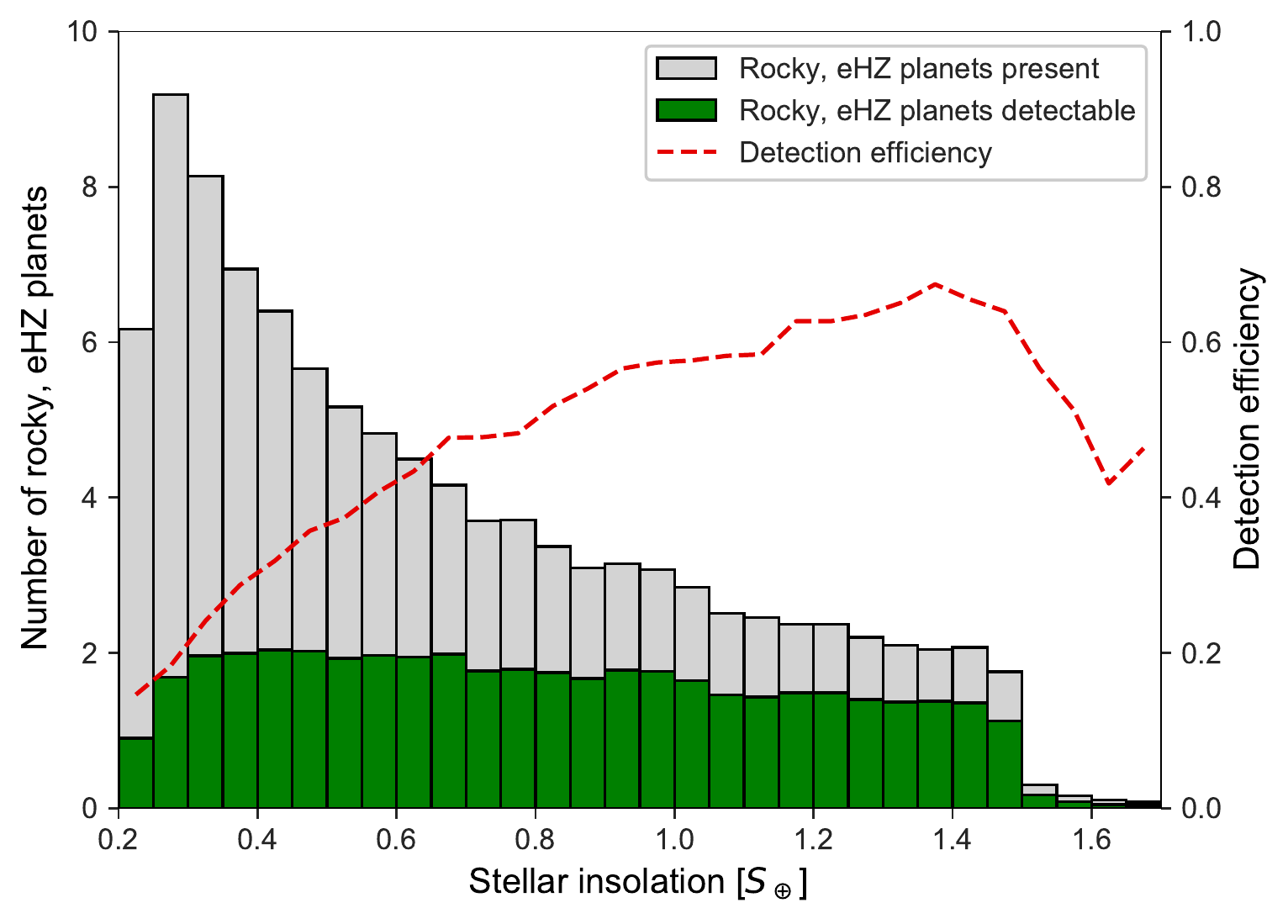}
    \caption{Detection efficiency for rocky eHZ exoplanets for our scenario 2. Top panel: Equilibrium temperature distribution of all exoplanets present in the surveyed sample (gray) and all detected exoplanets (blue). Bottom panel: Same as above, but as a function of stellar insolation. In both panels the detection efficiency (y axis on the right-hand side) is shown with the dashed red line.}
    \label{fig:hab_planets_efficiency}
\end{figure}

Another important parameter to look at is the detection efficiency, that is, the number of detectable rocky eHZ exoplanets relative to the total number of such exoplanets that were generated in our simulations. This is illustrated in Fig.~\ref{fig:hab_planets_efficiency}, which is based on the results for scenario 2. One can see that, depending on the received stellar insolation (or, to a first approximation, the resulting equilibrium temperature), only a certain fraction of the exoplanets is detected. As indicated above, the main reason is the required sensitivity rather than the spatial resolution. Still, some of the simulated exoplanets are indeed at an orbital phase where they escape a detection with the interferometer. However, 
it is reassuring that $\geq$50\% detection efficiency is achieved for exoplanets with $T_{\textrm{eq}}\geq225$ K, or insolations within $0.8\;S_{\oplus} \leq S_{\textrm p}\leq 1.5\;S_{\oplus}$. This number could be further increased by implementing a multi-visit search phase. Work on quantifying the gain in detection efficiency (and survey completeness) as a function of number of visits in currently ongoing.

\begin{figure*}[t!]
    \centering
    \includegraphics[width=0.48\linewidth]{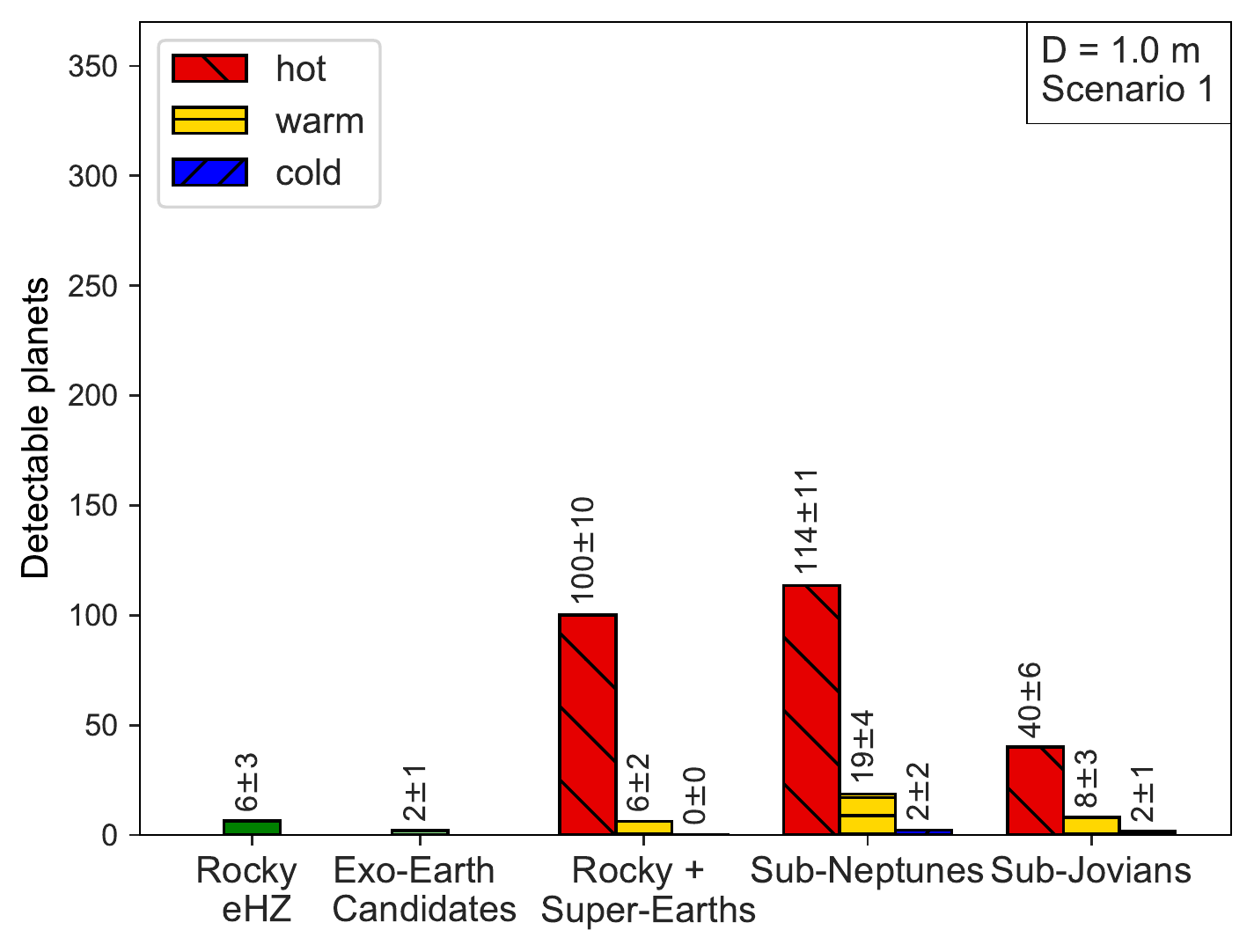}
    \includegraphics[width=0.48\linewidth]{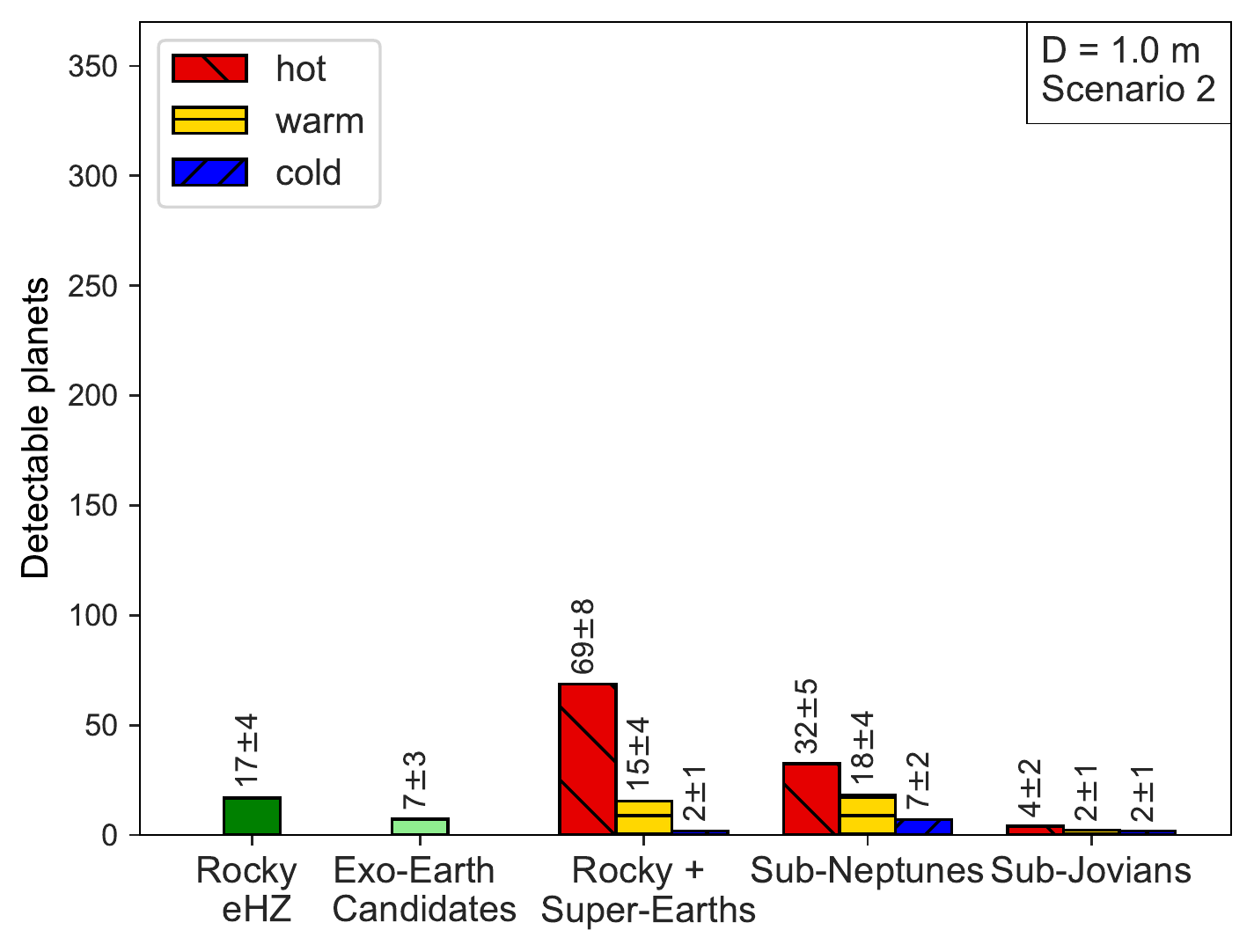}
    \caption{Same as Fig.~\ref{fig:baseline_yields_bars}, but now for $D=1.0$ m.}
    \label{fig:1m_yields_bars}
\end{figure*}

\begin{figure*}[t!]
    \centering
    \includegraphics[width=0.48\linewidth]{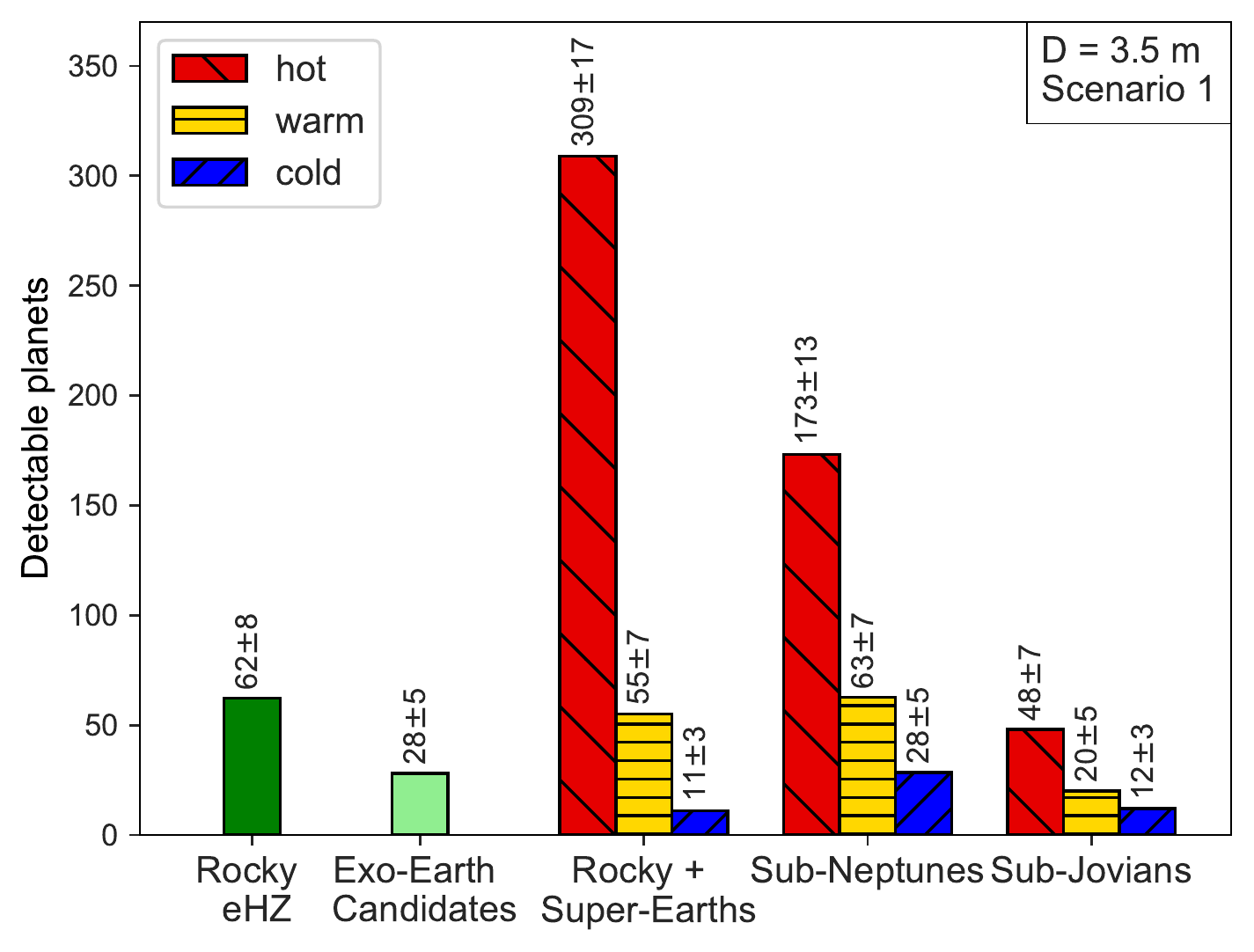}
    \includegraphics[width=0.48\linewidth]{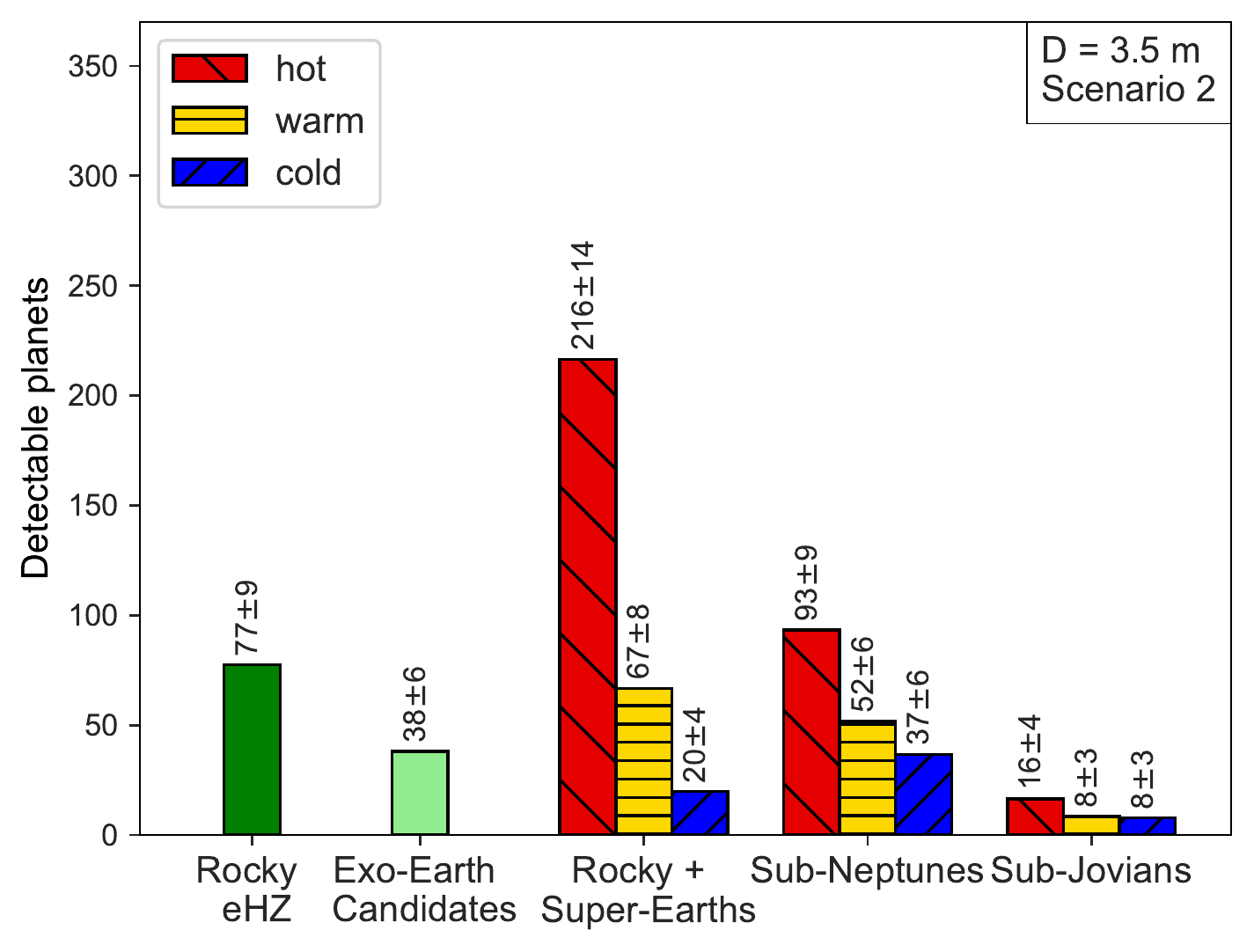}
    \caption{Same as Fig.~\ref{fig:baseline_yields_bars}, but now for $D=3.5$ m.}
    \label{fig:3.5m_yields_bars}
\end{figure*}

\subsection{Impact of aperture size and wavelength range}
\label{sec:res_apertures}

\label{sec:imp_apt_size}
\begin{figure}[t!]
    \centering
    \includegraphics[width=0.98\linewidth]{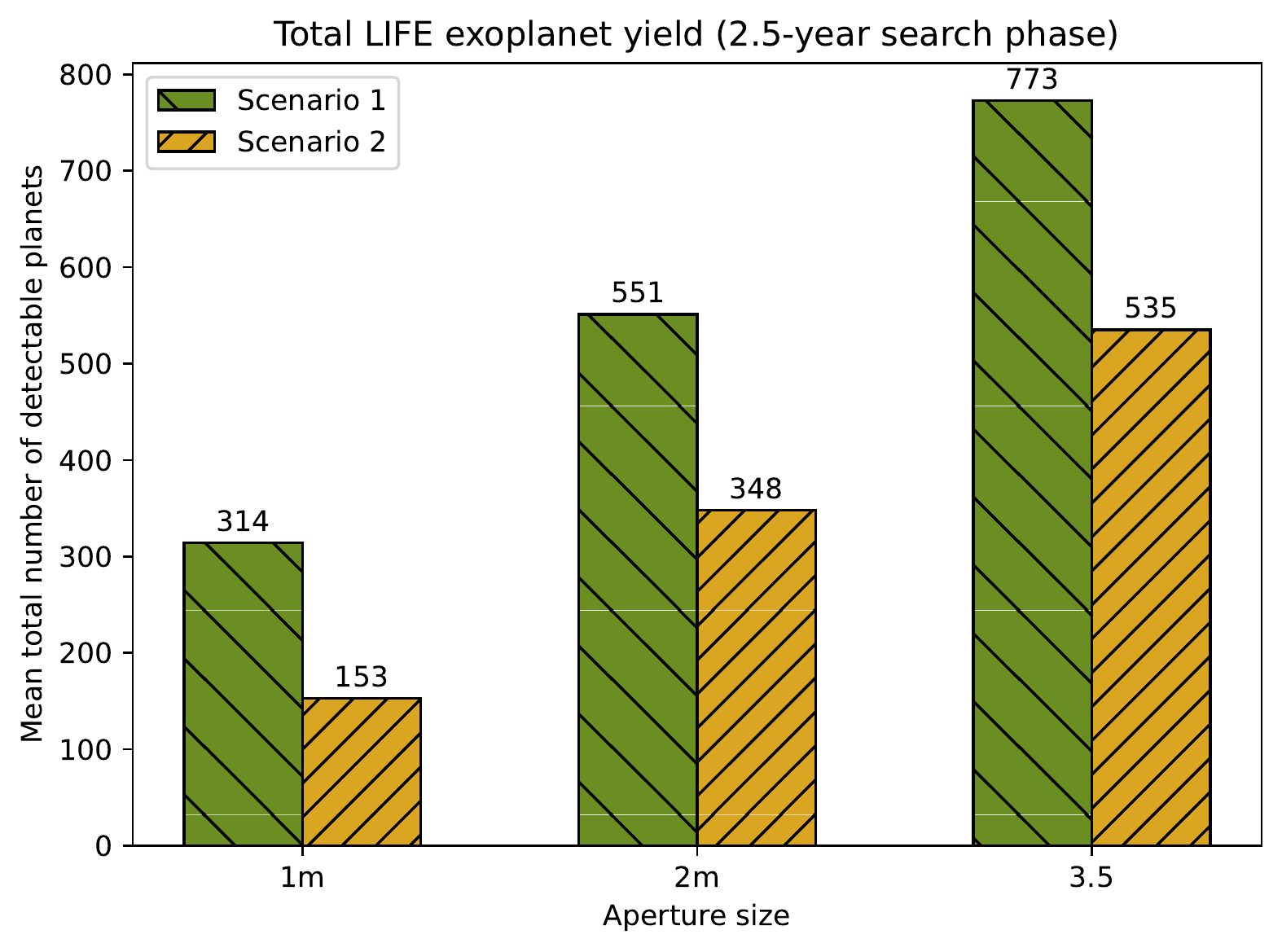}
    \caption{Total expected exoplanet detection yield as a function of aperture size and for each of the two scenarios. The numbers shown here are the sum of the mean numbers shown in the grid cells in Figs.~\ref{fig:baseline_yields_grid},~\ref{fig:1m_yields_grid}, and ~\ref{fig:3.5m_yields_grid}.}
    \label{fig:LIFE_yield_aperture_comparison}
\end{figure}

In Figs.~\ref{fig:1m_yields_bars} and~\ref{fig:3.5m_yields_bars} we show the expected detection yield for apertures with $D=1$ m and $D=3.5$ m, respectively. The format is the same as for the reference case shown in Fig.~\ref{fig:baseline_yields_bars}; the plots corresponding to Fig.~\ref{fig:baseline_yields_grid} are available in Appendix~\ref{sec:appendix_plots}, where we also show in Fig.~\ref{fig:LIFE_yield_aperture} the relative changes in yield compared to the reference case.  Figure~\ref{fig:LIFE_yield_aperture_comparison} provides a summary of the impact of the aperture size on the total \emph{LIFE} exoplanet detection yield during the 2.5-year search phase. It shows that, as expected, aperture size strongly affects the number of detectable exoplanets and it is important to point out that the gain (loss) when going to larger (smaller) apertures is most significant for small exoplanets of all temperatures and cool exoplanets of all sizes. Specifically, Figs.~\ref{fig:1m_yields_bars} and~\ref{fig:3.5m_yields_bars} show that in the case of $D=3.5$~m, the number of rocky eHZ exoplanets and EECs would increase to $\approx$63 (+132\%) and $\approx$28 (+161\%), respectively, in scenario 1. The corresponding numbers in scenario 2 are $\approx$78 (+78\%) and $\approx$39 (+88\%). The relative smaller gain in scenario 2 compared to scenario 1 is explained by the higher number of exoplanets already detected with the reference aperture size of $D=2$~m. In case of $D=1$~m, the number of rocky eHZ exoplanets and EECs would decrease to $\approx$6 (-76\%) and $\approx$2 (-81\%), respectively, in scenario 1. In scenario 2, the numbers would go down to $\approx$17 (-61\%) and $\approx$7 (-64\%) for rocky eHZ exoplanets and EECs, respectively.

\begin{figure}[]
    \centering
    \includegraphics[width=0.98\linewidth]{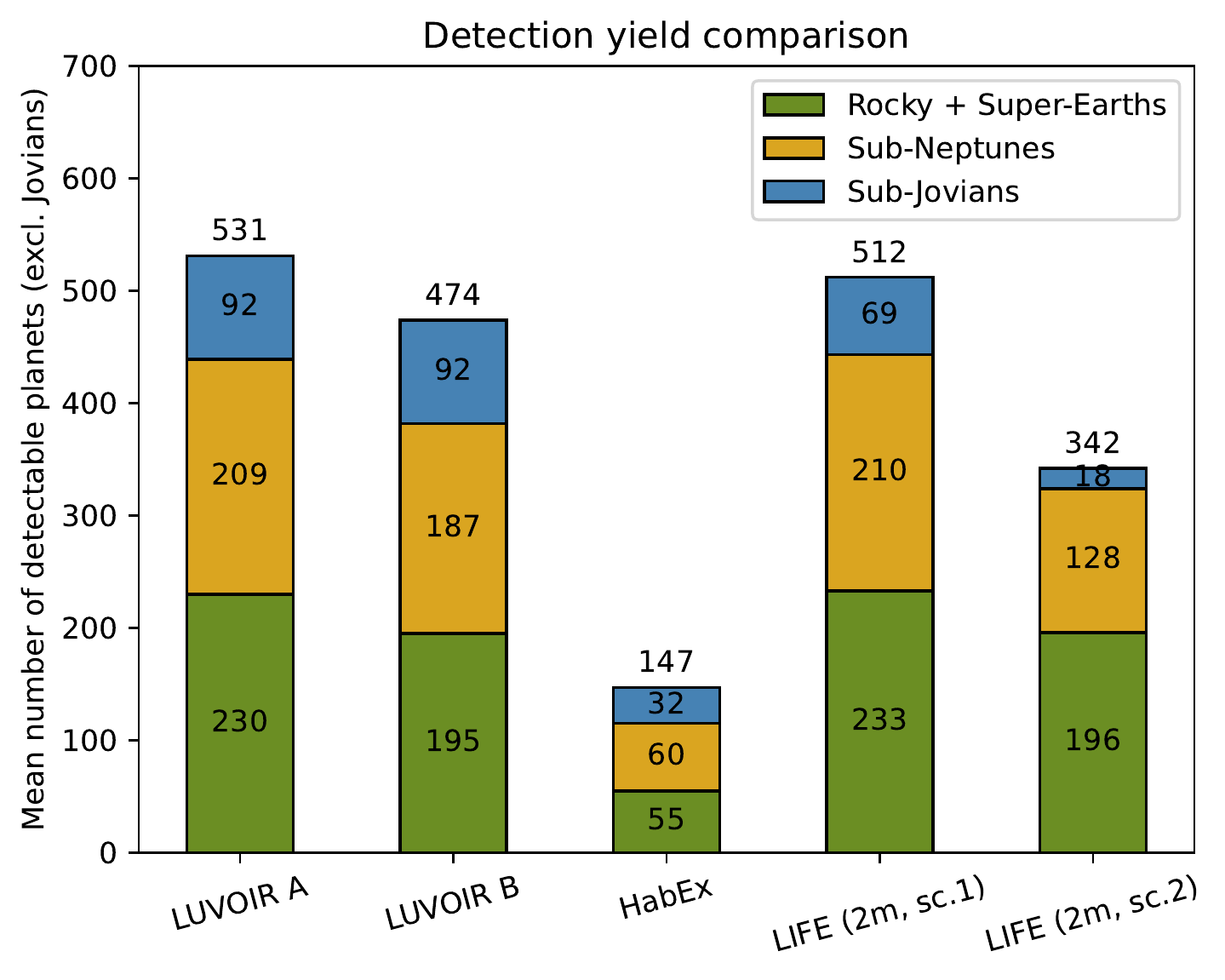}
    \caption{Detection yield comparison between \emph{LUVOIR A/B}, \emph{HabEx}, and \emph{LIFE}. For \emph{LIFE} we show the numbers for the $D=2$ m reference case and for \emph{HabEx} the numbers from the baseline 4-meter concept. Jovian planets are not shown, because they were not included in the \emph{LIFE} simulations (cf. Sect.~\ref{sec:planet_pop}; see text for important details).}
    \label{fig:yield_comparison_missions}
\end{figure}

The effect of changing the wavelength range is much weaker by comparison, and generally the number of detectable planets only increases or decreases by a few percent. Figures~\ref{fig:3-20_yields_grid} and \ref{fig:6-17_yields_grid} in Appendix~\ref{sec:appendix_plots} show the results in the same format as Fig.~\ref{fig:baseline_yields_grid}, and the changes relative to the reference case with $\lambda=4 - 18.5\,\mu m$ are shown in Fig.~\ref{fig:LIFE_yield_lambdarange}.
\section{Discussion}
\label{sec:discussion}
\subsection{Total exoplanet yields}

\begin{figure*}[h!]
    \centering
    \includegraphics[width=0.49\linewidth]{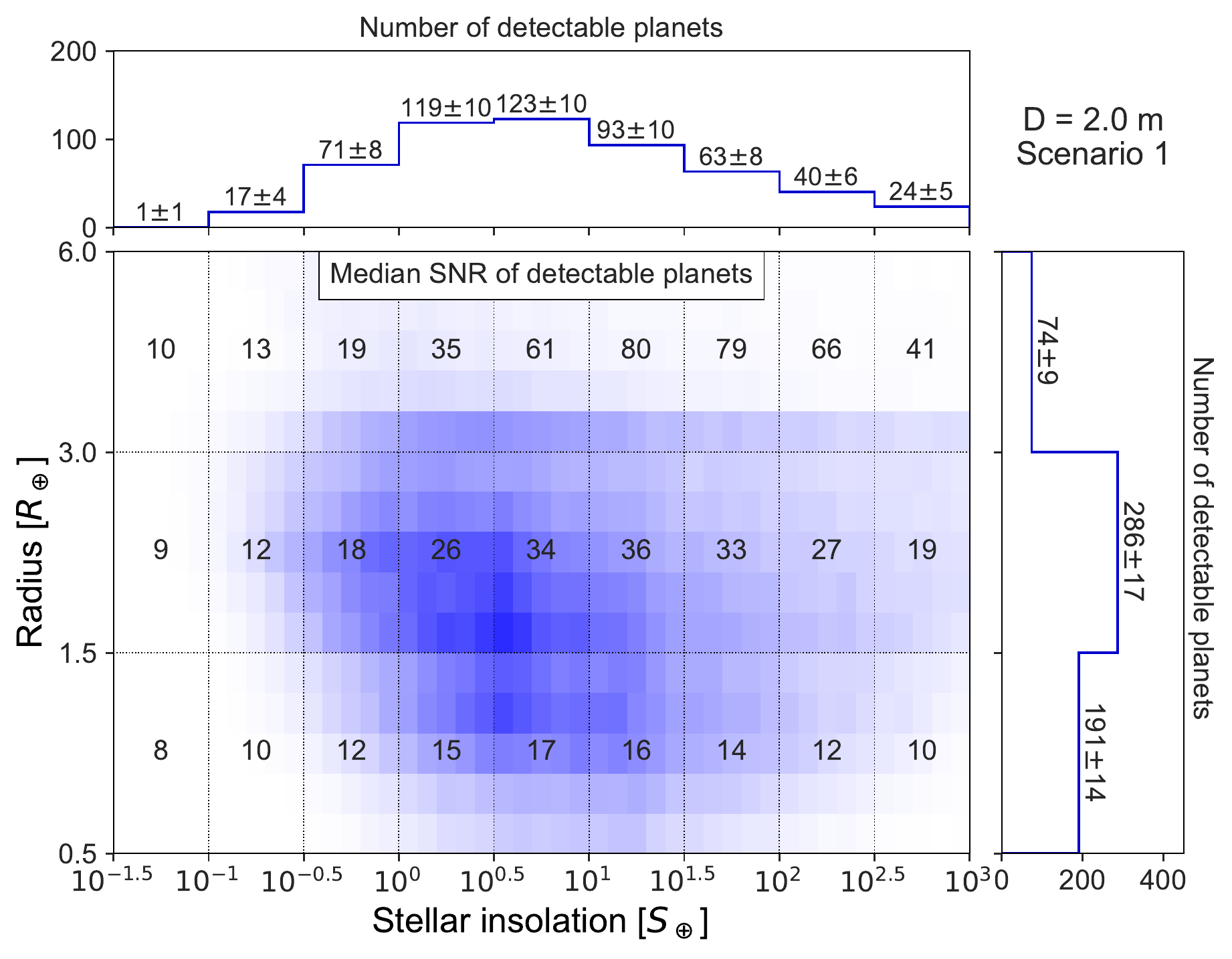}
    \includegraphics[width=0.49\linewidth]{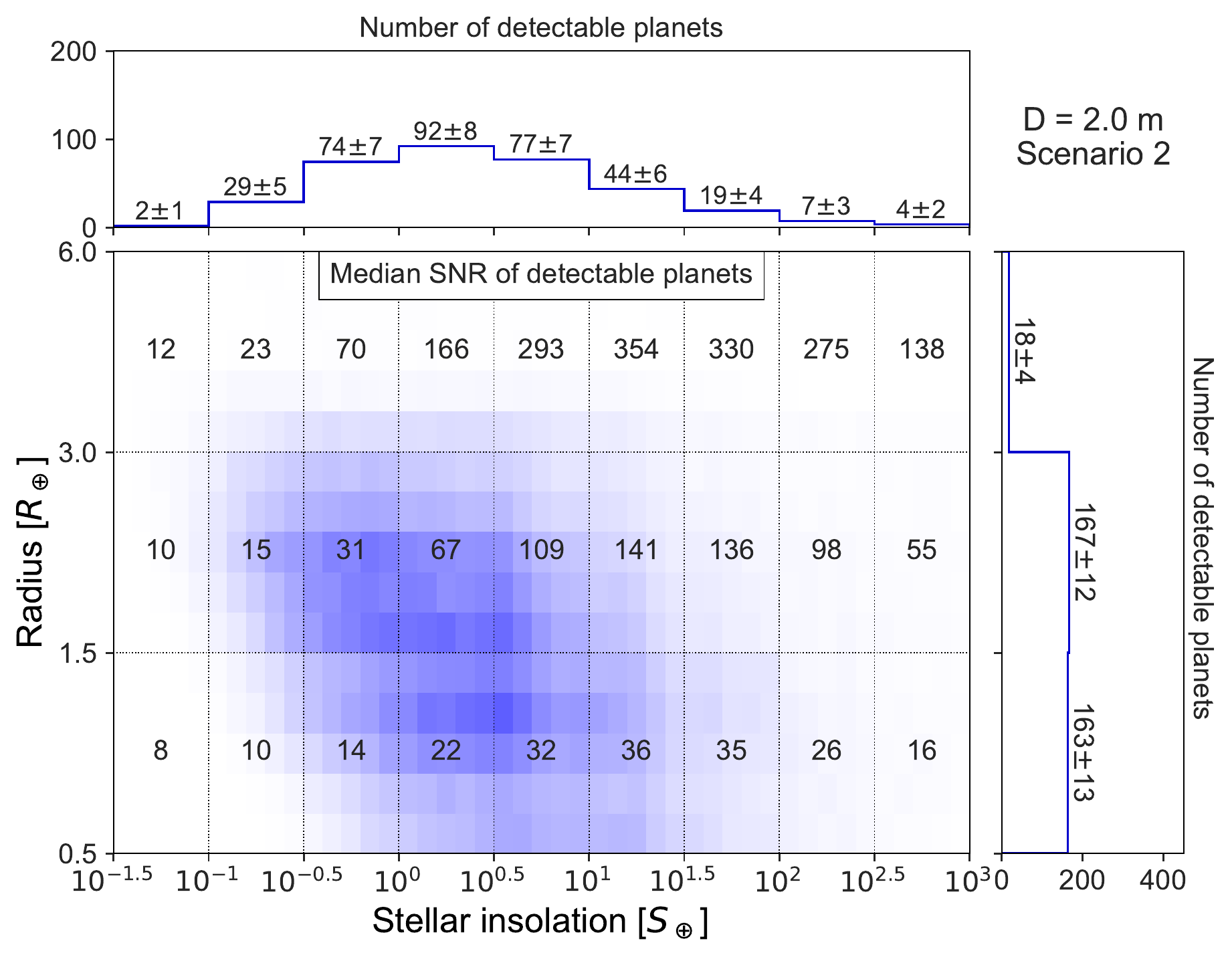}
    \caption{Median S/N of the detected exoplanets in our reference case scenario simulations in the radius vs. stellar insolation plane (left panel: Scenario 1; right panel: Scenario 2). We note that the 1D distributions on top and to the right of the grids (as well as colored area) represent the numbers of detected exoplanets, including the 1-$\sigma$ uncertainties shown in Fig.~\ref{fig:baseline_yields_grid}, and not the marginalized distributions of the S/N.}
    \label{fig:noise_grid}
\end{figure*}

Looking at the total number of detectable exoplanets and their properties reveals how diverse the expected \emph{LIFE} exoplanet yield will be. This sample spans about four orders of magnitudes in planet insolation and about a factor of 10 in planet radius. In addition to investigations concerning the habitability of a subset of the sample, \emph{LIFE} has the potential to address a number of scientific questions related to the formation and evolution of exoplanets and their atmospheres. 

\label{sec:dis_yields}

In Fig.~\ref{fig:yield_comparison_missions} we provide a comparison with the detection yields published in the \emph{HabEx} and \emph{LUVOIR} study reports \citep{habex2019,luvoir2019}\footnote{See \cite{stark2019} for details on the yield calculations for the reflected light missions.}. This plot suggests that for exoplanets with radii up to 6 R$_\oplus$, \emph{LIFE}, with four times $D=2$ m apertures, can achieve overall detection yields comparable to those of the \emph{LUVOIR-A} (15-meter aperture) and \emph{LUVOIR-B} (8-meter aperture) concepts; \emph{HabEx}, with a 4-meter primary mirror, is predicted to yield fewer detections. It needs to be noted, though, that while in our simulations planets with radii $>$6 R$_\oplus$ were not included, \emph{LUVOIR A} and \emph{B} and \emph{HabEx} are predicted to detect $\approx$117, $\approx$102 and $\approx$31 of these Jovian planets, respectively. It is also important to mention that the numbers for \emph{LIFE} are the sum of numbers for the various planet types shown in Fig.~\ref{fig:baseline_yields_bars}. The resulting overall numbers of detectable planets differ slightly from those shown in Fig.~\ref{fig:LIFE_yield_aperture_comparison} because some detectable planets fall outside the insolation ranges defined in Table~\ref{table:1}.

Overall, these results show that in principle both approaches, large, single-aperture reflected light missions and interferometric MIR missions (under the assumptions laid out in Sect.~\ref{sec:simulations}), offer unprecedented opportunities for the direct detection and detailed investigations of hundreds of nearby exoplanets. Going forward, it will be important to investigate possible scientific synergies between missions such as \emph{HabEx} or \emph{LUVOIR} and \emph{LIFE} because at least a subset of the exoplanets detected by one approach is likely also detectable by the other. 

\begin{figure*}[t]
    \centering
    \includegraphics[width=0.95\linewidth]{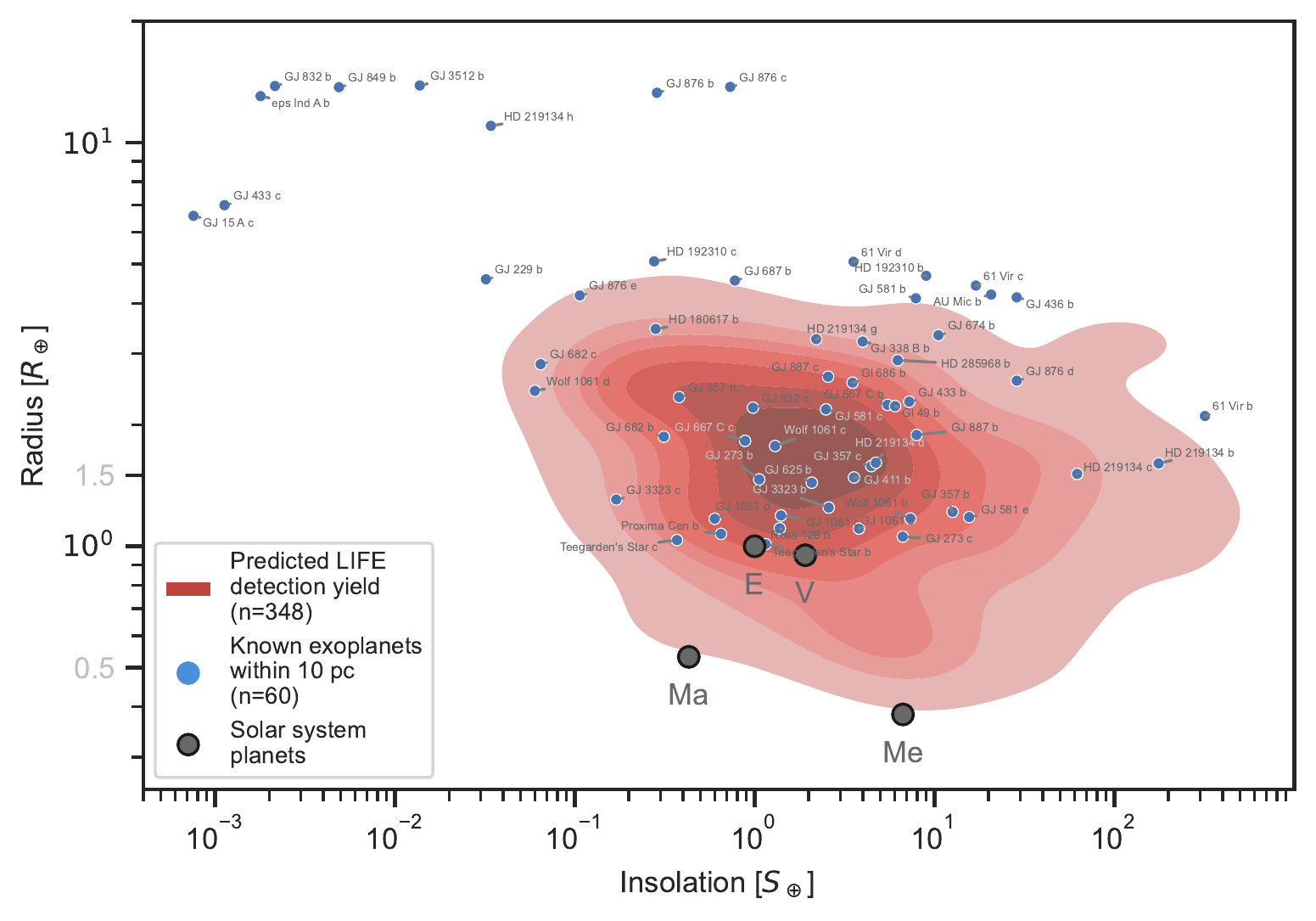}
    \caption{Comparison of exoplanet detections with \emph{LIFE} to known Solar System planets and exoplanets. The \emph{LIFE} yield for the reference case (scenario 2) is shown in red contours using a kernel density estimate of the detected sample. Every shaded contour level corresponds to 50 exoplanets detected in the respective parameter space. Blue points represent 60 out of the 79 known exoplanets within 10 pc of the Sun for which we could estimate the radius and insolation level. Gray points represent the four rocky planets in the Solar System (E=Earth, V=Venus, Ma=Mars, and Me=Mercury).}
    \label{fig:known_planets}
\end{figure*}

The single most important parameter related to the number of detectable exoplanets is, unsurprisingly, the aperture size of the collector spacecraft. While here we focus on a 2-D array architecture for the interferometer with four collector spacecraft with aperture sizes between $D=1$ m and $D=3.5$ m, \citet{dandumont2020} recently presented a similar yield analysis based on 4 different implementations of a two-aperture Bracewell interferometer: 2 CubeSat options (with a 0.5 or 1 m baseline and 0.08 m apertures), a Proba mission option (with a 5 m baseline and 0.25 m apertures), and the \emph{Fourier Kelvin Stellar Interferometer} concept presented in \citet{danchi2008} and \citet{danchi2010} (with a 12 m baseline and 0.5 m apertures). The trend shown here continues down to CubeSat apertures and the detection of at least $\approx$10--15 rocky eHZ exoplanets requires an aperture size of at least $D=1$ m. The strong dependence of the detection yield on the aperture size results from the fact that, in the vast majority of cases, the local zodiacal dust emission is an important noise term and the effective FoV of the collector spacecraft scales with $\lambda/D$ (cf. Sect.~\ref{sec:noise_sims}). In Appendix~\ref{sec:appendix_plots2} we provide an overview of the relative contributions of the various noise terms to the total noise for planets detected in the reference case scenarios (Fig.~\ref{fig:noise_contributions}).

Another key result from our analyses is that, depending on how the observing time is distributed amongst the stellar targets, both the number of detected exoplanets and the type of detected exoplanets can vary significantly. This illustrates a strong need for the community to clearly define and  prioritize the scientific objectives of such a mission in order to derive the appropriate observing strategy. An additional parameter that needs to be considered in this context is the completeness of the survey, that is, how important it is to have detected, with a certain level of confidence, all (or at least most) exoplanets from a specific subset of the exoplanet population in the solar neighborhood. Higher completeness requires more observing time per target star (including multiple visits) and hence a lower number of detectable exoplanets overall. 

The results from the search phase, and also its duration, have an immediate impact on the follow-up strategy during the characterization phase of the mission. It is hence important to understand how well the fundamental properties of the detected exoplanets (such as radius and temperature, but also their orbital position) can be constrained from single-epoch data. 
In  Fig.~\ref{fig:noise_grid} we present some first indications by showing the median S/N of the detected exoplanets as a function of their radius and insolation. Because of the much longer average integration time per star in scenario 2 (cf. Sect.~\ref{sec:res_yields}), the median S/N is, in many cases, significantly higher than in scenario 1. Interestingly, 
the exoplanets in most grid cells (and certainly warm and hot exoplanets with radii $>1.5~R_{\oplus}$) are detected with sufficiently high S/N that some first-order estimate of their radius and effective temperature and maybe even a rough analysis of their SED based on (very) low-resolution spectroscopy appears feasible. This aspect needs to be investigated further as the possibility to obtain spectral information already from single-epoch data allowing for a first characterization and classification of the exoplanets has an impact on the follow-up  strategy during the characterization phase. For completeness we note that the largest, hottest planets receiving the highest levels of insolation do not show the highest median S/N. This is because the S/N is related to the location of the exoplanets in the transmission map of the interferometer, which maximizes the throughput for exoplanets located in the eHZ and not for close-in exoplanets (cf. Sect.~\ref{sec:noise_sims}).

\begin{figure*}[t!]
    \centering
    \includegraphics[width=0.33\linewidth]{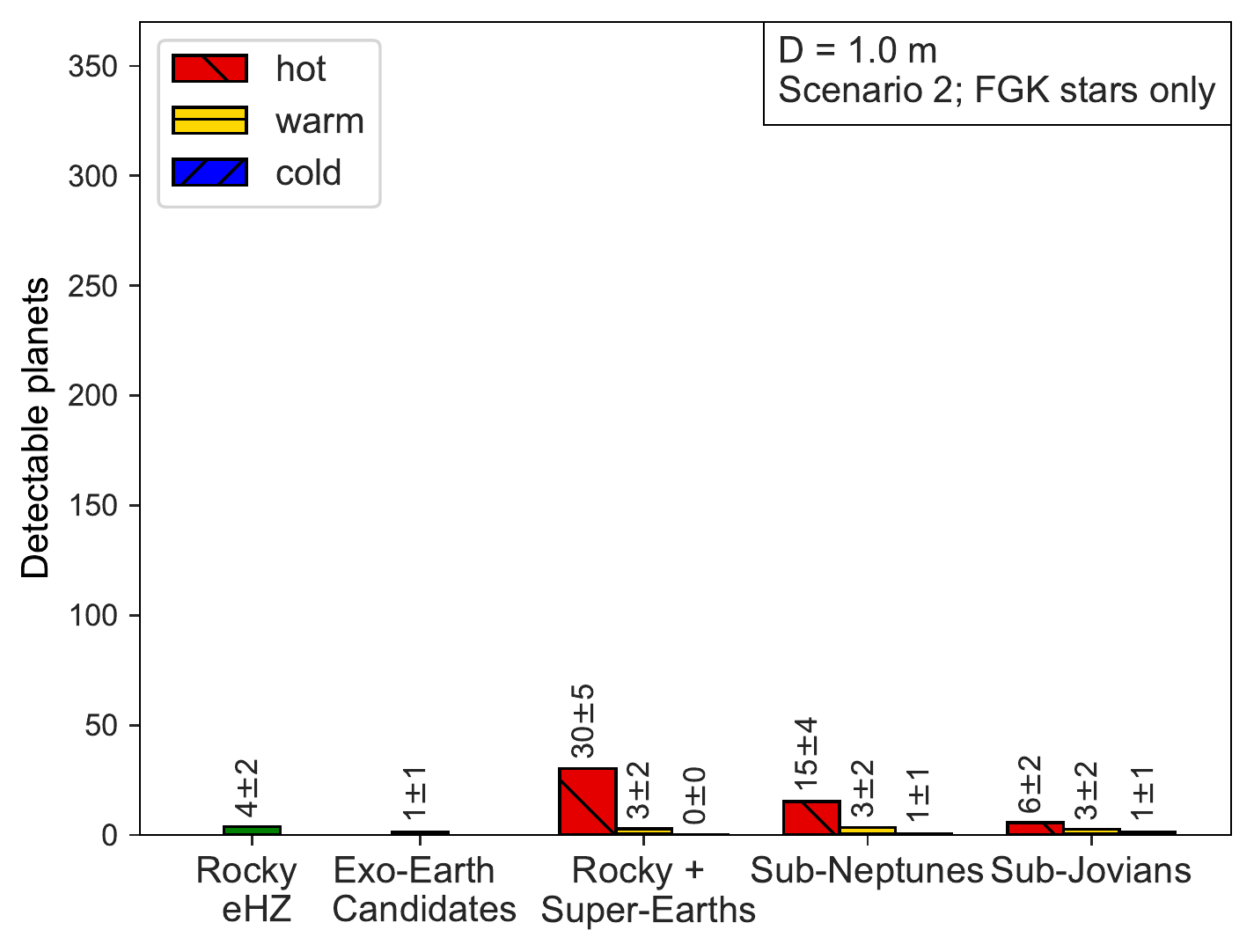}
    \includegraphics[width=0.33\linewidth]{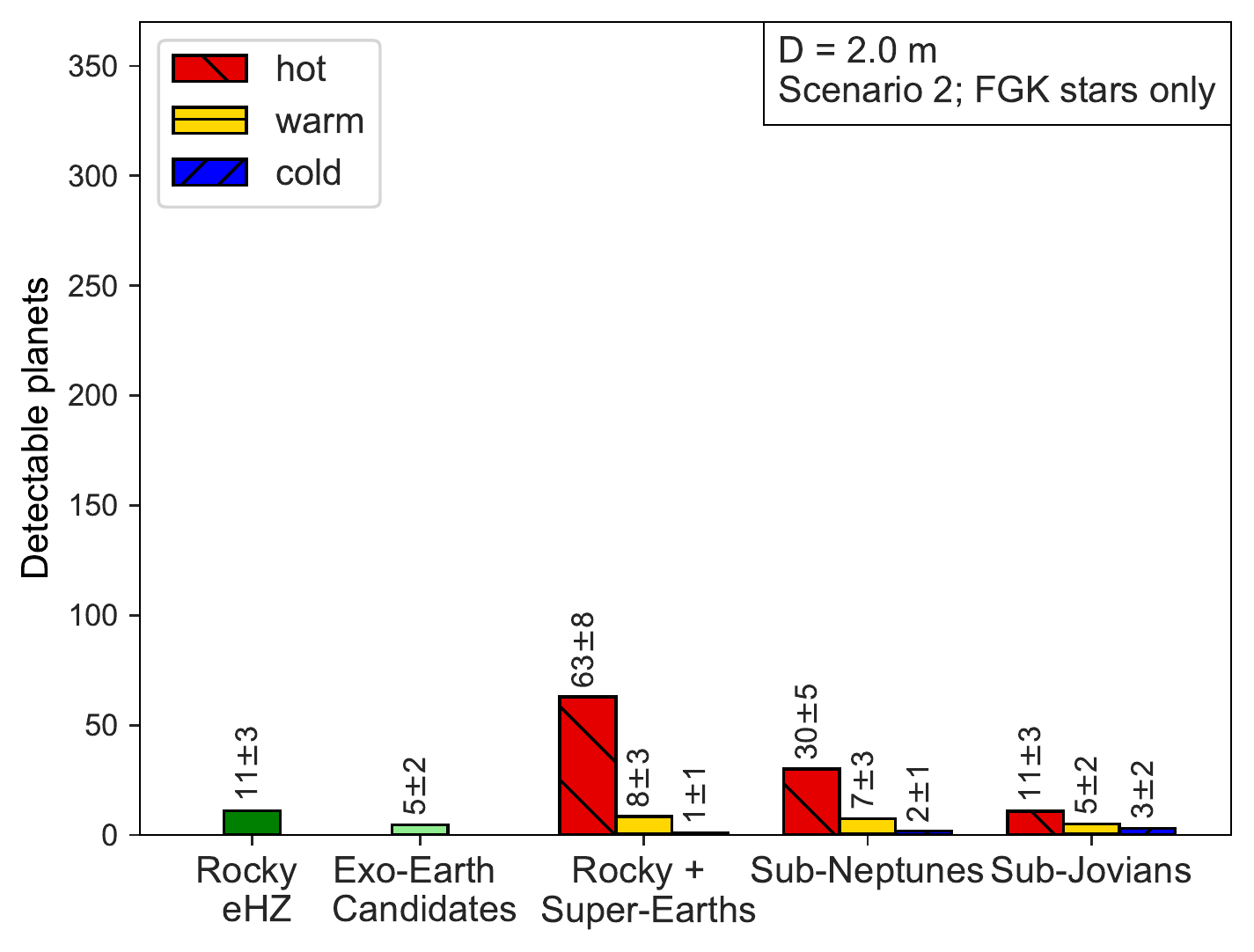}
    \includegraphics[width=0.33\linewidth]{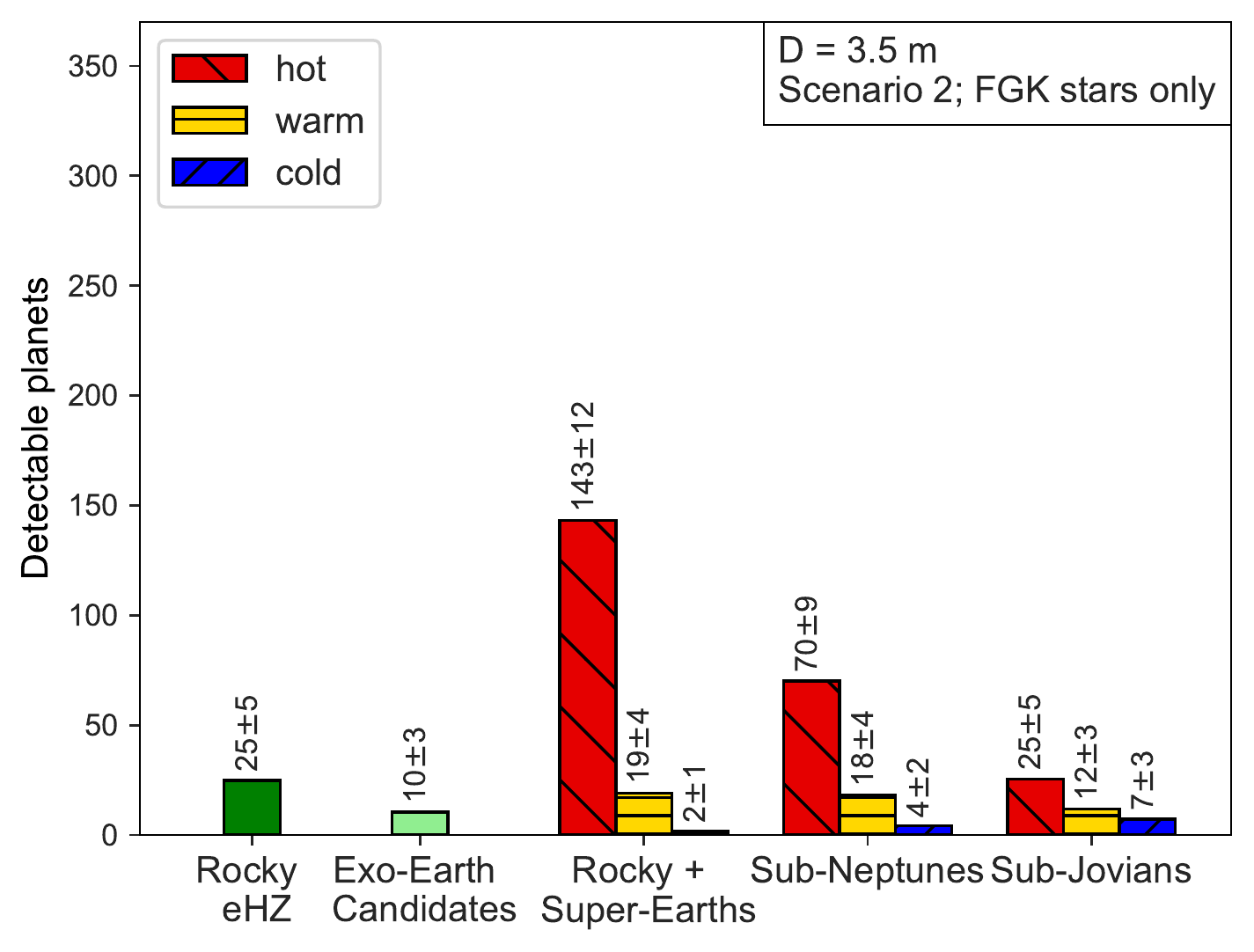}
    \caption{Same as the right panel in Fig.~\ref{fig:baseline_yields_bars}, but ignoring all M-type dwarfs in the target catalog and spending the full search phase on FGK stars only. The panels show the results for $D=1.0$~m, $D=2.0$~m, and $D=3.5$~m from left to right, respectively.}
    \label{fig:yields_no_m_stars}
\end{figure*}

Finally, whether or not future exoplanet imaging space missions will have to carry out a somewhat extended search phase, will also depend on the progress and results of ongoing and future ground-based RV surveys. In Fig.~\ref{fig:known_planets} we show a comparison between the expected \emph{LIFE} detection yield (reference case; scenario 2) and currently known exoplanets within 10 pc of the Sun drawn from the NASA Exoplanet Archive\footnote{\url{https://exoplanetarchive.ipac.caltech.edu}}. If the insolation is not provided for planets in the archive, it is calculated via $L_\star[\mathrm{L_\odot}] / a[\mathrm{AU}]^2$, with $L_\star$ the host star luminosity, $L_\odot$ the solar luminosity, and $a$ the semimajor axis of the exoplanet orbit. A missing radius measurement is estimated using \texttt{forecaster}\footnote{\url{https://github.com/chenjj2/forecaster}} \citep{chen2016}. This leads to 60 out of the 79 confirmed exoplanets within 10 pc for which we can assign both radius and insolation. Fig.~\ref{fig:known_planets} shows on the one hand that there is an interesting sample of planets already known within 10 pc from which a preliminary target list could be compiled. On the other it demonstrates that one expect a factor of 5 more planets to be found within 10 pc with \emph{LIFE}. New (or the continuation of) systematic RV exoplanet searches in the solar neighborhood will be fundamentally important to either provide future imaging missions with a predefined exoplanet target list or at least provide them with stringent constraints on the existence of nearby planetary systems. The same is true for systematic searches of exozodi disks around nearby stars. As mentioned above, already during the search phase typical integration times are easily on the order of days. This means that S/Ns $>$5 per spectral channel\footnote{First science requirements for the spectral resolution and wavelength coverage of \emph{LIFE} are presented im \citep{konrad2022}. Previous works in this direction and in the context of atmospheric characterization of terrestrial exoplanets at MIR wavelengths suggested spectral resolutions of up to $R\approx40$ \citep[e.g.,][]{desmarais2002,leger2019}.} are costly and knowing interesting or promising targets beforehand saves valuable observing time.

\subsection{Rocky HZ exoplanets: M-star preference and detection efficiency}
\label{sec:dis_preference}

As shown in Fig.~\ref{fig:baseline_yields_bars}, \emph{LIFE} could detect $\approx$25--45 of rocky exoplanets located within the  eHZ of their host stars and $\approx$10--20 EECs following the definition of \citet{kopparapu2018}. However, these numbers are strongly affected by the aperture size of the collector spacecraft. As shown in Figs.~\ref{fig:1m_yields_bars} and~\ref{fig:3.5m_yields_bars}, decreasing (increasing) the aperture size yields a significant decrease (increase) in the number of rocky temperate exoplanets. These findings will be of great importance during upcoming trade-offs between mission cost, where aperture size will be an important parameter, and science return. We stress that this is not only important for the search phase, but it is even more relevant for the characterization phase that aims at investigating a subsample of the detected exoplanets in greater detail with high-S/N spectra \citep{konrad2022}.

Following similar arguments presented in \citet{stark2014,stark2015}, \citet{quanz2021} argued that in order to obtain statistically robust results on the fraction of rocky HZ exoplanets that are indeed habitable, at least 30 (better 50) exoplanets in that part of parameter space need to be studied. According to the numbers presented above, this appears to be achievable with \emph{LIFE}. However, the vast majority of these planets is found around M stars and at this point in time it is unknown whether exoplanets orbiting within the HZ around M stars are able to retain (secondary) atmospheres because of the high activity of M-type stars, in particular at young ages \citep[e.g.,][]{feng2015,luger2015,lingam2018,godolt2019,atri2021}. It has been shown that under certain circumstances such exoplanets, which are very likely tidally locked, may still provide habitable conditions \citep[e.g.,][]{leconte2015,ribas2016,turbet2016,boutle2017}, but empirical data are still lacking. There is, however, hope that \emph{JWST} will be able to address this fundamentally important question and, for a few cases, investigate the existence of atmospheres of rocky exoplanets transiting M stars \citep[e.g.,][]{koll2019}. Also, a deep characterization effort of the potential M-star targets should be carried out, including the high-energy radiation budget and its past history, in order to understand which stars may have provided a more quiescent environment for their expected planets. If rocky exoplanets orbiting M stars can retain atmospheres, then \emph{LIFE} is in an excellent position to robustly characterize a significantly larger sample. If not, then one may want to reconsider the observing strategy and possibly de-prioritize M stars in the stellar input catalog. Figure~\ref{fig:yields_no_m_stars} shows the results for the most extreme case, where all M stars are ignored and the full search phase is spent on FGK stars. In this case, $\approx$25 rocky eHZ exoplanets can be expected assuming an aperture size of $D=3.5$~m; with $D=2.0$~m this number would drop to $\approx$11, limiting the statistical power of the analysis. In this context, two  points are important to mention: (a) in order to further increase the number of detected rocky eHZ planets around FGK stars, it will be important to investigate how a search phase with a multi-visit strategy would affect the results. We remind the reader that a detection efficiency $\geq$50\% is currently achieved for exoplanets with $T_{\textrm{eq}}\geq225$ K or insolations within $0.8\;S_{\oplus} \leq S_{\textrm p}\leq 1.5\;S_{\oplus}$. Ignoring the M stars and repeating the analysis shown in Fig.~\ref{fig:hab_planets_efficiency} for FGK stars only reveals that the overall detection efficiency is indeed lower (see Fig.~\ref{fig:detection_efficiency_FGK} in Appendix~\ref{sec:appendix_plots}). Hence, we can expect to gain additional detections when multiple visits per star are applied; (b) As we discuss in Sect.~\ref{occurrence_rates}, our underlying occurrence rates for rocky, temperate planets around FGK stars may be on the rather conservative side.

Similar to the expected total detection yield, also the number of predicted EECs can be compared to those published in the \emph{HabEx} and \emph{LUVOIR} study reports \citep{habex2019,luvoir2019}. \emph{HabEx}, with its 4-meter baseline concept, is expected to detect $\approx$8 EECs, while \emph{LUVOIR-A} and \emph{LUVOIR-B} are predicted to directly image $\approx$54 and  $\approx$28 EECs, respectively. Considering the  $\approx$20 EECs that \emph{LIFE} is expected detect (assuming $D=2$ m and scenario 2), one has to keep in mind its preference for planets around M stars, while \emph{HabEx} and \emph{LUVOIR} have a strong detection bias for planets around solar-type stars. This suggests that there is only limited overlap in primary discovery space for EECs between the missions if they were to carry out independent search phases. Hence, it will be important to check the potential overlap assuming that one mission is following-up after the other (e.g., \emph{LIFE} following after \emph{LUVOIR/HabEx}). In addition, looking more carefully at the underlying EEC occurrence rates, it shows that the numbers cannot be directly compared (see Sect.~\ref{occurrence_rates} below).

\subsection{Remaining limitations and uncertainties in the simulations}
\label{sec:dis_limitations}

\subsubsection{Treatment of noise sources}
Compared to previous works the yield simulations presented here are based on a more realistic treatment of the observing technique and include all major astrophysical noise terms. One of the next crucial steps is to continue the development of an instrument concept including a noise breakdown structure so that quantitative instrumental noise estimates can be included in the simulations. The calculations from \cite{lay2004} indicate that the noise contributions from photon noise and instrumental noise may indeed not be very different. However, these calculations were done for a specific example and how the relative contributions scale with stellar and planet properties and exozodi brightness needs to be investigated. Also, recent work by \citet{dandumont2020} and other previous analyses in the context of \emph{TPF-I} \cite[e.g.,][]{lay2006} or \emph{Darwin} \citep[e.g.,][]{defrere2010} can serve as starting points. In addition to the noise budget, important instrument parameters such as overall throughput and detector quantum efficiency need to be further validated. For the interested reader we provide a summary of the status of some key technologies relevant for \emph{LIFE} in Appendix~\ref{sec:appendix_technologies}.

\subsubsection{Treatment of exozodiacal and zodiacal dust}\label{zodidiscussion}

While we do take into account emission from potential exozodiacal dust disks using the nominal distribution from the HOSTS survey \citep{ertel2020}, it needs to be acknowledged that there is still considerable uncertainty in the median exozodi level: while in the nominal distribution the median zodi level is $\bar{z}$$\approx$3.2, \citet{ertel2020} show that one can only be confident at the 1$\sigma$ level that the median is below 9 zodis and at the 2$\sigma$ level that it is below 27 zodis. In order to quantify the impact of these uncertainties on the detection yield, we did the following experiment for the reference case scenarios: the exozodi level distribution was shifted by adding multiples of the median absolute deviation (MAD) of the distribution (MAD($z$)$\approx$2.7) to each individual exozodi level in the sample. In the most extreme case we analyzed the distribution was shifted by 9$\cdot$MAD, resulting in a median $\bar{z}$$\approx$27 corresponding to the 2$\sigma$ level mentioned above. In this case, the total number of detectable planets decreased by $\lesssim$6\% and the number of rocky, HZ planets changed even less. One reason for this somewhat limited impact is that, compared to noise from stellar leakage and local zodiacal dust emission, noise from exozodiacal dust disks contributes only little to the total noise budget of detected planets (see Fig.~\ref{fig:noise_contributions}). Hence, in a statistical sense, the current uncertainties may not have a strong impact on the overall results. Still, additional observational efforts determining exozodi levels would further improve upon the current statistical uncertainties and, maybe even more importantly, would also help prioritize the most promising individual targets for future space missions. In addition, at the moment, the HOSTS survey does not show a correlation between spectral type and the level of exozodi emission \citep{ertel2020}, and hence we apply the same underlying distribution of exozodi levels to all target stars irrespective of spectral type. A larger data set would be required to further confirm this current finding. 

Furthermore, the inclination of exozodiacal dust disks and possible spatial offsets and substructures \citep[e.g., ``clumps,'' such as\ those seen in the zodiacal light;][]{reach2010,krick2012} are not considered in our simulations. As long as exozodiacal dust disks are centrally symmetric and optically thin, their contribution to the photon noise in a \emph{LIFE} measurement is to first order independent from their inclination; hence, changing the inclination of the disks has no measurable impact on the results. Spatial offsets and disk substructures would, however, have an impact on the S/N calculations. \citet{defrere2010} looked at the specific case of an Earth-Sun twin located at 15 pc. They modeled planet induced resonant structures in exozodi disks with varying dust density and inclination and investigated up to what exozodi level the planet would still be detectable. They concluded that around 10 $\mu$m wavelength, up to $\sim$15 and $\sim$7 zodis are acceptable for disks with inclinations between 0-30$^\circ$ and up to 60 $^\circ$, respectively. For edge-on systems this limit drops to $\sim$1.4 zodis. In order to further refine the results presented here, analyses as the ones presented in \citet{defrere2010} could to be carried out, possibly enlarging the covered parameter space. However, as already noted by the authors, advanced signal processing approaches may further relax the constraints in terms of acceptable zodi levels mentioned above. Also, if one considers the nominal distribution of exozodi levels published by the HOSTS team, the above mentioned zodi level limits appear to be on the high-end side: about two-thirds of our simulated systems with randomly assigned zodi levels based on the nominal distribution have disks with $\le$7 zodis. Hence, we do not expect exozodi dust disks to be a show-stopper for \emph{LIFE}. However, systems that are seen (close to) edge-on may pose a real challenge and have to be investigated in more detail, and a coordinated effort to reduce the existing uncertainties in the median exozodi level of nearby stars remains certainly important.

 \citet{defrere2010} also addressed the question of disk offsets, where the geometric center of the exozodi disk is shifted away from the center of the star, which would lead to additional flux through the modulation map of the interferometer. They focused on a Sun-like star at 15 pc and considered only face-on systems. When looking at the modulated signal covering the full wavelength range, systems with up to $\sim$50 zodis and offsets as large as $\sim$0.5 au were considered acceptable. As the offset increases, the level of acceptable zodis decreases, but for offsets as large as 1 au, $\sim$5 zodis could still be tolerated. It hence seems that, apart from potentially extreme cases, disk offsets are not a major concern for \emph{LIFE}. For reference: in the Solar System, the center of the zodiacal cloud is shifted by only about 0.013 AU from the Sun \citep{landgraf2001}.

Similarly to the exozodi disks, also our zodiacal dust model does not contain any substructures. As mentioned above, and shown in Fig.~\ref{fig:noise_contributions}, the MIR emission from the zodiacal dust is an important astrophysical noise source in a typical \emph{LIFE} observation and the model should hence be further refined to correct for the current overestimation of the emission shortward of 6 $\mu$m.

\subsubsection{Occurrence rates of small, temperate exoplanets}
\label{occurrence_rates}

An additional uncertainty in our results is related to the underlying exoplanet population. While in some parts of the parameter space the occurrence rate of exoplanets was robustly measured by the Kepler mission, there remains significant uncertainty related to the completeness and reliability for the occurrence rates of rocky, temperate exoplanets around Sun-like stars \citep[e.g.,][]{bryson2020}. We note that recent estimates for $\eta_\oplus$, which is the fraction of stars with terrestrial exoplanets within their HZ, are higher than the ones resulting from our underlying distributions: \cite{bryson2020b} provided two values, $\eta_\oplus^{\rm o}=0.58^{+0.73}_{-0.33}$ and $\eta_\oplus^{\rm o}=0.88^{+1.28}_{-0.51}$, for the occurrence rate of planets with radii between 0.5 and 1.5 $R_\oplus$ orbiting in the eHZ of stars with effective temperatures between 4800 and 6300 K\footnote{Our definition of the eHZ is identical to their ``optimistic'' HZ case; the superscript ``o'' in $\eta_\oplus^{\rm o}$ refers to the word ``optimistic''.}. These bounds represent two extreme assumptions about the extrapolation of completeness beyond orbital periods where the Kepler DR25 completeness data are available. For EECs around solar-type stars, \citet{bryson2020b} found  a lower bound of $\eta_\oplus^{\rm EEC}=0.18^{+0.16}_{-0.28}$ and an upper bound of $\eta_\oplus^{\rm EEC}=0.28^{+0.30}_{-0.09}$. In our simulations, $\eta_\oplus^{\rm o}\approx 0.37$ and $\eta_\oplus^{\rm EEC}\approx0.16$ for FGK dwarfs (and $\approx$0.56 and  $\approx$0.31 for M dwarfs, respectively; see Table~\ref{table:rocky_planets}). Hence, in particular for solar-type stars, we might be underestimating the number of EECs and rocky, eHZ planets. Also, our value for $\eta_\oplus^{\rm EEC}$ is lower than the one used in the \emph{HabEx} and \emph{LUVOIR} concept studies, which was $\eta_\oplus^{\rm EEC}=0.24^{+0.46}_{-0.16}$. One reason for this difference is that we did not keep this parameter constant throughout our stellar sample, but we let it vary between various spectral types (see notes in Table~\ref{table:rocky_planets}). Hence, at least for this specific subset of planets, a direct quantitative comparison between our results and the other mission studies is not immediately straightforward. In a future study,  we will further investigate the impact of the various values of $\eta_\oplus$ and their statistical and systematic uncertainties on the resulting detection yield  \citep[cf.][]{leger2015}.

Overall, it is clear that reanalyses of the \emph{Kepler} data, in combination with additional results from \emph{K2}, \emph{TESS}, and the upcoming \emph{PLAnetary Transits and Oscillations of stars (PLATO)} mission \citep{rauer2014} are extremely important to provide a more robust empirical basis for future updates of the analyses presented here. In particular \emph{PLATO} is designed to detect Earth-like planets in the HZ of solar-like stars and will improve our knowledge of the occurrence rate and formation mechanism of these targets. 
\section{Summary and conclusions}
\label{sec:summary}

We have presented new and more realistic results for the exoplanet detection yield of a space-based MIR nulling interferometer based on the exoplanet statistics observed by the \emph{Kepler} mission and targeting main-sequence FGKM stars within 20 pc of the Sun. Taking into account all major astrophysical noise terms and adding some margin for the not yet included instrumental noise effects (we require a S/N$\ge$7 for a detection), we find that an interferometer array consisting of four 2~m apertures and covering the 4--18.5~$\mu$m wavelength range with a total throughput of 5\% will yield, depending on the observing strategy, between $\approx$350 and $\approx$550 directly detected exoplanets with radii between 0.5 and 6 R$_\oplus$ within a 2.5-year search phase. Between $\approx$160 and $\approx$190 of these exoplanets have radii between 0.5 and 1.5 R$_\oplus$. As there is some freedom in how to assign observing time to the stellar targets, one can attempt to maximize the number of detected planets in certain areas of parameter space. We demonstrated this with two scenarios where either the total number of exoplanets or the number of rocky planets within the empirical HZ is maximized. The observing strategy must be adapted to the overall scientific objectives of the \emph{LIFE} mission since it influences the number (and types) of detected exoplanets. 

Keeping the instrument throughput fixed at 5\%, we find the number of detectable exoplanets to be a strong function of aperture size. In our current analysis, the wavelength range has a negligible impact on the exoplanet yield. We have shown that $\approx$25--45 rocky exoplanets within the empirical HZ of their host stars are expected to be detectable with four 2~m apertures, but this number could go up to $\approx$60--80 if the aperture size were increased to 3.5~m. In this case, the total number of detectable planets could go up to $\approx$770. With four 1~m apertures, the number of rocky exoplanets within the empirical HZ would be $\le$20 and the total detection yield $<$320. Irrespective of aperture size, the vast majority of rocky exoplanets orbiting within the empirical HZ are detected around M dwarfs. It will be important to further investigate if these planets could in principle possess atmospheres despite the high-energy UV flux and flaring activity these stars display. To further increase the number of detected rocky HZ planets around FGK stars, multiple visits per star during the search phase need to be considered in future work.

All numbers presented here (i.e., the total number of detected planets and the number of rocky planets within the HZ) are competitive with those predicted for current mission concepts searching for exoplanets in reflected light. 
Further studies investigating potential scientific and operational synergies between a reflected light and a thermal emission mission should be considered. Such efforts are particularly important for small temperate planets, such as EECs, because reflected light missions have a strong bias for detecting these objects primarily around solar-type stars, while \emph{LIFE} has a strong bias for planets around M stars. We note, however, that when comparing the numbers of detectable EECs with those predicted for the \emph{LUVOIR} and \emph{HabEx} missions, the underlying occurrence rates are not identical. The simulations presented here use lower values for EECs around FGK stars. This shows that care must be taken when comparing predicted detection yields of future missions, and additional efforts, such as obtaining new data and investigating new data analysis approaches, are needed to further refine the statistical occurrence rates that form the basis for all yield calculations.
 
Comparing the predicted primary discovery space of \emph{LIFE} with known exoplanets within 10 pc of the Sun shows that there are $>$40 objects, $\approx$15 of which have predicted radii $<$1.5 R$_{\oplus}$, that could be added to a target list today. To minimize the time that future exoplanet imaging space missions have to devote to an initial search phase, continuing ground- and space-based detection surveys is crucial.

We note that both the overall exoplanet detection yield and the observing time required to robustly characterize the atmospheric properties of rocky, temperate exoplanets with an MIR interferometer are strong functions of apertures size, which must be considered in future trade-off studies. The MIR regime is particularly rich in molecular absorption bands of the main constituents of terrestrial exoplanet atmospheres, including major biosignatures \citep[e.g.,][]{schwietermann2018,catling2018}. Also, thermal emission spectra provide more information about the atmospheric structure and allow for a more direct measurement of the planetary radius than reflected light data \citep[e.g.,][]{line2019}. The relatively high S/N (integrated over the full wavelength range) that most detectable planets in our simulations have suggests that decent estimates for their radii and effective temperatures -- and in some cases even rough SEDs -- seem possible. In this case, the data from a single-epoch observation obtained during the search phase will already provide crucial information for categorizing and prioritizing the planets for the follow-up characterization phase \citep[for further details, see][]{dannert2022}.
 
Our results show that when investigating and selecting future large exoplanet imaging space missions, for instance in the context of ESA's Voyage 2050 program, a concept such as \emph{LIFE} should be considered a serious contender and may be required to ultimately assess the habitability of exoplanets\footnote{See, e.g., "Exoplanet Strategy Report 2018" from the National Academies of Sciences, Engineering, and Medicine available at \url{https://www.nap.edu/catalog/25187/exoplanet-science-strategy}.}.
 Taking into account the heritage from the \emph{Darwin} and \emph{TPF-I} studies and more recent progress based on various local activities, new coordinated efforts to further understand and increase the technological readiness level of key components have started as part of the \emph{LIFE} initiative \citep[e.g.,][]{gheorghe2020}.

\label{sec:discussion-and-conclusion}
    \begin{acknowledgements}
    We thank the anonymous referee for a critical and constructive review of the original manuscript which helped improve the quality of the paper. This work has been carried out within the framework of the National Centre of Competence in Research PlanetS supported by the Swiss National Science Foundation. SPQ, EA and HSW. acknowledge the financial support of the SNSF. SK acknowledges funding from an ERC Starting Grant (grant agreement No. 639889). DE acknowledges funding from the European Research Council (ERC) under the European Union's Horizon 2020 research and innovation programme (project {\sc Four Aces}; grant agreement No 724427). JL-B acknowledges financial support received from ``la Caixa'' Foundation (ID 100010434) and the European Union's Horizon 2020 research and innovation programme under the Marie Sklodowska-Curie grant agreement No 847648, with fellowship code LCF/BQ/PI20/11760023. RA is a Trottier Postdoctoral Fellow and acknowledges support from the Trottier Family Foundation. This work was supported in part through a grant from FRQNT. TL acknowledges funding from the Simons Foundation (SCOL award No. 611576). Part of this work was conducted at the Jet Propulsion Laboratory, California Institute of Technology, under contract with NASA. This research has made use of the SIMBAD database, operated at CDS, Strasbourg, France, and of the Washington Double Star Catalog maintained at the U.S. Naval Observatory. This research has made use of the NASA Exoplanet Archive, which is operated by the California Institute of Technology, under contract with the National Aeronautics and Space Administration under the Exoplanet Exploration Program.
    This research has made use of the following Python packages: 
    \texttt{astropy} \citep{Astropy_2013, Astropy_2018},
    \texttt{matplotlib} \citep{Hunter_2007}, 
    \texttt{numpy} \citep{VanDerWalt_2011},
    and
    \texttt{scipy} \citep{Virtanen_2020}.\newline\newline
    
    \emph{Author contributions:} SPQ initiated the project, devised the analyses and wrote the manuscript. MO and FD wrote the \textsc{LIFESim} tool and created the figures. AGh and EF contributed to the \textsc{LIFESim} tool. JK simulated the exoplanet populations. FM created the \emph{LIFE} target star catalog. 
    All authors discussed the results and commented on the manuscript.
    \end{acknowledgements}


\bibliographystyle{aa}

\begin{appendix}
\section{Stellar sample}
\label{sec:appendix_stars}

The stellar sample was compiled from querying the SIMBAD database\footnote{\url{http://simbad.cds.unistra.fr/simbad/}} \citep{wenger2000} for objects within 20 pc. We removed the substellar objects (planets and brown dwarfs) using the object type parameter. For the remaining stellar objects we focused on main-sequence stars as indicated by the luminosity class of the objects (in case no luminosity class was given we assumed the objects were main-sequence objects). Based on the spectral type of the object we then assigned effective temperature, radius and mass using the relation published in Table 5 of \cite{pecaut2013}, which is based on empirical data. In order to assess which objects are members of binary or multiple systems we used SIMBAD's hierarchical link. This feature connects an object with its parent and child objects.
To keep the complexity of our multiple star sample low we decided to only use wide binaries where planetary orbits are possible around both components and thus are most similar to orbits around single
stars. We therefore excluded all systems with more than two components and also those that had binary subtypes as SIMBAD object type. This was necessary because not all binary components had their own SIMBAD entry (for example, if the objects cannot be observed individually due to
too small separations). In that case the system has no children and we cannot distinguish it from a single star. We also removed objects with incomplete information for the stellar parameters (this step also included binaries, where one component was not listed as a main-sequence star). To obtain an estimate for the separation between the remaining binary systems, we cross-matched the system position with the Washington Visual Double Star Catalog \citep[WDS;][]{mason2001} by drawing a circle with a radius of 1 arcsec around the system position. If the position of a WDS object was close enough
to fall within the circle we assumed that the two objects were the same physical system. In the case the WDS catalog had
more components listed per system than SIMBAD (e.g., because of background stars) we took  the separation between the two main components. The separation is normally given for two different observations. We  kept the smaller one as the assumed physical separation of the binary system. In order to ensure that all stellar components of the remaining binary systems could harbor stable planetary systems between 0.5 and 10 AU we used the stability criterion from \citet{holman1999} that takes into account the stellar masses, their separation and the eccentricity of the binary orbit (which we assumed to be zero). Systems that did not fulfill this stability criterion were removed. 

\begin{figure}[th!]
    \centering
    \includegraphics[width=\linewidth]{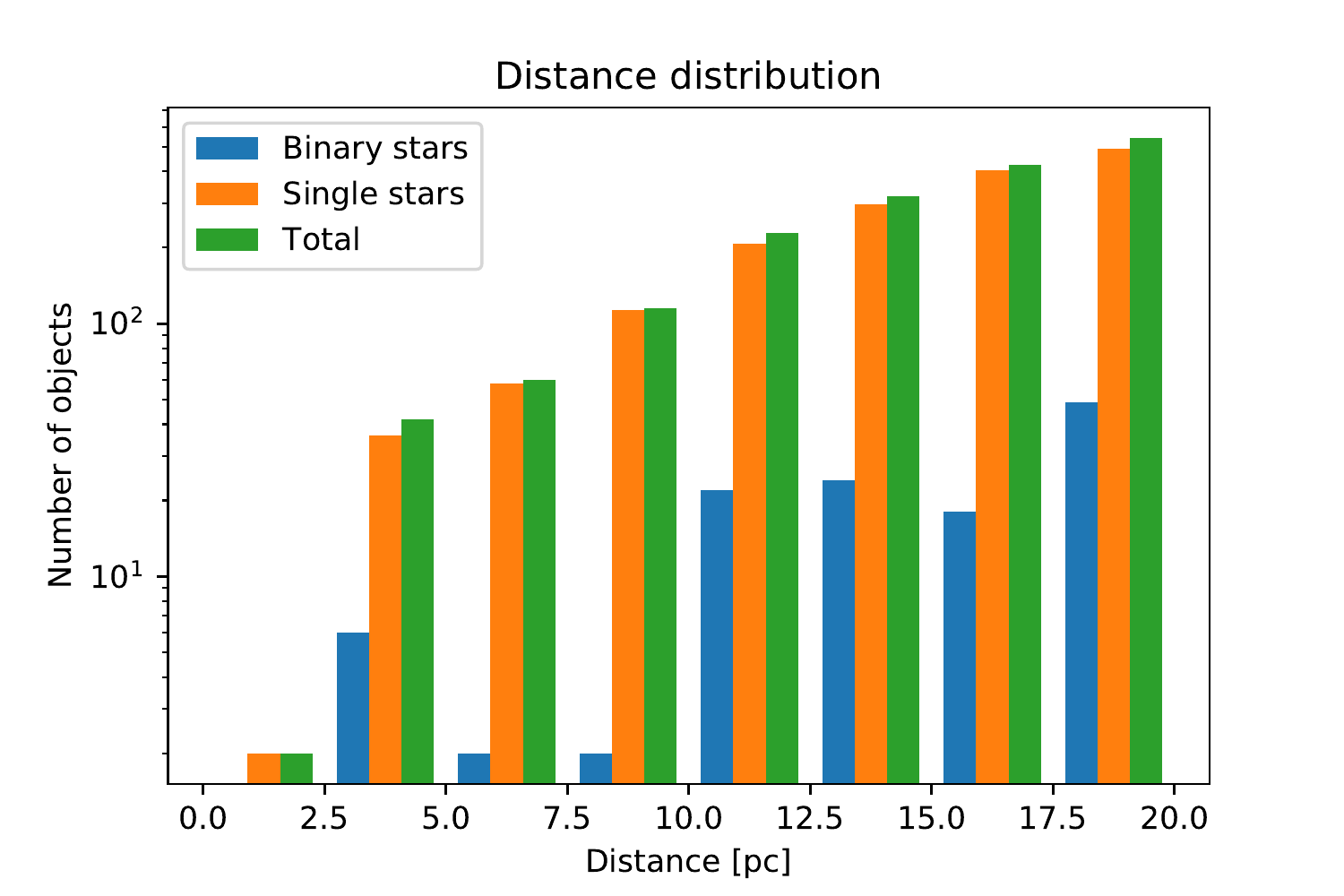}
    \includegraphics[width=0.9\linewidth]{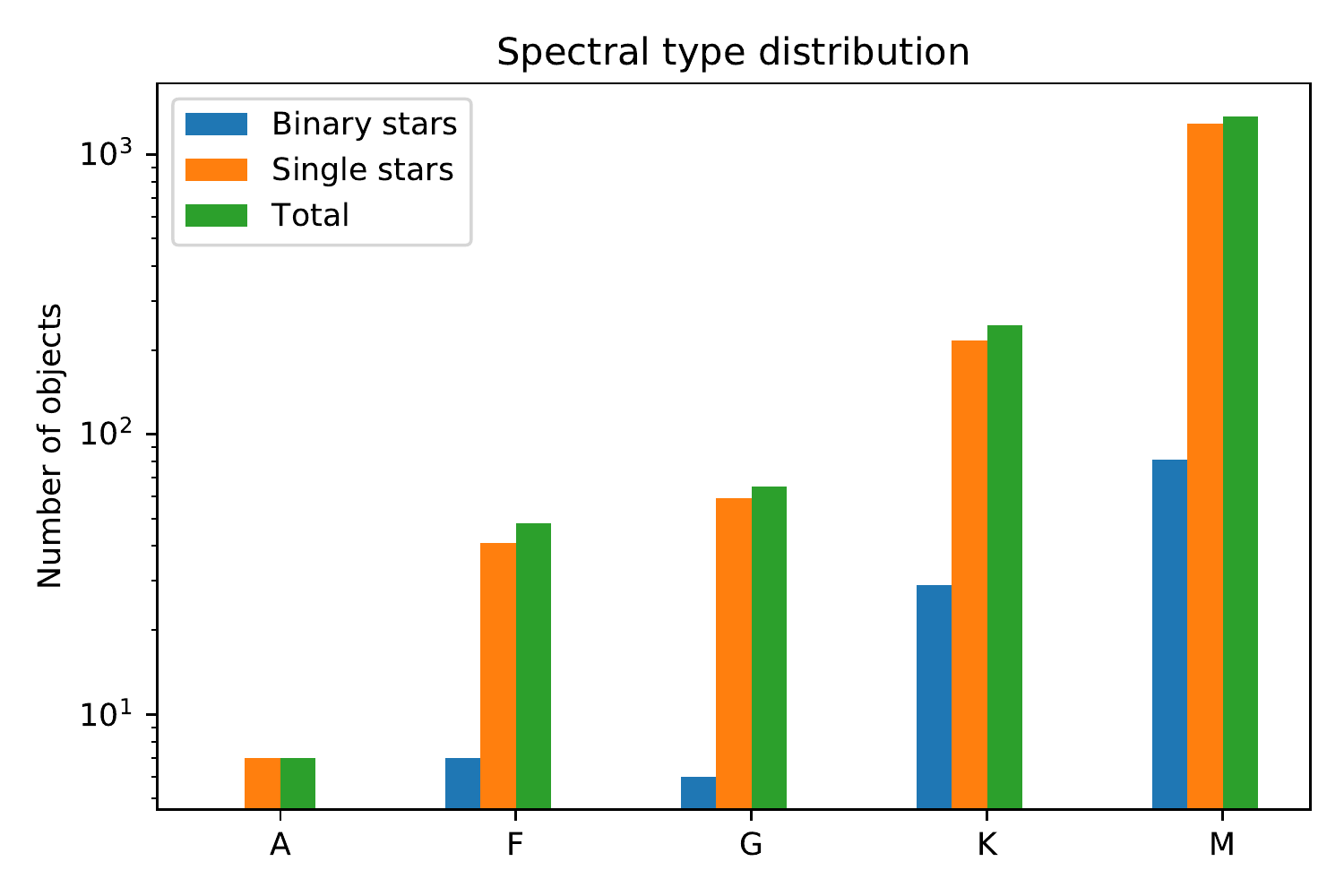}
    \caption{Properties of target stars considered in our study. Top panel: Distance distribution of stars from the stellar input catalog in bins of 2.5 pc. Single stars are shown in orange, wide binaries (see text) in blue, and the sum in green. Bottom panel: Spectral type distribution of stars from the stellar input catalog. Colors are the same as above.
    }
    \label{fig:stellar_sample}
\end{figure}

The catalog that was the basis for our simulations consists of 1732 stars in total (123 wide binary components and 1609 single stars). The distance and spectral type distributions are shown in Fig.~\ref{fig:stellar_sample}. The catalog is available upon request. We are continuously improving the catalog and plan for an updated version to include the results from \emph{Gaia} data release 3. 

\section{Technological readiness}
\label{sec:appendix_technologies}

The readiness of key technologies relevant for a space mission such as \emph{LIFE} are summarized in \citet{defrere2018a} and \citet{quanz2021}, but given their importance for the mission success, we discuss three aspects in the following:
  
\begin{description}
    \item{\textbf{Nulling interferometry:}}
    It is important to understand that the \emph{LIFE} measurement principle has been demonstrated successfully in the lab at ambient temperatures \citep{martin2012}. One of the next steps is to build a corresponding experiment, but fully under cryogenic conditions, for a broad wavelength coverage and with sensitivity as one of the key drivers (as needed for a space mission). This is underway in the form of the \emph{Nulling Interferometry Cryogenic Experiment} at ETH Zurich \citep{gheorghe2020}. In addition, new nulling instruments have been proposed for the \emph{Very Large Telescope Interferometer}, including one working in the 3--5~$\mu$m range \citep{defrere2018b}. These efforts join previous successful projects related to N-band ground-based nulling interferometry, with the Keck Nuller \citep{colavita2009} and the LBTI \citep{hinz2014}. For \emph{LIFE} it will be important to leverage the experience from these projects and realize possible synergies.
    
    \item{\textbf{Autonomous formation flying:}}
    To meet the assumptions made for our simulations, a high level of autonomous formation flying of all spacecraft will be required. Specifically, the baselines of the array should be rearranged for every new target star in order to maximize the transmission of photons from a specific distance from the star, and, during an observation, the array should rotate in order to modulate the signal from potential exoplanets. ESA's Proba-3 mission (current launch date: 2023) will demonstrate many critical aspects of formation flying for \emph{LIFE}. Proba-3 will feature two cube-sized spacecraft with lengths of $\sim$1.5 m and masses of 200-300 kg and will maintain a virtual ``rigid structure'' with millimeter and arcsecond relative precision. In addition, Proba-3 aims specifically at demonstrating manoeuvres relevant for \emph{LIFE}, such as station-keeping at distances from 25 m up to 250 m, approaching and separating in formation without losing millimeter precision, repointing the formation, and the combination of station-keeping, resizing, and re-targeting manoeuvres. Details can be found in \cite{penin2020}. We note that some relevant technology related to formation flying was developed in both France and Germany in the context of the \emph{Darwin} mission and has been flying on the Swedish \emph{Prototype Research Instruments and Space Mission technology Advancement} space experiment\footnote{\url{https://earth.esa.int/web/eoportal/satellite-missions/p/prisma-prototype}}.
    
    \item{\textbf{High quantum-efficiency, low-noise MIR detectors:}}
    The low photon rate of exoplanets (cf. Fig.~\ref{fig:noise_model}) and the need to integrate for many hours, if not days, will put strong requirements on the detector technology in terms of quantum efficiency (in our simulations we assumed 70\%; see Sect.~2.3), low read-out noise and dark current, and high stability. In addition, a wavelength coverage of at least $\sim$4--18.5~$\mu$m is required based on atmospheric retrieval analyses quantifying the characterization potential of \emph{LIFE} for Earth-like atmospheres \citep[cf.][]{konrad2022}. In this overall context, the technology development plan for the 5.9-meter \emph{Origins Space Telescope}\footnote{\url{https://asd.gsfc.nasa.gov/firs/docs/OriginsVolume2TechDevelopmentPlanREDACTED.pdf}}, another mission concept proposed in the context of NASA's 2020 astrophysics decadal survey \citep{ost2019}, is of great importance as its proposed MISC instrument \citep{sakon2018} would cover the same wavelength range as \emph{LIFE}. While a detailed noise budget and requirements breakdown for \emph{LIFE} are still being worked on, it is clear that two types of detector technologies can be considered (and eventually traded): HgCdTe detectors (as, for instance, used in NEOCam) and Si:As detectors (as used in \emph{JWST/MIRI}). Currently, the 3-11 $\mu$m range is better covered by HgCdTe than Si:As since the latter is basically transparent below 10 $\mu$m wavelengths. First efforts to go up to 15 $\mu$m with HgCdTe detectors  were already reported \citep{cabrera2020} and new arrays with cutoff wavelength $>$16 $\mu$m have been grown, hybridized, and packaged and are undergoing testing (cf. \emph{OST} Technology Development Plan). A key parameter for HgCdTe detectors will be the achievable dark current. 
 For Si:As detectors, in addition to potential challenges related to dark current and 1/f noise, a general problem is the question of availability. Industrial fabricators that have built these detectors in the past (e.g., for \emph{Spitzer}, \emph{WISE}, and \emph{JWST}) have stopped their production of low background detectors in the relevant wavelength range and it is unclear under what conditions and on what timescales restarting the production would be an option. 
\end{description}

\newpage
\section{Additional figures for non-reference cases}
\label{sec:appendix_plots}
\begin{figure*}[t!]
    \centering
    \includegraphics[width=0.4\linewidth]{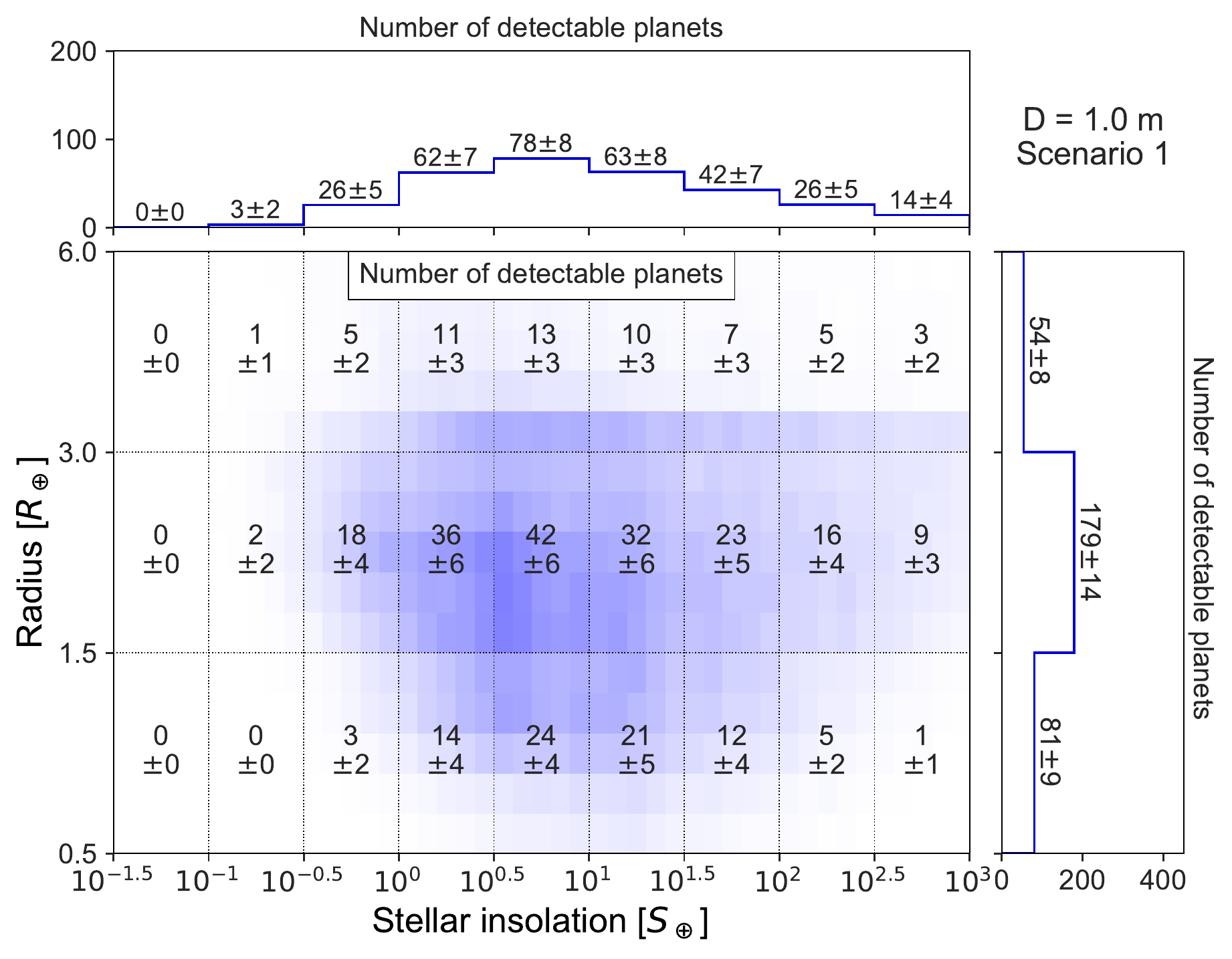}
    \includegraphics[width=0.4\linewidth]{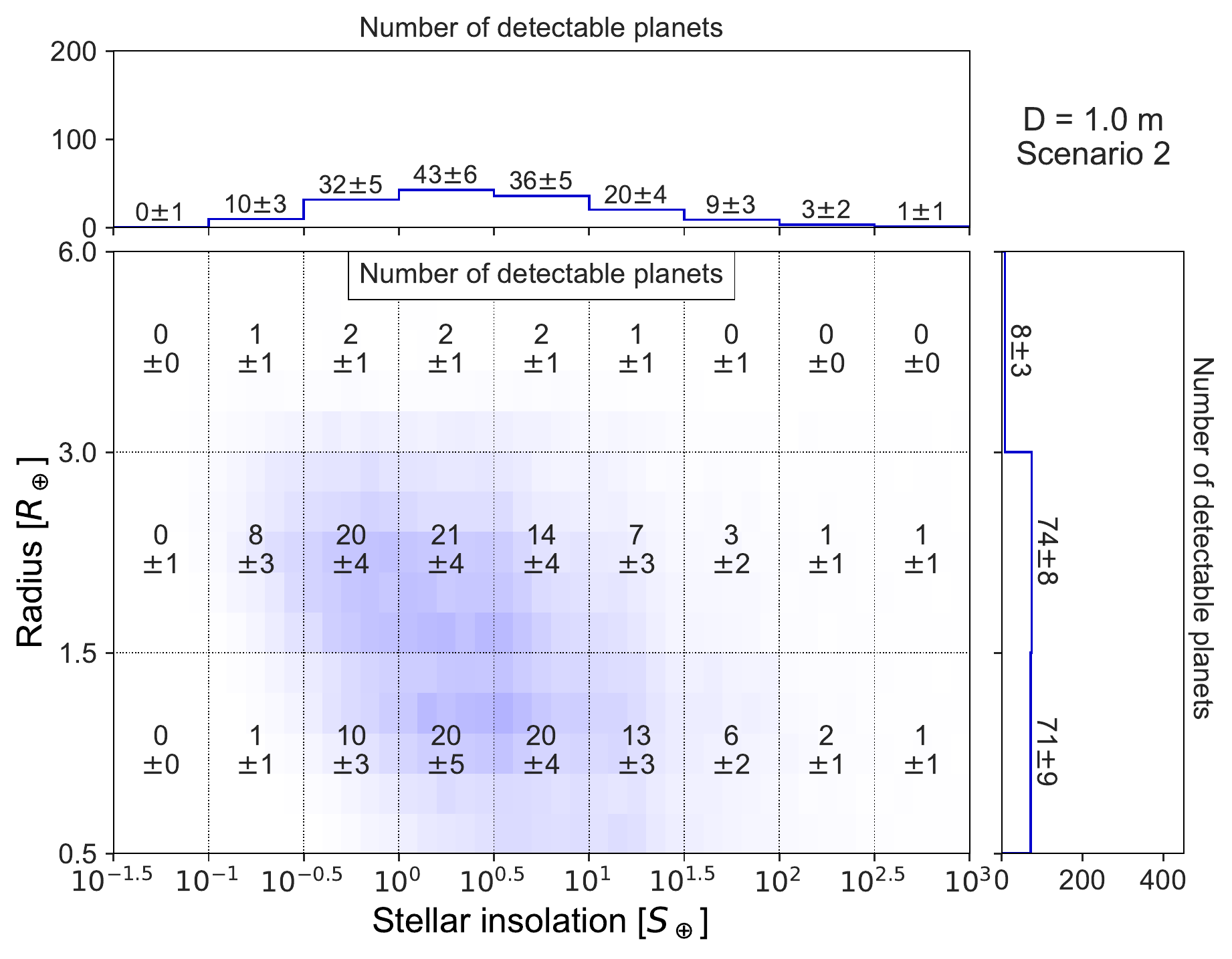}
    \caption{Same as Fig.~\ref{fig:baseline_yields_grid}, but now for $D=1.0$ m.}
    \label{fig:1m_yields_grid}
\end{figure*}

\begin{figure*}[t!]
    \centering
    \includegraphics[width=0.4\linewidth]{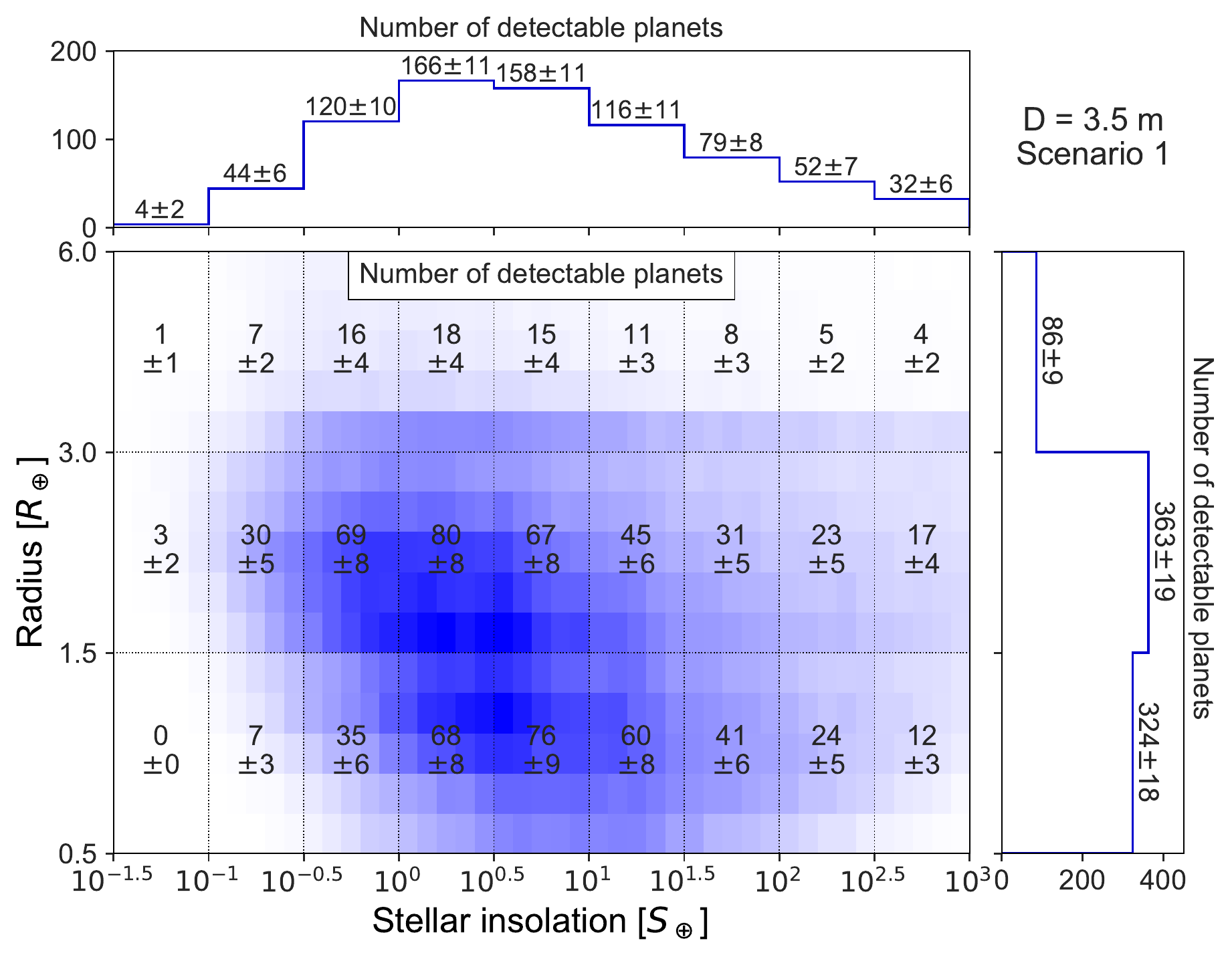}
    \includegraphics[width=0.4\linewidth]{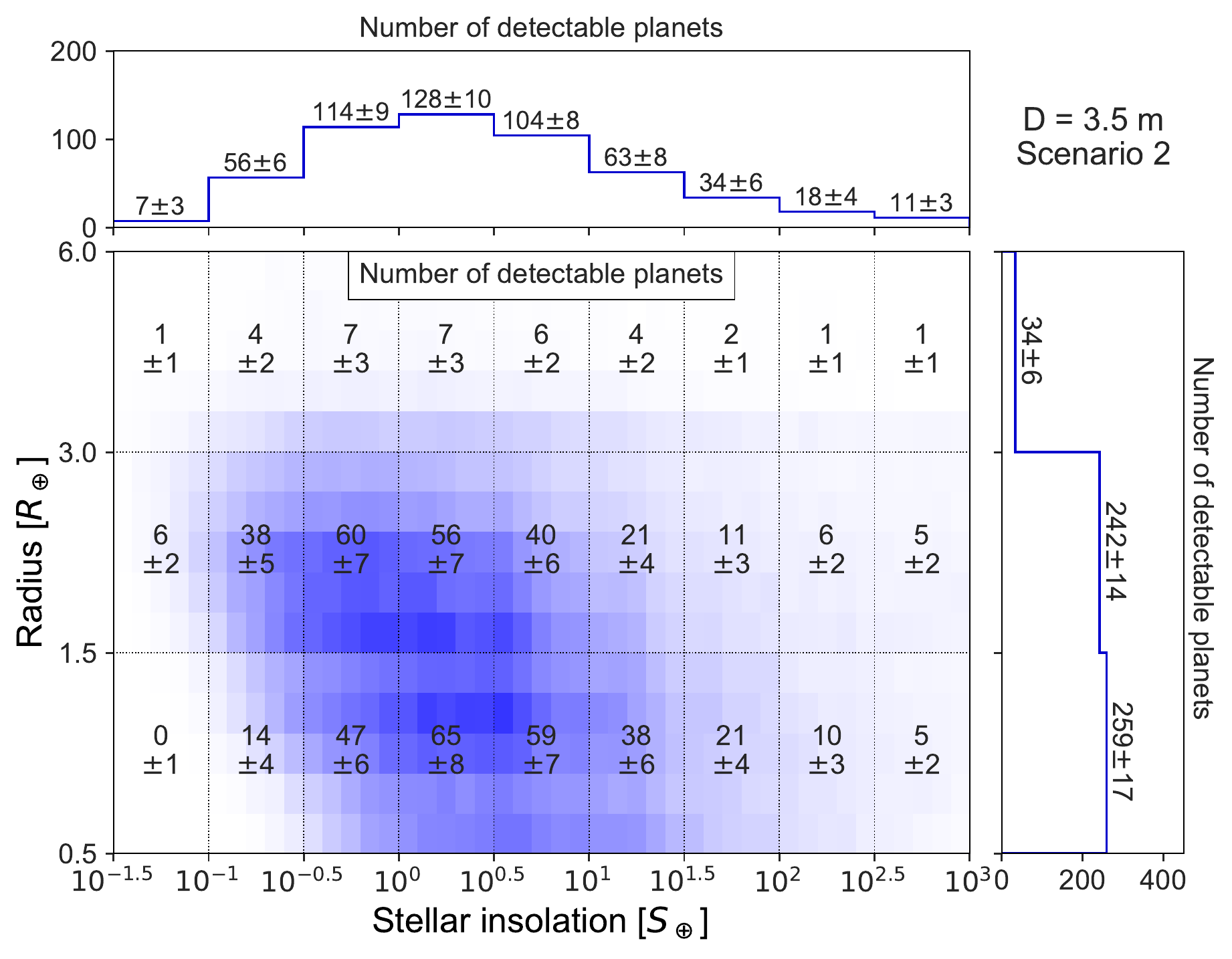}
    \caption{Same as Fig.~\ref{fig:baseline_yields_grid}, but now for $D=3.5$ m.}
    \label{fig:3.5m_yields_grid}
\end{figure*}

\begin{figure*}[h!]
    \centering
    \includegraphics[width=0.4\linewidth]{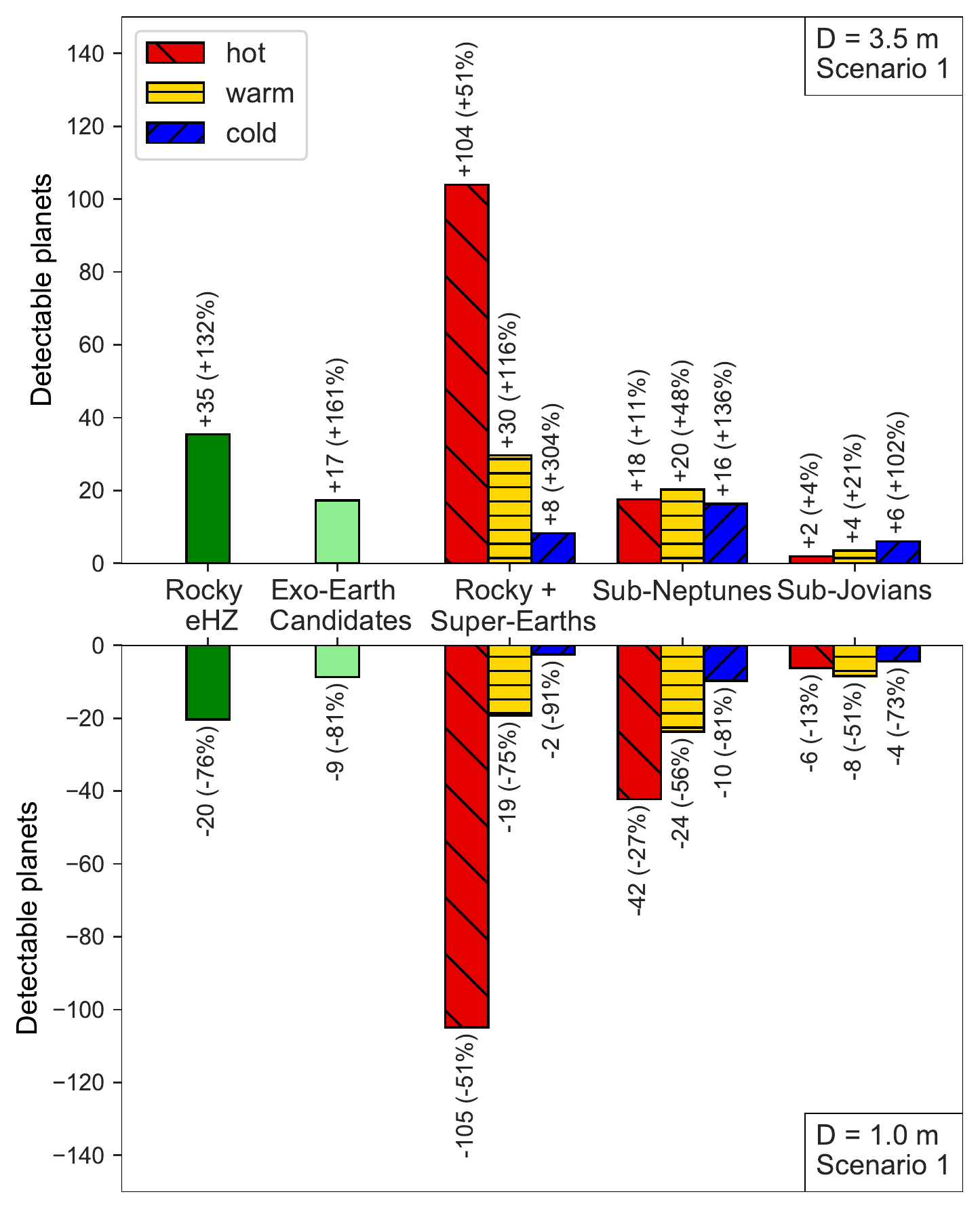}
    \includegraphics[width=0.4\linewidth]{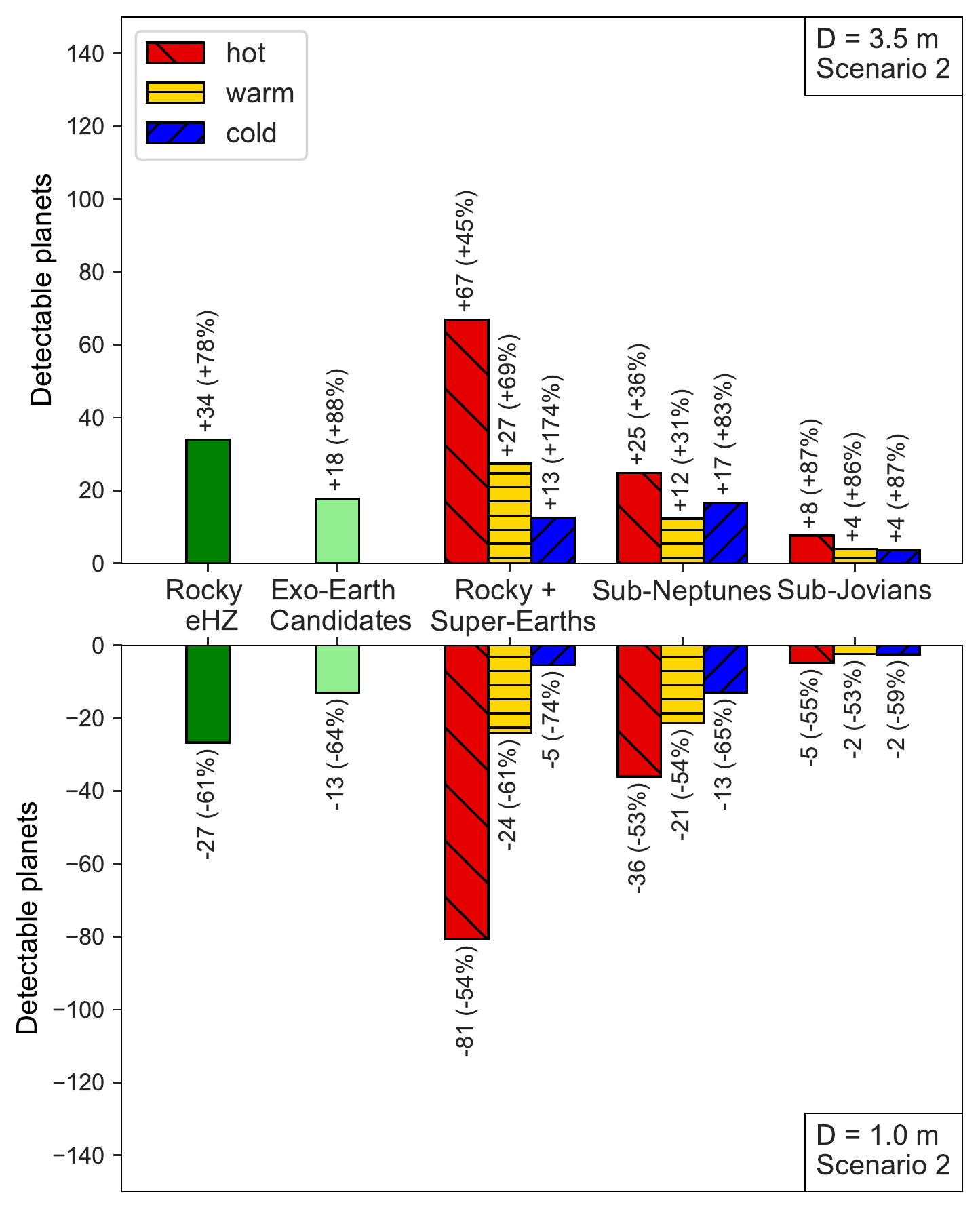}
    \caption{Impact of the aperture size on the exoplanet detection yield. The numbers are relative to those shown for the reference scenarios with $D=2$~m in Fig.~\ref{fig:baseline_yields_bars}. Left: Scenario 1, with $D=3.5$ m in the top panel and $D=1.0$ m in the bottom panel. Right: Scenario 2, with $D=3.5$ m in the top panel and $D=1.0$ m in the bottom panel.}
    \label{fig:LIFE_yield_aperture}
\end{figure*}

\begin{figure*}[t!]
    \centering
    \includegraphics[width=0.4\linewidth]{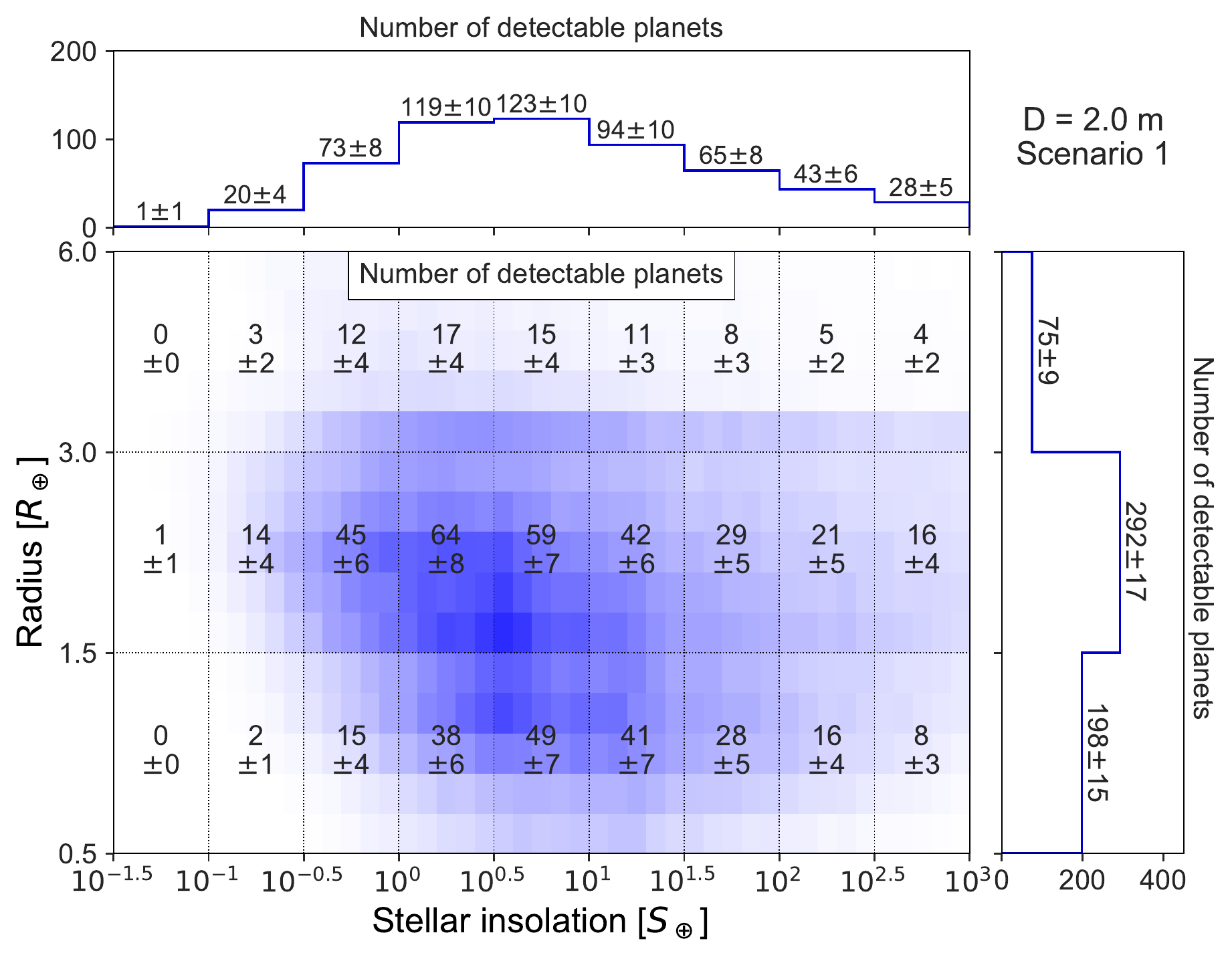}
    \includegraphics[width=0.4\linewidth]{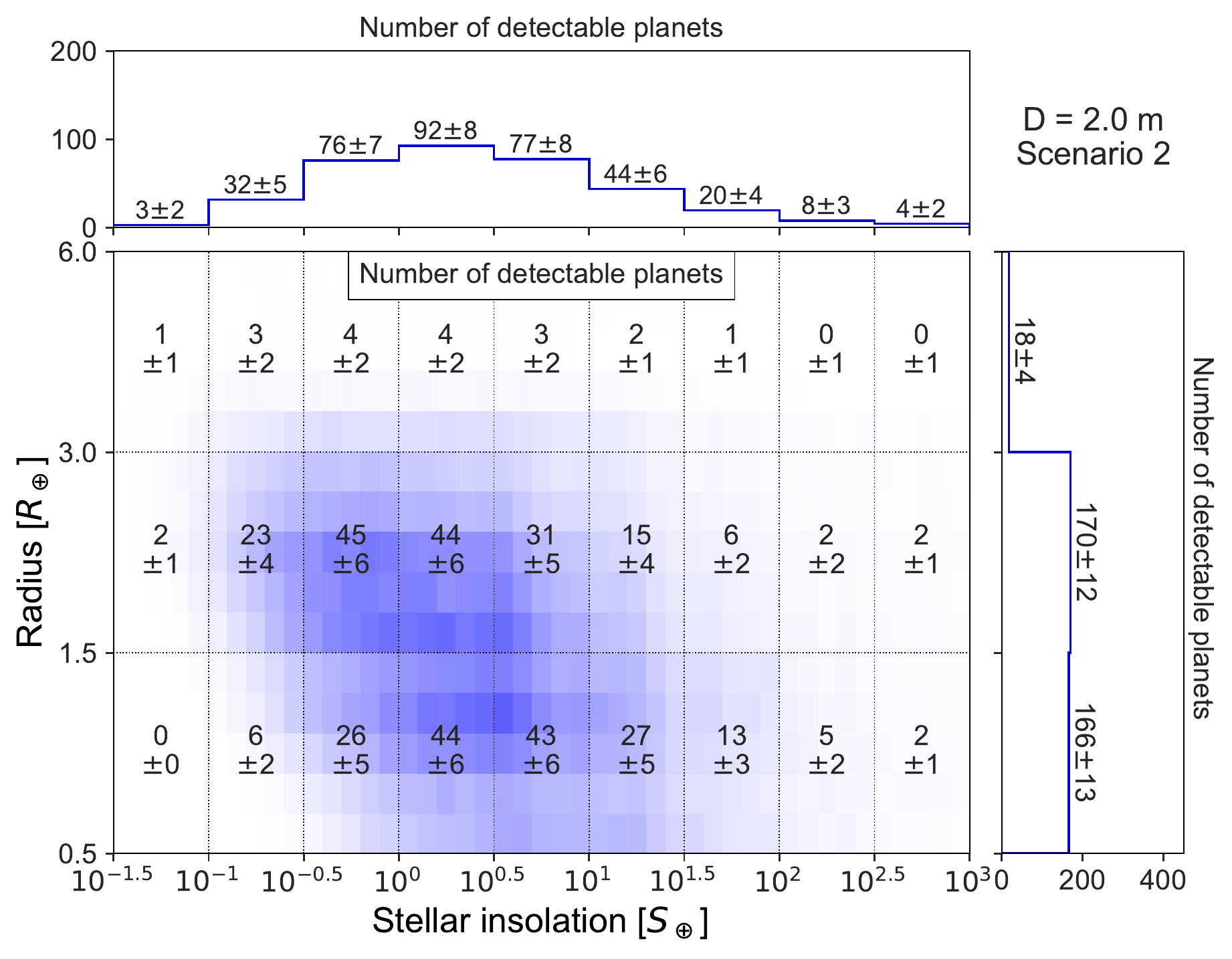}
    \caption{Same as Fig.~\ref{fig:baseline_yields_grid}, but now for a wavelength range of 3--20 $\mu$m.}
    \label{fig:3-20_yields_grid}
\end{figure*}

\begin{figure*}[t!]
    \centering
    \includegraphics[width=0.4\linewidth]{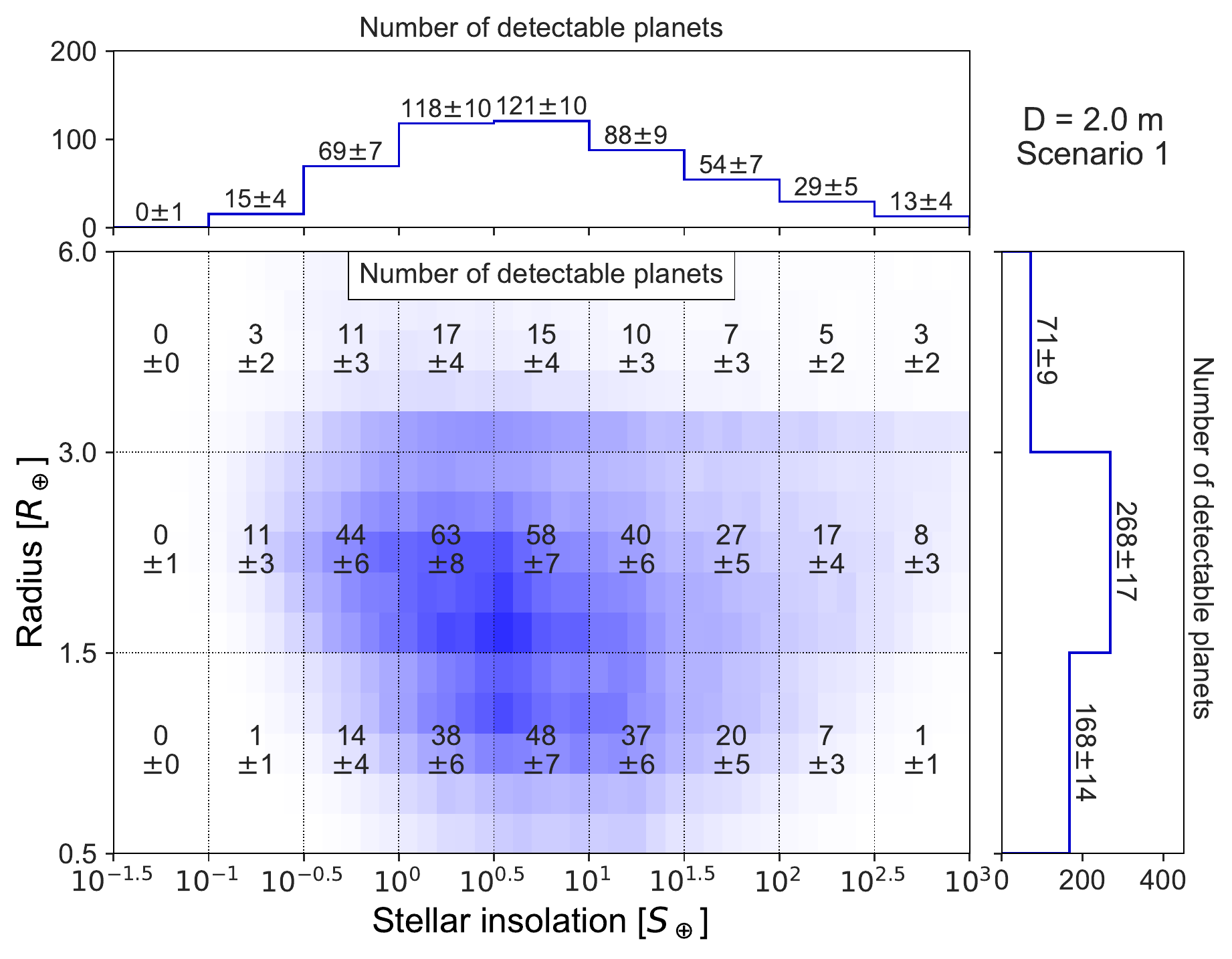}
    \includegraphics[width=0.4\linewidth]{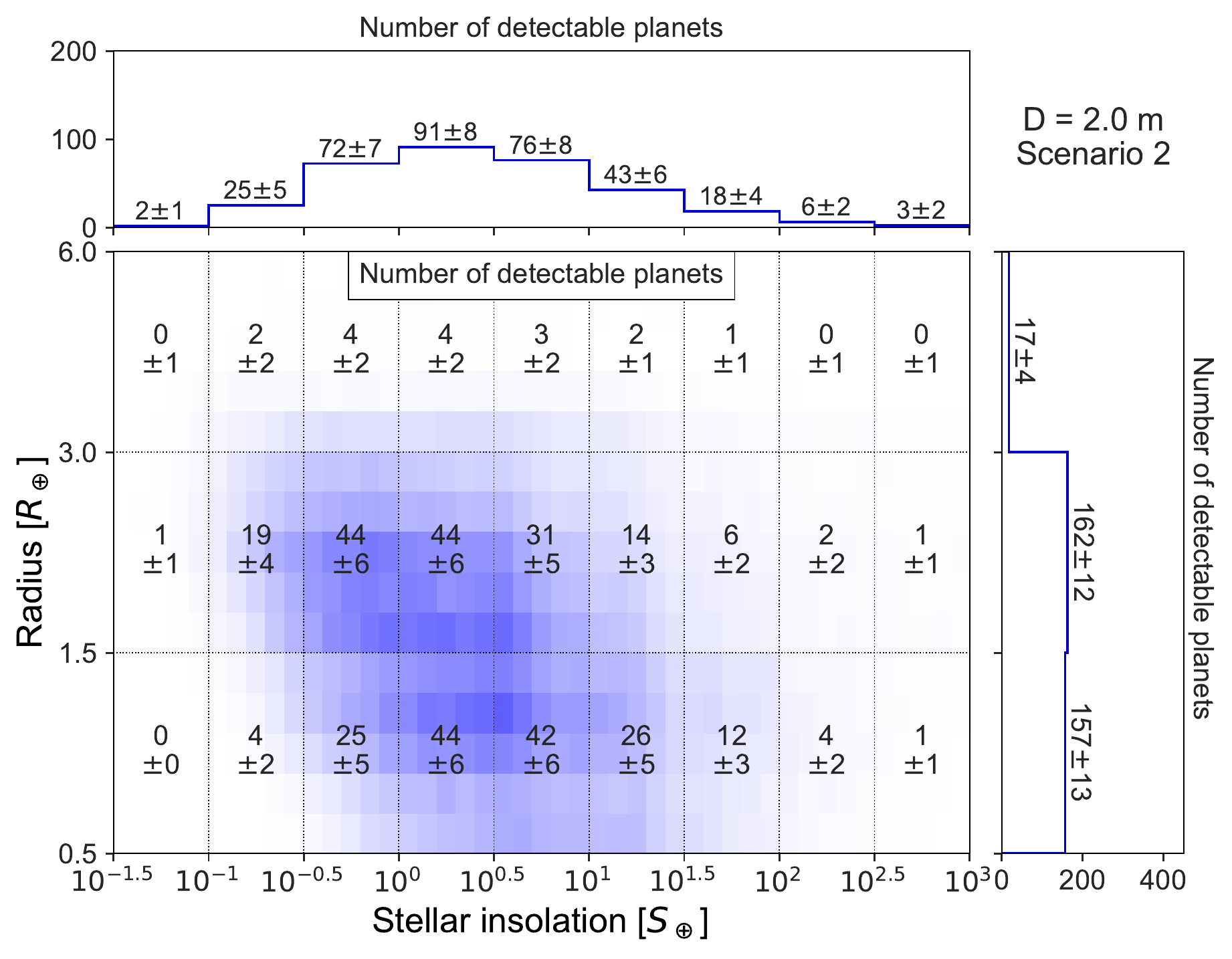}
    \caption{Same as Fig.~\ref{fig:baseline_yields_grid}, but now for a wavelength range of 6--17 $\mu$m.}
    \label{fig:6-17_yields_grid}
\end{figure*}

\begin{figure*}[h!]
    \centering
    \includegraphics[width=0.4\linewidth]{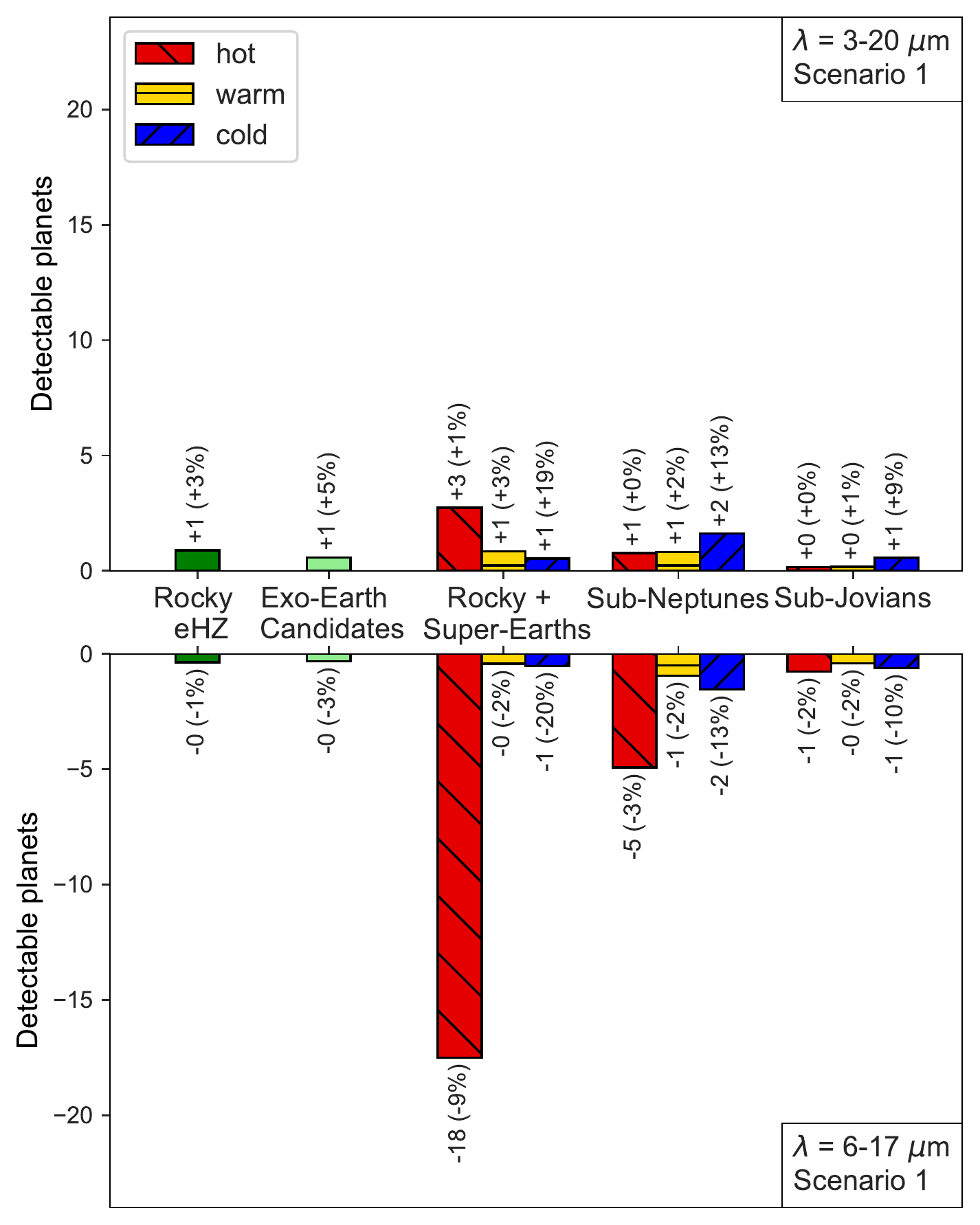}
    \includegraphics[width=0.4\linewidth]{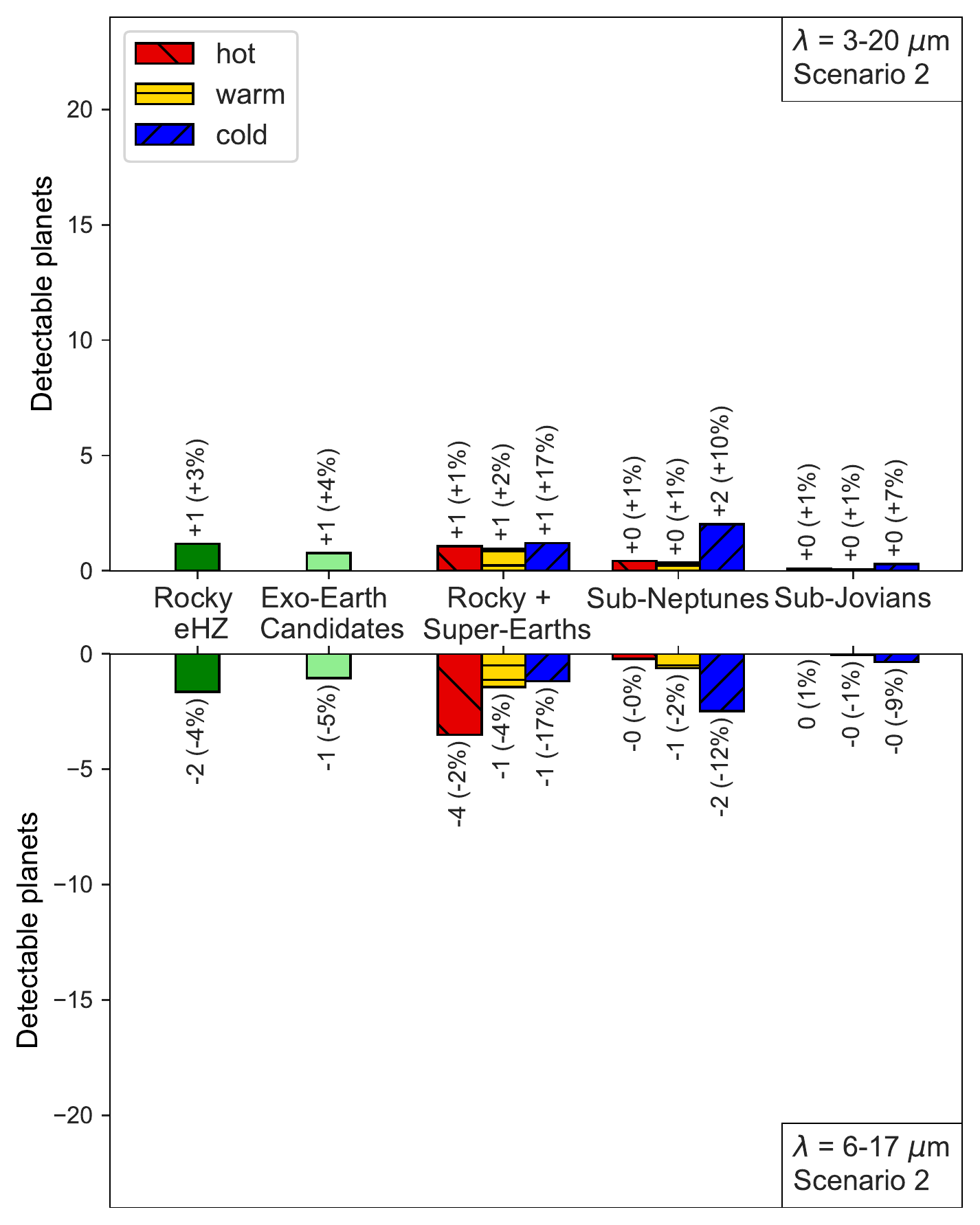}
    \caption{Impact of the wavelength coverage on the exoplanet detection yield. The numbers are relative to those shown for the reference scenarios with $\lambda=4-18.5\,\mu m$ in Fig.~\ref{fig:baseline_yields_bars}. Left: Scenario 1, with $\lambda=3 - 20\,\mu m$ in the top panel and $\lambda=6-17\,\mu m$ in the bottom panel. Right: Scenario 2, with $\lambda=3 - 20\,\mu m$ in the top panel and $\lambda=6-17\,\mu m$ in the bottom panel.}
    \label{fig:LIFE_yield_lambdarange}
\end{figure*}

\begin{figure*}[h!]
    \centering
    \includegraphics[width=0.48\linewidth]{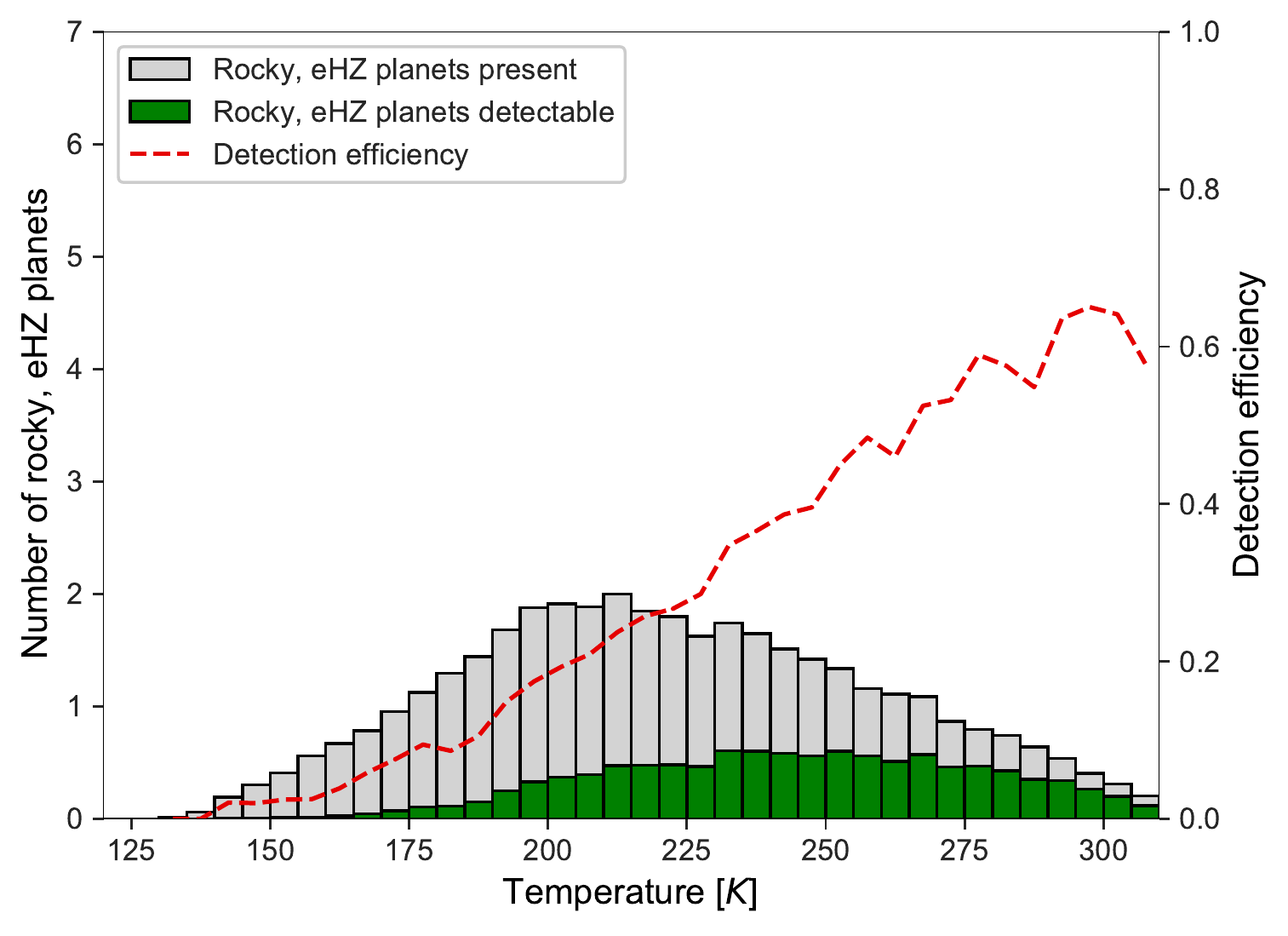}
    \includegraphics[width=0.48\linewidth]{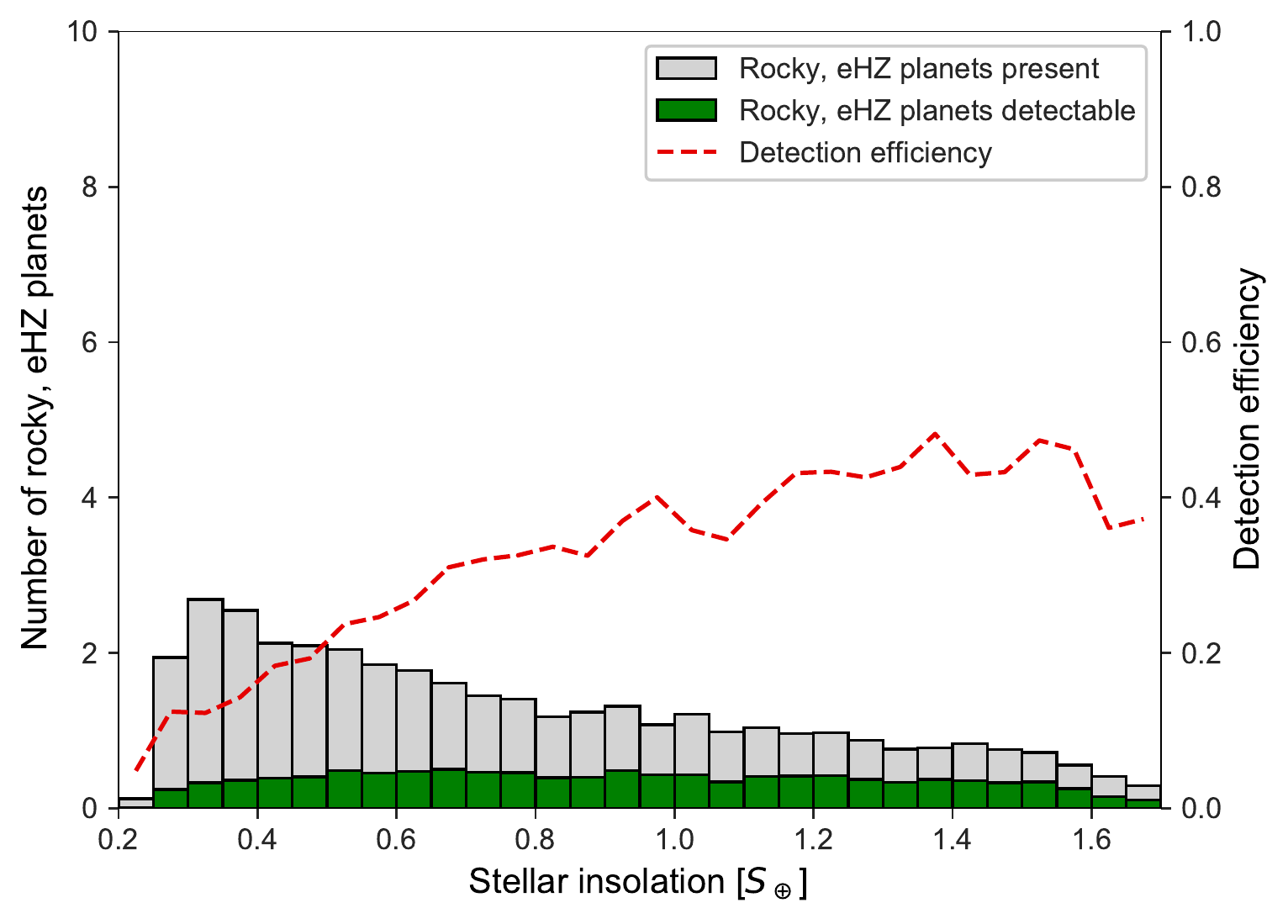}
    \caption{Same as Fig.~\ref{fig:hab_planets_efficiency}, but ignoring M stars and only considering FGK stars.}
    \label{fig:detection_efficiency_FGK}
\end{figure*}

Figures~\ref{fig:1m_yields_grid}, \ref{fig:3.5m_yields_grid}, \ref{fig:LIFE_yield_aperture}, \ref{fig:3-20_yields_grid}, \ref{fig:6-17_yields_grid}
, and \ref{fig:LIFE_yield_lambdarange} show the expected detection yield for the non-reference cases. Figure~\ref{fig:detection_efficiency_FGK} shows the detection efficiency for FGK-dwarf host stars (i.e., ignoring M dwarfs). 

\section{Distribution of noise contributions for detected planets in the reference case scenarios}
\begin{figure*}
    \centering
    \includegraphics[width=0.9\linewidth]{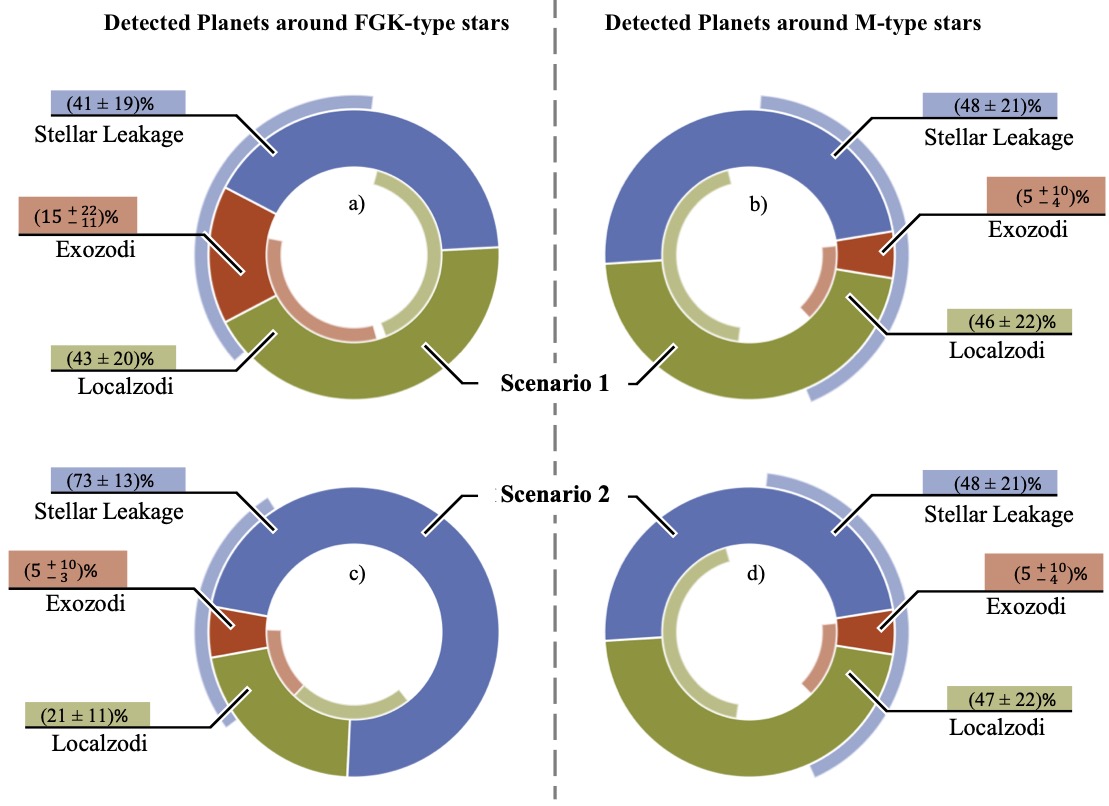}
    \caption{Donut charts illustrating how the main noise terms considered in our simulations (stellar leakage and noise from the thermal emission of the local and exozodiacal dust disks) contribute to the overall noise budget for the detected planets. The left column shows the results for planets detected around FGK stars and the right column for planets detected around M stars. The top row is for reference case scenario 1 and the bottom row for reference case scenario 2. The numbers correspond to the mean relative contribution of the various noise terms to the total noise per detected planet averaged over all planets. The quoted uncertainties are the corresponding standard deviations (graphically indicated by the colored arcs inside and outside of the donuts). For M stars, noise from exozodiacal dust disks is basically negligible, and stellar leakage and noise from the zodiacal dust disk contribute equally to the total noise in both scenarios. This trend is generally the same for FGK stars in scenario 1, even though the relative share of exozodi noise is larger. For scenario 2, however, stellar leakage clearly dominates the noise budget of planets around FGK stars.}
    \label{fig:noise_contributions}
\end{figure*}

Figure~\ref{fig:noise_contributions} summarizes the contribution of the main noise terms to the overall noise budget for detected planets orbiting FGK stars or M stars.
\label{sec:appendix_plots2}
\end{appendix}

\end{document}